\documentclass[lettersize,journal]{IEEEtran}
\usepackage{makecell}
\usepackage{amsmath,amsfonts}
\usepackage{amssymb}
\usepackage{algorithmic}
\usepackage{algorithm}
\usepackage{array}
\usepackage[caption=false,font=normalsize,labelfont=sf,textfont=sf]{subfig}
\usepackage{textcomp}
\usepackage{stfloats}
\usepackage{url}
\usepackage{verbatim}
\usepackage{graphicx}
\usepackage{cite}
\usepackage{bbm}
\usepackage[table,xcdraw]{xcolor}
\usepackage{multirow}
\usepackage{caption}
\usepackage{footmisc}
\usepackage{hyperref} 

\usepackage[most]{tcolorbox}
\newcounter{findingcounter}
\renewcommand{\thefindingcounter}{\Roman{findingcounter}} 

\definecolor{boxframe}{HTML}{FFE7C8}

\newtcolorbox{promptbox}{
    colback=boxframe,
    colframe=gray!30,
    sharp corners,
    boxrule=0.5pt,
    left=5pt, right=5pt,
    top=5pt, bottom=5pt,
    before skip=10pt,
    after skip=10pt
}

\newtcolorbox{summarybox}{
    colback=gray!10,
    colframe=gray!30,
    sharp corners,
    boxrule=0.5pt,
    left=5pt, right=5pt,
    top=5pt, bottom=5pt,
    before skip=10pt,
    after skip=10pt
}

\newcommand{\summary}[1]{
    \stepcounter{findingcounter}
    \begin{summarybox}
        \textbf{Summary \thefindingcounter:} #1
    \end{summarybox}
}

\newcommand{\contributions}[1]{%
    \begin{summarybox}%
        \textbf{Contributions:} #1%
    \end{summarybox}%
}

\begin{document}
\title{Fine-grained Approaches for Confidence Calibration of LLMs in Automated Code Revision}

\author{Hong~Yi~Lin,~\IEEEmembership{Member,~IEEE,}
        Chunhua~Liu,~\IEEEmembership{Member,~IEEE,}
        Haoyu~Gao,~\IEEEmembership{Member,~IEEE,}
        Patanamon~Thongtanunam,~\IEEEmembership{Member,~IEEE,}
        Christoph~Treude,~\IEEEmembership{Member,~IEEE,}
        }

\maketitle

\begin{abstract}
In today’s AI-assisted software engineering landscape, developers increasingly depend on large language models (LLMs) that are highly capable, yet inherently imperfect.
The tendency of these models to produce incorrect outputs can interrupt a developer’s workflow and reduce developer productivity.
One approach to improving developer interactions with these imperfect models is to provide greater transparency at the instance-level.
This is often achieved by complementing the model generated output with a well-calibrated confidence score that faithfully reflects the likelihood correctness.
Such information allows users to make immediate decisions regarding output acceptance, abstain error-prone outputs, and better align their expectations with the model’s capabilities.
Since post-trained LLMs do not inherently produce well-calibrated confidence scores, researchers have developed post-hoc calibration methods, with global Platt-scaling of sequence-level confidence scores proving effective in many generative software engineering tasks but remaining unreliable or unexplored for automated code revision tasks such as program repair, vulnerability repair, and code refinement.
We hypothesise that the coarse-grained nature of this conventional method makes it ill-suited for automated code revision tasks, where correctness is often determined by local edit decisions and miscalibration can be sample-dependent, thereby motivating fine-grained confidence calibration.
To address this challenge, our study proposes local Platt-scaling applied separately to three different fine-grained confidence scores.
Through our experiments across three separate tasks and correctness metrics, as well as 14 different models of various sizes, we find that fine-grained confidence scores consistently achieve lower calibration error across a broader range of probability intervals compared to sequence-level confidence scores, and this effect is further amplified when global Platt-scaling is replaced with local Platt-scaling.
Our replication package can be found here: \url{https://github.com/hongyi-tom/FinegrainedCalibrationACR}

\end{abstract}


\section{Introduction}
\IEEEPARstart{L}{arge} Language Models have become an indispensable tool for today’s software developer.
Their proficiency in both natural language and source code allows them to be instructed to perform a wide range of daily code revision activities, such as repairing buggy programs~\cite{repairllama}, patching security vulnerabilities~\cite{appatch}, and resolving code review comments~\cite{chatgpt_acr}.
Whilst these models have shown the potential to enhance developer productivity, their susceptibility to generating incorrect outputs can also create friction that limits their synergy with human developers~\cite{Liang0M24}.
More specifically, incorrect code generations can disrupt a developer's workflow due to the time wasted in debugging and rewriting implementations~\cite{Vaithilingam0G22}.
When occurring repeatedly, these inaccuracies can even induce frustration, annoyance, and distrust, eventually causing some developers to abandon the tool altogether~\cite{montes2025emotional}.

To this end, prior research have shown that interactions with imperfect models can be enhanced through increased model transparency and information integrity at the instance-level~\cite{MarusichB0K24, BenzR23, montes2025emotional,fat0004CF024}. 
This is often operationalised as a quantifiable measure—namely, the confidence score assigned by the model to a given output~\cite{fat0004CF024,BenzR23,MaLWZSYM23}.
A well-calibrated confidence score which faithfully reflects the model’s empirical likelihood of correctness, enables model users to make immediate decisions about whether to disregard its output~\cite{fat0004CF024}. 
Such a measure also supports abstention mechanisms, whereby outputs that have been assigned a confidence score that is below a certain pre-selected threshold are suppressed~\cite{ZablotskaiaPMN023}.
Tangential to this, explicitly communicating a trustworthy measure of the model's uncertainty can help align the expectations with the model’s capability, thereby mitigating over reliance, or excessive scepticism~\cite{SabouriEZZMLC25,PrabhudesaiYAHL23,ChoudhuriTPKSGS25,ZhangLB20}.
For a software developer, such information can support routine decision-making regarding when the boundaries of the model's coding capabilities are exceeded and human-written implementations are needed.
However, these benefits are contingent on the availability of well-calibrated confidence scores, which post-trained LLMs are generally unable to provide out-of-the-box~\cite{training_dynamics,LiuBW24,LiuK024,ShenDGSWG24}, giving rise to the long-standing field of confidence calibration~\cite{neuralnetwork_calibration}.

Although prior research on confidence calibration for software engineering tasks have reported promising results across a variety of generative tasks, including line completion, function synthesis~\cite{Spiess0PPRAJDA25}, and code summarisation~\cite{VirkDA25}, these methods have either not been examined or proven ineffective for automated code revision (ACR) tasks, which form an important class of real-world software quality assurance activities.
In particular, existing post-hoc calibration approaches such as Platt-scaling~\cite{platt1999probabilistic} i.e., a single logistic regression that maps sequence-level confidence scores to correctness probabilities, have been found to be ineffective for program repair~\cite{Spiess0PPRAJDA25} and have not yet been evaluated on core tasks like vulnerability repair~\cite{appatch} and automated code refinement~\cite{codereviewqa}. 
We hypothesise that the coarse-grained nature of this conventional method is unsuitable for the ACR scenario.
First, we posit that sequence-level confidence scores fail to capture meaningful uncertainty in ACR tasks because the most salient signals are concentrated in specific edit decisions, whereas the vast majority of tokens in a code revision correspond to trivial model decisions.
Second, using a single calibrator assumes a uniform relationship between uncalibrated confidence scores and true probabilities of correctness. 
However, there is no evidence that calibration errors across all samples within a task consistently follow the same relationship.
Given the potential for streamlining LLM-augmented developer workflows, this study explores fine-grained calibration approaches for eliciting well-calibrated confidence scores in ACR tasks.

We address this issue by proposing two fine-grained confidence calibration approaches, namely the use of fine-grained confidence scores and local Platt-scaling.
Firstly, we propose three fine-grained confidence scores: minimum token probability, lowest-$K$ token probability, and attention-weighted uncertainty, and compare them with two traditional sequence-level confidence scores: normalised sequence likelihood and average token probability~\cite{Spiess0PPRAJDA25}.
Secondly, we replace conventional Platt-scaling, which shares a single calibrator amongst all samples of a given task (global Platt-scaling), with local Platt-scaling, a sample-aware method that selects from an ensemble of local calibrators according to embedding-based cluster assignment.
Through our experiments, we demonstrate that fine-grained confidence scores consistently achieve lower calibration error across a wider range of probability intervals than sequence-level scores. 
Specifically, minimum token probability is the best-performing confidence score across all tasks, correctness metrics, and models considered. 
Our results also show that local Platt-scaling outperforms the global implementation, particularly in automated code refinement and, to a lesser extent, in program and vulnerability repair.

\contributions{
\begin{itemize}
    \item The proposal of three fine-grained confidence scores as calibration targets to capture localised uncertainty from code editing decisions in ACR tasks.
    \item The proposal of local Platt-scaling as a post-hoc calibration approach to correct sample-dependent miscalibration patterns in ACR tasks.
    \item A comprehensive analysis of five confidence scores and two calibration methods across three ACR tasks and correctness metrics for 14 different models.
\end{itemize}
}

The paper is structured as follows: 
Section~\ref{sec:background_related} introduces the background and related work.
Section~\ref{sec:approach} introduces our proposed approach.
Section~\ref{sec:study_design} outlines the study design and methodology.
Section~\ref{sec:experiment_setup} details the experiment setup.
Section~\ref{sec:results} reports the main results of the study.
Section~\ref{sec:discussion} facilitates discussions on the findings.
Section~\ref{sec:case_study} explores a case study on downstream utility.
Section~\ref{sec:threats} addresses the potential threats to validity.
Section~\ref{sec:conclusion} draws the conclusion.

\section{Background and Related Work}
\label{sec:background_related}
This work spans the areas of automated code revision and confidence calibration in machine learning, including its applications to AI for software engineering.
We now introduce the formal problem, background and related work in the field.

\subsection{Automated Code Revision}
Following rapid gains in the general effectiveness of LLMs, the field of AI for software engineering has largely converged on these types of models for generative tasks~\cite{llm_review}.
Specifically, they have demonstrated state-of-the-art capabilities in automating code revision-based software quality assurance tasks.
In this study, we focus on three of the most extensively researched ACR tasks for software quality assurance: automated program repair, vulnerability repair, and automated code refinement.
Automated program repair~\cite{repairbench,repairllama,apr_zhang} aims to generate bug-fixing patches for erroneous code, typically addressing functional defects that cause incorrect program behaviour.
Vulnerability repair~\cite{appatch,vr_survey} aims to generate security patches, fixing weaknesses in code that expose the software to potential exploits.
Automated code refinement~\cite{chatgpt_acr, TufanoDMCB24, codereviewqa} seeks to resolve code review comments in pull requests.
They cover a wide variety of issues, ranging from critical defects to maintainability concerns~\cite{mantyala}.
Whilst prior work emphasises improving the correctness of LLM-generated code revisions, our study instead addresses how to accurately communicate their likelihood of correctness, enabling trustworthy and efficient use in their current, imperfect state.

\subsection{Confidence Calibration in Machine Learning}

We now formally introduce the problem of confidence calibration.
To be considered well-calibrated, a model's confidence should approximate their empirical likelihood of correctness~\cite{neuralnetwork_calibration}.
Formally, given input $X$, a perfectly calibrated probabilistic model should satisfy the following condition:

\begin{equation}
P(\hat{y} = y \mid P(\hat{y} \mid X) = \hat{p} ) = \tilde{p}, \forall p \in [0,1] \label{eq:1}
\end{equation}

In this perfect parity, the model's confidence in the output $\hat{p}$ should be exactly equal to the empirical fraction of correct examples at that confidence level $\tilde{p}$, for all levels of model confidence.
In practice, it is impossible to achieve this continuous parity $\forall p \in [0,1]$ with finite samples, therefore achieving a close empirical approximation, such that $\hat{p} \approx \tilde{p}$ in a target test-set should suffice for downstream applications.

The field of confidence calibration has a long-standing history in machine learning, where it plays a key role in ensuring trustworthy applications ranging from natural language processing~\cite{training_dynamics,chen-etal-2023-close,wang-etal-2020-inference} to computer vision~\cite{liu2023model,tomani2021post,joy2023sample}.
Prior work on LLMs indicate that larger model sizes and longer pretraining are associated with improved confidence calibration when performance is far from saturation~\cite{chen-etal-2023-close}, whereas extensive post-training often degrades it~\cite{training_dynamics}. 
In general, deep learning models are uncalibrated by default and therefore require post-hoc calibration techniques to align the model's confidence with their likelihood of correctness~\cite{neuralnetwork_calibration}.

Traditional post-hoc calibration techniques include statistical methods, e.g., Platt-scaling~\cite{platt1999probabilistic}, temperature-scaling~\cite{neuralnetwork_calibration}, histogram binning~\cite{zadrozny2001obtaining}, and isotonic regression~\cite{zadrozny2002transforming}, which only require the original inference step from the underlying LLM, whereas more recent techniques, e.g., P(True)~\cite{kadavath2022language,tian-etal-2023-just,lin2022teaching} and stochastic sampling~\cite{chung2024sampling}, require additional rounds of inference.
Amongst traditional post-hoc calibration techniques, histogram binning and isotonic regression involve a higher degree of parameterisation and are therefore prone to overfitting, making them ill-suited for the typical sizes of software engineering benchmarks~\cite{Spiess0PPRAJDA25}.
Since Platt-scaling is a strict generalisation of temperature-scaling, owing to its additional bias term, we focus on the former technique.
Amongst the more recent techniques, P(True) prompts a well-calibrated LLM to estimate the probability that its previously generated output is correct, whereas stochastic sampling quantifies uncertainty through repeated sampling of it.
These techniques are not considered in our study as they belong to a different inference-time complexity where the latency is incomparable.

\subsection{Confidence Calibration in AI for Software Engineering}

In the field of AI for software engineering, confidence calibration have been investigated for both code classification tasks~\cite{calibrate_cm} e.g., exception type, defect detection, clone detection, and code generation tasks~\cite{Spiess0PPRAJDA25,VirkDA25} e.g., code summarisation, function synthesis, automated program repair.
Prior studies on calibration for code generation have shown that conventional Platt-scaling with sequence-level confidence scores is sufficient for most tasks, with the notable exception of program repair~\cite{Spiess0PPRAJDA25}.
We posit that code revision differs from general code generation, where solutions are produced from scratch rather than focused adaptations of existing code.
As such, salient uncertainty information is dependent on local contexts, rendering sequence-level confidence scores and conventional Platt-scaling overly coarse-grained.
Motivated by this hypothesis, our study investigates whether fine-grained calibration approaches can address this specific class of tasks.
Table~\ref{Tab:study_comparison} compares our study with past work on confidence calibration for code generation tasks in software engineering.
Whilst the notion of fine-grained uncertainty has been explored in code generation, it has not yet been applied in real-world code revision tasks or to the context of calibration. 
Additionally, these approaches either rely on auxiliary deep learning models trained on supplementary code-editing data that is seldom available~\cite{vasconcelos2025generation} or repeated sampling techniques~\cite{johnson2023ru} that may not be feasible in actual deployment due to latency.

\begin{table}[!htbp]
\caption{Comparison with Past Work on Confidence Calibration for Code Generation Tasks in Software Engineering}
\centering
\resizebox{\columnwidth}{!}{%
\begin{tabular}{!{\vrule width 1.2pt}l!{\vrule width 1.2pt}l|l|l!{\vrule width 1.2pt}}
\Xhline{2\arrayrulewidth}
\rowcolor[HTML]{F0C88C}
\textbf{Study} & \textbf{Task} & \textbf{Single-Inference Confidence Score} & \textbf{Post-hoc Scaler} \\ \Xhline{2\arrayrulewidth}
\rowcolor[HTML]{FFF5E3} 
\cellcolor[HTML]{F0C88C}\textbf{Spiess et al.~\cite{Spiess0PPRAJDA25}} & {\color[HTML]{000000} \begin{tabular}[c]{@{}l@{}}Function Synthesis\\ Line-level Completion\\ Automated Program Repair\end{tabular}} & {\color[HTML]{000000} \begin{tabular}[c]{@{}l@{}}Average Token Probability\\ Sequence Likelihood\end{tabular}} & {\color[HTML]{000000} Global Platt-Scaling} \\ \hline
\rowcolor[HTML]{FFF5E3} 
\cellcolor[HTML]{F0C88C}\textbf{Virk et al.~\cite{VirkDA25}} & {\color[HTML]{000000} Code Summarisation} & {\color[HTML]{000000} \begin{tabular}[c]{@{}l@{}}Average Token Probability\\ Normalised Sequence Likelihood\end{tabular}} & {\color[HTML]{000000} Global Platt-Scaling} \\ \hline
\rowcolor[HTML]{FFF5E3} 
\cellcolor[HTML]{F0C88C}\textbf{(Ours)} & {\color[HTML]{000000} \begin{tabular}[c]{@{}l@{}}Automated Program Repair\\ \textbf{Vulnerability Repair}\\ \textbf{Automated Code Refinement}\end{tabular}} & {\color[HTML]{000000} \begin{tabular}[c]{@{}l@{}}Average Token Probability\\ Normalised Sequence Likelihood\\ \textbf{Minimum Token Probability}\\ \textbf{Lowest-$K$ Token Probability}\\ \textbf{Attention-Weighted Uncertainty}\end{tabular}} & {\color[HTML]{000000} \begin{tabular}[c]{@{}l@{}}Global Platt-Scaling\\ \textbf{Local Platt-Scaling}\end{tabular}} \\ \Xhline{2\arrayrulewidth}
\multicolumn{4}{l}{\footnotesize\textbf{Bold:} Contribution from this Study} \\
\end{tabular}
}
\label{Tab:study_comparison}
\end{table}

\begin{figure}[!htbp]
    \centering
    \includegraphics[width=0.9\columnwidth]{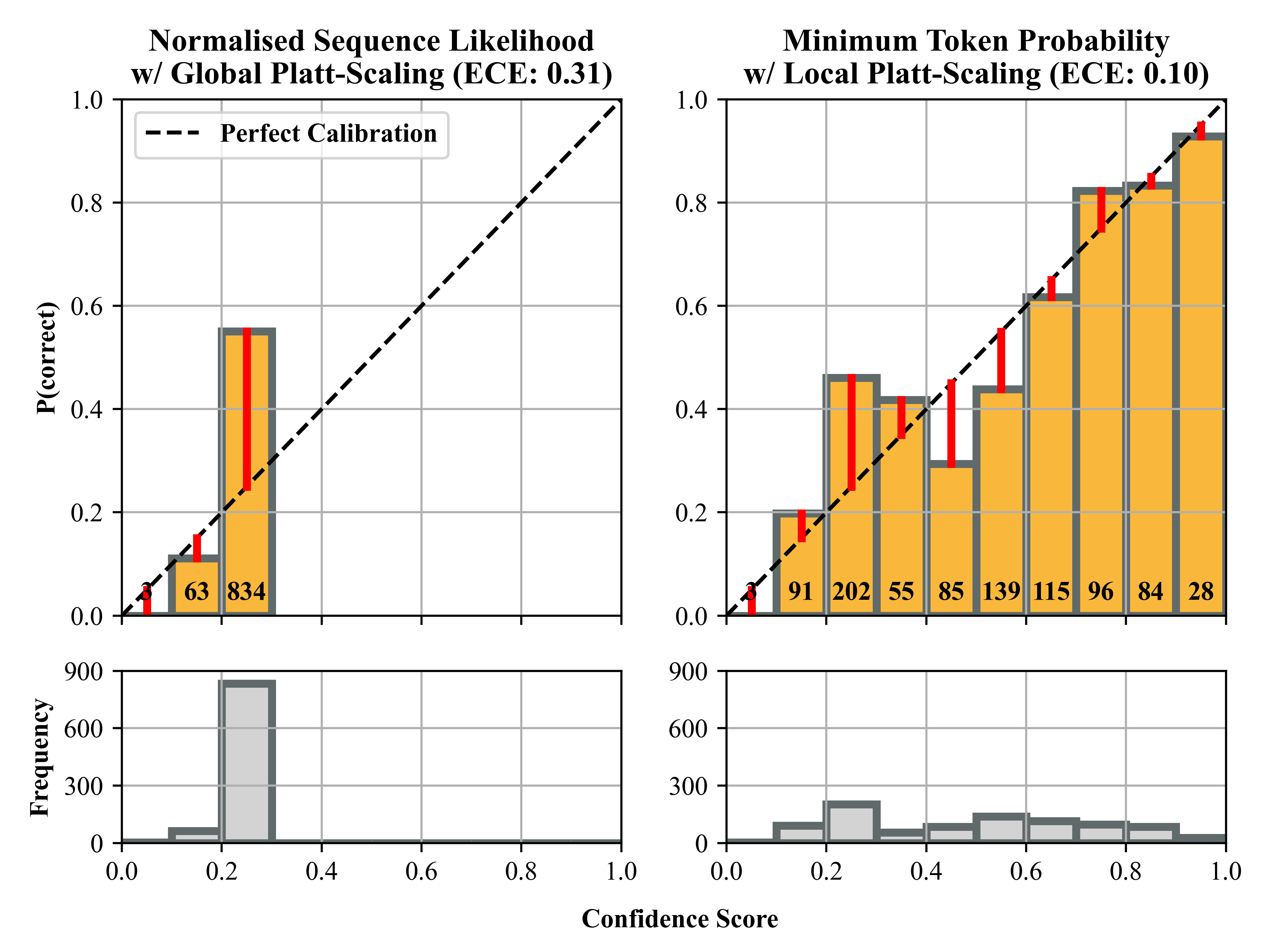} 
    \captionsetup{justification=centering} 
    \caption{Miscalibrated (Left) and Well-Calibrated (Right) Confidence Scores for Code Revisions Generated by CodeLlama-70B in Automated Code Refinement}
    \label{fig:reliability_plot}
\end{figure}

Figure~\ref{fig:reliability_plot} illustrates the characteristics of poorly calibrated and well-calibrated confidence estimates using an example reliability plot for the task of automated code refinement. 
The poorly calibrated confidence scores are produced by the conventional coarse-grained method, whereas the well-calibrated confidence scores are produced by the fine-grained confidence calibration approach proposed in this study.
As demonstrated, improved calibration yields confidence estimates that more accurately reflect the true likelihood that LLM-generated code revisions are correct across the probability space, whereas poorly calibrated confidence estimates remain misaligned and are therefore unable to reliably support decision-making.

\section{The Proposed Approach}
\label{sec:approach}
In this section, we begin by introducing conventional approaches to confidence estimation and calibration, followed by our proposed fine-grained confidence calibration methods.

\subsection{Confidence Scores}
This study focuses on intrinsic single inference measures of confidence due to practical considerations for model serving.
Specifically, we restrict our consideration to methods with low inference-time complexity to ensure minimal computational overhead and latency.
Since these methods rely on access to model internals, they are considered white-box methods.

\subsubsection{Sequence-Level Confidence Scores}
We discuss the two separate sequence-level confidence scores commonly used in past calibration studies~\cite{wang-etal-2020-inference,Spiess0PPRAJDA25,VirkDA25}.
They represent the baselines that our study aims to improve upon.

\textbf{Normalised Sequence Likelihood.}
This measure represents the model assigned likelihood for the generated output as a whole.
It is the product of the conditional probabilities assigned to each token in the output sequence.
The conditional probabilities are from the softmax layer of the auto-regressive LLM.
We adopt the formulation from prior work on confidence calibration for software engineering~\cite{Spiess0PPRAJDA25}.
The sequence likelihood can be formalised by the following equation:

\begin{equation}
P_{sl}(\hat{y}) = \prod_{t=1}^{T} P(\hat{y}_t \mid \hat{y}_{<t}, X) 
\end{equation}

Where $\hat{y}_t$ is the current token being generated, $\hat{y}_{<t}$ are the previously generated tokens, and $T$ is the sequence length.
Since this measure penalises longer sequences, we consider the length normalised version.
The length normalised sequence likelihood can be formalised by the following equation:
$P_{sl\_norm}(\hat{y}) = P_{sl}(\hat{y})^{\frac{1}{T}} $.
This measure ensures that longer sequences which involve more token probabilities are not disproportionately penalised compared to shorter sequences that involve less token probabilities.

\textbf{Average Token Probability.}
This measure represents an average of the model assigned likelihoods to each token in the sequence.
This simply computes the arithmetic mean of the conditional token probabilities.
The average token probability can be formalised by the following equation:

\begin{equation}
P_{avg}(\hat{y}) = \frac{1}{T} \sum_{t=1}^{T} P(\hat{y}_t \mid \hat{y}_{<t}, X) 
\end{equation}

Whilst the normalised sequence likelihood uses a geometric mean that is sensitive to small probabilities, average token probability is instead sensitive to large probabilities due to the additive properties of the arithmetic mean.

\subsubsection{Fine-Grained Confidence Scores}
Rather than consider the probabilities of all output tokens, we hypothesise that correctness in ACR tasks is best captured by the model assigned probabilities to local edit decisions.
Hence, we propose three fine-grained confidence scores that are also intrinsic and single inference: minimum token probability, lowest-K token probability, and attention-weighted uncertainty. 
Each represents a different approach to determining the locality of salient decisions.
The concept of minimum token probability is well aligned with broader work in uncertainty quantification~\cite{manakul2023selfcheckgpt} under different formulations and domains, yet remains unexplored for automated software engineering tasks.
Meanwhile, lowest-$K$ token probability and attention-weighted uncertainty are novel formulations of confidence introduced in this study.

\textbf{Minimum Token Probability.}
This measure represents the minimum likelihood assigned by the model to any token in the output sequence.
In ACR, edit decisions are often granular but critical~\cite{clm_impact,codereviewqa}, where any single incorrect token may invalidate the entire implementation.
We therefore use this measure to reflect the potential fragility introduced by the model’s least confident decision.
The minimum token probability can be formalised by the following equation:
\begin{equation}
P_{\min}(\hat{y}) = \min_{t \leq T} P(\hat{y}_t \mid \hat{y}_{<t}, X)
\end{equation}

\textbf{Lowest-$K$ Token Probability.}
Unlike average token probability which indiscriminately averages probabilities over the entire output sequence, this measure restricts the averaging to tokens with the lowest-$K$ probabilities.
Compared to minimum token probability, this measure should be more robust to individual outlier tokens that are mistakenly under-confident.
We utilise the arithmetic mean rather than the geometric mean due to the fact that these tokens can be disjoint.
The lowest-$K$ token probability can be formalised by the following equation:
\begin{equation}
\begin{aligned}
P_{\mathrm{low}\text{-}K}(\hat{y}) &= 
\frac{1}{K} \sum_{t \in \mathcal{I}_K} 
P\big(\hat{y}_t \mid \hat{y}_{<t}, X\big), \\[4pt]
\mathcal{I}_K &=
\operatorname*{arg\,sort}_{t}^{(K)} 
P\big(\hat{y}_t \mid \hat{y}_{<t}, X\big)
\end{aligned}
\end{equation}

Where $\mathcal{I}_K$ represents the set of token positions with the lowest-$K$ token probabilities.
We use the Kneedle algorithm~\cite{SatopaaAIR11} to dynamically select $K$ for each sample.
Following our rationale on informative local edit decisions, if the majority of tokens lie in the high probability region whilst a minority exhibit low probabilities, they will form a sharply increasing concave with a long plateau when sorted in ascending order.
The algorithm identifies the local maximum of the difference from the $y=x$ diagonal in the concave curve, corresponding to the $K^{th}$ token where the curve of token probabilities begins to plateau.
Similar to the original implementation~\cite{SatopaaAIR11}, we set a sensitivity of $S=0$ and use the offline setting.

\textbf{Attention-Weighted Uncertainty.}
This measure extends the lowest-$K$ token probability by incorporating attention-based, token-level weighting.
Our lowest-$K$ token probability measure carries an assumption that all of the lowest token probabilities are equally informative.
However, it is possible that correctly generated tokens arise from diffuse softmax distributions.
Rather than reflecting the model’s uncertainty about correctness, these lower probabilities may instead arise from syntactic flexibility in the token completion i.e., multiple correct alternatives, or from the correct token being statistically rare in the training distribution~\cite{LinT024} (e.g., unconventional syntax, formatting, expressions).
As such, we want to reduce their impact on the calculation of the confidence score, since they don’t meaningfully affect the overall correctness of the revision.
To address this, attention-weighted uncertainty scales each token’s probability according to their model assigned saliency in the generated code revision.

We quantify a token’s saliency using the attention mass assigned to it by subsequently generated tokens.
The intuition is that uncertainty in tokens with strong downstream influence should matter more than the aforementioned trivial tokens. 
If such a token is assigned a low softmax probability, the likelihood of cascading errors in the rest of the code revision increases, and its uncertainty should therefore be more heavily weighted.
The attention-weighted uncertainty can be formalised by the following equation:

\begin{equation}
\begin{aligned}
P_{attn\text{-}\omega}(\hat{y}) &= 
\frac{1}{K} \sum_{t \in \mathcal{I}_K} 
P\big(\hat{y}_t \mid \hat{y}_{<t}, X\big), \\[6pt]
\mathcal{I}_K &=
\operatorname*{arg\,sort}_{t}^{(K)} 
\left[ w_t \cdot \left( 1 - P\big(\hat{y}_t \mid \hat{y}_{<t}, X\big) \right) \right], \\[4pt]
\text{where } w_t &= \sum_{j=t}^{T} \tilde{A}_{j,t} + 1 \\[12pt]
\end{aligned}
\end{equation}

Where $\mathcal{I}_K$ represents the set of token positions with the highest-$K$ weighted token uncertainties.
Similar to lowest-$K$ token probability, we use the Kneedle algorithm~\cite{SatopaaAIR11} to dynamically select  $K$  for each sample.
Unlike the lowest-$K$ token probability setup, which orders softmax probabilities in ascending, concave order, we instead adopt a decreasing, convex ordering, as weighted uncertainties represent an inverse quantity and should be prioritised in descending order.
We compute $\tilde{A}_{j,t}$ using the rollout method~\cite{AbnarZ20}, which estimates layer-wise attention propagation with residual stream effects.

\subsection{Calibration Techniques}
Given that deep learning-based language models traditionally do not optimise with a calibration-aware loss function, the token probabilities they emit do not naturally approximate the aforementioned parity that is expected from a well-calibrated model~\cite{training_dynamics}.
Thus, a conventional solution is to rescale the model's own confidence in its generated sequence using post-hoc calibration techniques.
This study focuses on the statistical technique of Platt-scaling. We choose this method because it has minimal parameterisation, which makes it less prone to overfitting on typical software engineering benchmark sizes~\cite{Spiess0PPRAJDA25}. 
Additionally, it has lightweight inference-time complexity i.e., introduces minimal computational overhead and latency.
Together, these properties make the method well-suited for practical deployment in software engineering tasks.

\subsubsection{Global Platt-scaling}

This conventional method trains a univariate logistic regression model to map a target model’s real-valued outputs (e.g., decision scores) onto binary correctness outcomes using a held-out training set~\cite{platt1999probabilistic}.
Formally, this can be represented by the following equation:

\begin{equation}
P(y=1 \mid X) = \sigma(w \hat{p} + \beta)
\end{equation}

Where $\sigma(z) = \frac{1}{1 + e^{-z}}$ and $w, \beta \in \mathbb{R}$ are learnable parameters.
This method takes a global approach, where the single calibrator is shared across all samples for a given task.

\subsubsection{Local Platt-scaling}

Since we will be proposing a localised version of the conventional method, we identify the baseline as global Platt-scaling, and our proposed approach as local Platt-scaling.
Conventional Platt-scaling assumes calibration errors are uniform across a task by learning only a single global calibrator. 
However, the variability in code and specifications within ACR tasks can induce heterogeneous errors~\cite{wald2021calibration, detommaso2024multicalibration, yu2022robust}. 
Because global Platt-scaling uses a single set of fixed parameters $w$ and $\beta$, it lacks the expressivity to account for sample-dependent features, which can result in suboptimal uncertainty estimates for instances that deviate from the global average.
To address this potential shortfall, we propose local Platt-scaling, a calibration method that is sample-specific, and therefore capable of providing fine-grained calibration for different types of scenarios within an ACR task.
In contrast to global Platt-scaling, local Platt-scaling trains distinct calibrators for specific clusters of samples. 
These clusters are partitioned according to a multidimensional space comprised of input/output embeddings, alongside their corresponding uncalibrated confidence scores.
Local platt-scaling can be formalised by the following equation:

\begin{equation}
P(y=1 \mid X)  = 
\begin{cases} 
\sigma(w_{k} \hat{p} + \beta_{k}) & \text{if } H(\phi(x)) = k \\ 
\psi(\hat{p}) & \text{if } H(\phi(x)) = -1 
\end{cases}
\end{equation}

Where $\phi(x)$ denotes the feature vector that is constructed by concatenating an $n$-dimensional UMAP projection~\cite{McInnes2018} of the input/output embeddings with the uncalibrated confidence score.
Following similar setups from past studies~\cite{detommaso2024multicalibration}, we used $n=20$ dimensions.
For text embeddings, we used the Qwen3-Embedding-8B~\cite{qwen3embedding} model, as it is considered state-of-the-art for the coding domain at the time of writing.~\footnote{As of March 2026, based to the MTEB leaderboard~\cite{muennighoff2023mteb}.}
We selected HDBSCAN~\cite{campello2013density} to be the $H(\cdot)$ clustering operator since it is a non-parametric method that can manage varying densities, detecting outliers, and automatically determining the optimal number of clusters.
The hyperparameters for HDBSCAN are \textit{minimum cluster size} which  defines the smallest number of points required to form a cluster and \textit{minimum samples}  which determines the density threshold required for a point to be considered a core part of a cluster rather than an outlier.

At training time, clusters and outliers are first identified based on a held-out training set, after which a separate local calibrator is fitted to each cluster.
During inference, the new instance’s cluster (or outlier) membership is determined based on its nearest neighbour.
If an instance is assigned to cluster $k$, the corresponding local calibrator is applied; otherwise, it is classified as an outlier, triggering our novel $\psi(\cdot)$ backoff function.
The selection of the backoff strategy is treated as a hyperparameter, which can either be the conventional global calibrator or simply the uncalibrated confidence score.

\section{Study Design and Methodology}
\label{sec:study_design}

We formulate three research questions (RQs) to examine the suitability and effectiveness of fine-grained confidence calibration approaches for LLMs in ACR tasks.

\textbf{RQ1. How well can fine-grained confidence scores separate correct from incorrect code revisions in ACR tasks?}
Motivated by the rationale that certain local edit decisions capture the most informative signals regarding correctness, we set out to explore three fine-grained confidence scores, i.e., minimum token probability, lowest-$K$ token probability, and attention-weighted uncertainty.
We compare their distributions with conventional sequence-level confidence scores i.e., normalised sequence likelihood, average token probability, to assess if they can better separate correct from incorrect code revisions.
This informs the extent to which fine-grained confidence scores are suitable for calibration.

\textbf{RQ2. How effective is Platt-scaling for ACR tasks when using fine-grained confidence scores?}
To evaluate whether targeting certain local edit decisions is effective in practice, we conduct global Platt-scaling with both fine-grained and sequence-level confidence scores to compare their effectiveness in confidence calibration across the ACR tasks.
Specifically, we are interested in the extent to which fine-grained confidence scores can produce informative and well-calibrated confidence estimates for ACR tasks, a feat previously unattainable with sequence-level confidence scores~\cite{Spiess0PPRAJDA25}.

\textbf{RQ3. To what extent can local Platt-scaling improve over conventional Platt-scaling in ACR tasks?}
Motivated by the rationale that calibration errors can be non-uniform and sample-specific, we investigate whether local Platt-scaling, a method that learns separate calibrators for distinct sample clusters, can provide the expressivity needed to better capture localised errors.
Specifically, we evaluate whether the fine-grained approach enhances calibration performance over the conventional method of global Platt-scaling for ACR tasks.

\subsection{Automated Code Revision Benchmarks}
To assess confidence calibration across the ACR tasks, we carefully select benchmarks that utilise real-world examples. 
These benchmarks are also deliberately curated with the intention to mitigate data leakage which may cause calibration results to appear deceptively optimistic~\cite{bug_memorization}.
Additionally, we require benchmarks to be substantially large such that we are able to reliably compute calibration error.
Below, we describe three ACR benchmarks used in this study.

\textbf{Automated Program Repair.}
We use the non-security partition of the DeepCode AI Fix benchmark (DCF-Bug)~\cite{deepcodeaifix}, as it is the only program repair dataset that satisfies our criteria: it contains real-world examples, is non-contaminated, and is sufficiently large.
Specifically, the task is formulated as $P(C_{fix} \mid C_{bug}, V_{nl})$, where $C_{fix}$ is the fixed code snippet to be inferred, $C_{bug}$ is the buggy code snippet submitted as input, and $V_{nl}$ is the violated static analysis rule that can be considered as the natural language specification.
This benchmark includes 777 non-security related bugs from 489 GitHub repositories written in JavaScript and TypeScript.
The static analysis rules check for issues such as missing tags, duplicate variable names, invalid dataflows, resource leaks, and incorrect method interaction.
Concerns regarding data leakage are mitigated by the projects’ non-permissive licenses, which explicitly prohibit LLM training whilst permitting evaluation.
To support Platt-scaling, we use the training set accompanying DCF-Bug, which includes 1,917 bugs from 1,123 different GitHub repositories that use permissive licenses.

\textbf{Vulnerability Repair.}
We use the security partition of the DeepCode AI Fix benchmark (DCF-Vul)~\cite{deepcodeaifix}, as it is the only vulnerability repair dataset that satisfies our criteria: it contains real-world examples, is non-contaminated, and is sufficiently large.
Specifically, the task is formulated as $P(C_{fix} \mid C_{vul}, V_{nl})$, where $C_{fix}$ is the fixed code snippet to be inferred, $C_{vul}$ is the vulnerable code snippet submitted as input, and $V_{nl}$ is the violated static application security testing (SAST) rule that can be considered as the natural language specification.
This benchmark includes 1,041 security fixes from 600 GitHub repositories written in JavaScript and TypeScript.
The static analysis rules check for issues such as unsecure API usage, security misconfigurations, and dataflow vulnerabilities.
Similarly, concerns regarding data leakage are mitigated by the projects’ non-permissive licenses.
To support Platt-scaling, we use the training set accompanying DCF-Vul, which includes 1,615 vulnerabilities from 1,008 different GitHub repositories that use permissive licenses.

\textbf{Automated Code Refinement.}
We use the CodeReviewQA benchmark~\cite{codereviewqa}, as it is the only automated code refinement test set that is manually curated; other datasets in this space are known to contain substantial noise~\cite{TufanoDMCB24}.
Specifically, the task is formulated as $P(C_{post} \mid C_{pre}, R_{nl})$, where $C_{post}$ is the code snippet revised to address the attached code review comment $R_{nl}$, and $C_{pre}$ is the code snippet under review.
This benchmark features 900 resolved code review comments from 199 GitHub repositories written in nine different programming languages, i.e., C, C++, C\#, Go, Java, JavaScript, PHP, Python, and Ruby.
Given that these code reviews occurred in 2022 which may be susceptible to data contamination, we employed Gemini 2.5 Pro to transform the examples in terms of surface-level text (CR-Trans).
This involved using semantic-preserving transformations (e.g., identifier renaming, statement reordering) to refactor both $C_{pre}$ and $C_{post}$ code snippets~\cite{vjbench}.
The natural language comments $R_{nl}$ were then adapted to align with the transformed code snippets, as well as lightly paraphrased into different forms of expression (e.g., formality shifts, syntactic restructure).
All examples were manually verified for validity after the transformations.

To support Platt-scaling, we leveraged the CodeReviewer training set~\cite{codereviewer}, which is widely used in prior work~\cite{TufanoDMCB24,LuYLYZ23}.
It contains 150k resolved GitHub code review comments.
Since this dataset has been shown to have high amounts of noise~\cite{TufanoDMCB24}, we first conducted LLM-based data cleaning~\cite{LiuLT25}.
After validating on 50 examples, we find that Gemini 2.5 Pro can label clean data with 88\% accuracy.
The remaining 12\% comprised an even mix of borderline false positive and false negative cases, in which $R_{nl}$ was contextually relevant to $C_{pre}$ but necessitated nuanced reasoning to assess its alignment with $C_{post}$.
We subsequently used the LLM to clean the entire dataset.
To ensure comparability across tasks, we sampled a subset of examples to match the size of DCF-Bug's training set.
The sampling was conducted in a stratified manner to maintain diversity of programming languages and repositories.
The final training set includes 1,917 code reviews from 140 different GitHub repositories resolved before 2022.
It includes the same nine programming languages mentioned above.

\subsection{Selected Models}
We consider state-of-the-art LLMs from model families that are frequently included in research for ACR tasks.
Specifically, we consider models that use the canonical decoder-only transformer architecture, with billion scale parameter counts.
Given that our study is based on compute-efficient single inference measures of confidence, we only consider open-source models that allow white-box access.
We use instruction-tuned model variants as we need to engage in zero-shot prompting to perform the ACR tasks.
From the Llama series, we include Llama-3.1~\cite{appatch,codereviewqa} and CodeLlama~\cite{appatch,repairllama}.
From the Qwen series, we include Qwen2.5~\cite{repairbench,codereviewqa} and Qwen2.5-Coder~\cite{appatch,repairbench}.
From the DeepSeek series (DS), we include DeepSeek-LLM~\cite{codereviewqa,intention}, DeepSeek-Coder~\cite{codereviewqa,intention}, DeepSeek-V2-Lite~\cite{repairbench,codereviewqa}, and DeepSeek-Coder-V2-Lite~\cite{appatch,codereviewqa}.
We include a variation of architecture sizes from each series to increase the generalisability of our findings.

\subsection{Correctness Metrics for ACR Tasks}
Selecting an appropriate correctness metric is critical as it determines the calibration target.
Specifically, these metrics evaluate whether a generated code revision is correct for a given ACR task.
In this section, we discuss the three separate correctness metrics considered in this study.

\textbf{Exact Match.}
To be considered correct, exact match $EM$ requires the LLM-generated code revision to be identical to the ground truth implementation in terms of surface-level text.
That is $\hat{C_{fix}} = C_{fix}$ and $\hat{C_{post}} = C_{post}$ are the only scenarios where the generated code revision can be considered as correct.
We implement $EM$ that is whitespace insensitive.

\textbf{Edit Progress.}
Whilst $EM$ ensures no false positives are counted as correct, it can be overly strict, treating all generated code revisions that are non-identical to the ground truth implementation as equally incorrect.
To alleviate this, we utilise edit progress $EP$, a relaxed distance-based text matching metric~\cite{edit_progress} that is specifically designed for ACR-like tasks~\cite{zhou23}.
From a productivity perspective, developers have deemed code generations that reduce the overall effort to complete the task as valuable~\cite{DibiaFBPLA23}, even if they have not exactly fulfilled the developer's intent.
Specifically, $EP$ can be formalised by the following equation: 

\begin{equation}
EP = \frac{|D_{\tilde{C}\rightarrow C}| - |D_{\hat{C}\rightarrow C}|}{|D_{\tilde{C}\rightarrow C}|}
\end{equation}

Where $D$ is the edit distance between two sequences~\cite{Levenshtein}, $\tilde{C}$ is the submitted code, $\hat{C}$ is the candidate code revision, and $C$ is the ground-truth code revision.
Therefore, $EP$ represents the percentage of progress in transforming $\tilde{C}$ to $C$, and ($1 - EP$) represents the remaining effort to correct $\hat{C}$ to $C$.
If a model generates more errors than correct edits, $EP$ can be negative.
Following prior work~\cite{edit_progress}, we operationlise $EP$ into a binary correctness metric, where any positive edit progress $EP^{+}$ is considered as a net improvement to developer productivity.

\textbf{Checks Passed.}
This is a reference-free metric that instead relies on static analysis tools to check for correctness.
More specifically, we consider an LLM-generated code revision to be correct if it resolves the violated $V_{nl}$ rule's alarm without triggering any other alarms, i.e., passes all the checks ran by the static analysis tool.
The checks passed $CP$ metric can capture semantically correct solutions with syntactically different implementations, i.e., a reduction in false negatives compared to $EM$. 
As such, $CP$ can determine if a solution is plausibly correct.
This particular metric is only applicable to DCF-Vul, as the relevant rules for DCF-Bug have been deprecated in the Snyk Code SAST engine~\cite{SnykCode}.

\subsection{Descriptive Statistics}
We now introduce the statistical measures used in RQ1 to support the preliminary analysis of token-level softmax probabilities and confidence scores before confidence calibration.

\textbf{Median Token-level Skewness ($\tilde{\gamma}_1$).} 
To explore whether fine-grained confidence scores have the potential to yield substantially different distributions compared to sequence-level confidence scores, we first analyse the skewness of token-level softmax probabilities in LLM-generated code revisions. 
For each ACR benchmark, the per-sample skewness values are aggregated by taking their median across the test set.
If our rationale holds, there should be strong negative skewness across each test set, which indicates that the majority of token probabilities in generated code revisions are clustered around higher probabilities, whilst a few outliers are assigned far lower token probabilities.
The median token-level skewness can be formalised by the following equation:

\begin{equation}
\tilde{\gamma}_1 = \operatorname{median}_{i=1}^{N} \left( \frac{1}{T_i} \sum_{t=1}^{T_i} \left( \frac{p_{i,t} - \bar{p}_i}{s_i} \right)^3 \right)
\end{equation}

Where $p_{i,t}$ is the softmax probability of the $t$-th generated token in the $i$-th sample.
For the $i$-th sample, $T_i$ is the number of generated tokens, $\bar{p}_i$ is the mean of the token probabilities, and $s_i$ is the standard deviation of the token probabilities.
For the outer operator, $N$ is the overall number of samples.

\textbf{Wasserstein Distance ($W_1$).} 
To compare sequence-level and fine-grained confidence scores in terms of their inherent ability to separate the distributions of correct and incorrect code revisions, we use the Wasserstein distance.
This statistic quantifies the minimal “effort” required to transform one probability distribution into another in metric space, where a higher $W_1$ reflects wider separation.
If our rationale holds, we expect $W_1$ of fine-grained confidence scores to be higher than sequence-level confidence scores.
The Wasserstein distance can be formalised by the following discrete equation:

\begin{equation}
W_1 = \min_{\pi \in \Pi(\mu, \nu)} \sum_{i=1}^n \sum_{j=1}^m |p_i - p_j| \pi_{ij}
\end{equation}

Where $p_i$ and $p_j$ are the confidence scores of individual samples drawn from distributions $\mu$ and $\nu$, respectively.
These two distributions represent the correct and incorrect code revisions, where $n$ and $m$ are their respective number of samples.
The most efficient way to morph one distribution into the other is captured by $\pi_{ij}$, the optimal transport plan.

\textbf{Kendall's $\tau$ Coefficient $(\tau_b)$.} 
To compare sequence-level and fine-grained confidence scores in terms of their inherent ability to preserve the ranking consistency between correct and incorrect code revisions, we use Kendall's $\tau$ coefficient.
This statistic measures each confidence score's ability to sort correct and incorrect code revisions, where a higher $\tau_b$ represents better rank correlation.
If our rationale holds, we expect $\tau_b$ of fine-grained confidence scores to be higher than sequence-level confidence scores.
The Kendall's $\tau$ coefficient can be formalised by the following equation:

\begin{equation}
\tau_b = \frac{n_c - n_d}{\sqrt{(n_0 - n_1)(n_0 - n_2)}}
\end{equation}

Were $n_c$ and $n_d$ are the numbers of concordant and discordant pairs.
The total number of pairs is denoted by $n_0$, whilst $n_1$ and $n_2$ are the number tie adjustments for each variable.
In our case, these variables are the chosen confidence score and correctness label.
For a given confidence score, stronger measures in both $W_1$ and $\tau_b$ indicate a structure that closely approximates a monotonic, calibratable relationship with the correctness label, and thus greater amenability to Platt-scaling.

\subsection{Calibration Metrics}
We now introduce the main metrics used to measure confidence calibration in RQ2 and RQ3.
If our rationales hold, we expect fine-grained confidence scores and local Platt-scaling to yield improvements in these metrics compared to sequence-level confidence scores with global Platt-scaling.

\textbf{Expected Calibration Error (ECE).} To measure confidence calibration, we require a metric that can capture the degree of mismatch between a model's confidence and its correctness using finite samples.
To this end, expected calibration error $ECE$~\cite{ece} has been the standard in past research~\cite{neuralnetwork_calibration,Spiess0PPRAJDA25,VirkDA25}.
As demonstrated in Figure~\ref{fig:reliability_plot}, this metric reflects the confidence-correctness mismatch based on model outputs that have been discretised into bins.
The expected calibration error can be formalised by the following equation:

\begin{equation}
ECE = \sum_{b=1}^{B} P(b) \cdot |o_{b} - e_{b}|
\end{equation}

Where $B$ is the number of bins, $o_{b}$ is the fraction of correct outputs within the $b^{\text{th}}$ bin, $e_{b}$ is the average confidence within the $b^{\text{th}}$ bin, and $P(b)$ is the fraction of all outputs that fall into the $b^{\text{th}}$ bin.
Following prior studies~\cite{Spiess0PPRAJDA25}, we use $B=10$ equal width bins to partition the entire confidence range. 

\textbf{Brier Score ($\mathcal{B}$).}
Since $ECE$ can be sensitive to the choice of binning mechanism~\cite{KumarLM19,NixonDZJT19,RoelofsCSM22}, particularly when small sample sizes lead to sparse bins, we additionally report the Brier score $\mathcal{B}$~\cite{glenn1950verification}, which avoids binning altogether.
The Brier score can be formalised by the following equation:

\begin{equation}
\mathcal{B}=\frac{1}{N}\sum_{i=1}^{N}(p_{i} -  \mathbbm{1}(f(X_{i})))^{2}
\end{equation}

Where $N$ is the overall number of samples, $p_{i}$ is the model's confidence in the $i^{\text{th}}$ output, and $\mathbbm{1}(f(X_{i}))$ is a binary correctness function that evaluates to value $\in{\{1,0\}}$ based on the $i^{\text{th}}$ output.
This metric has the advantage of being well-defined for any sample size $N$.
However, $ECE$ is widely regarded as a calibration-focused metric, whereas $\mathcal{B}$ captures both accuracy and confidence calibration, potentially conflating the two. Accordingly, we report both metrics as complementary perspectives for a more comprehensive evaluation.

\textbf{Bin Coverage (BC).} 
For any downstream application, a well-calibrated confidence score should also be informative over the wider probability space, i.e., reasonable coverage of confidence bins.
Figure~\ref{fig:reliability_plot} illustrates confidence scores under limited bin coverage (left) and high bin coverage (right).
This is a key component to the original goal of confidence calibration~\cite{neuralnetwork_calibration}, as stated by the condition in Equation~\ref{eq:1}.
The importance of inducing such a trustworthy confidence score lies in its ability to support more granular decision-making across a wider range of probability intervals~\cite{rethinking}. 
This enables finer distinctions in uncertainty, allowing different decisions or thresholds to be applied based on varying confidence scores rather than a limited set of cutoffs.
To measure this, we include bin coverage $BC$, which counts the number of bins that are covered by the calibrated confidence score.
In our case, the maximum bin coverage would be $B=10$, which means that every confidence level is represented.

In past experiments with automated program repair~\cite{Spiess0PPRAJDA25}, global Platt-scaling with sequence-level confidence scores induced single bin collapse i.e., $BC = 1$, effectively producing near identical confidence scores for all samples in the test set.
Since this type of degenerate confidence score lacks the capability of instance-level discrimination, it cannot support decision-making or thresholding.
Following past work~\cite{Spiess0PPRAJDA25}, $ECE$ and $\mathcal{B}$ results are ignored for this type of scenario.

\section{Experiment Setup}
\label{sec:experiment_setup}
We now discuss the implementation details of our experiments, as well as model performances on the ACR tasks. 

\subsection{Implementation Details}
To ensure consistency within the evaluation of each benchmark, we use the same prompt setups for all models.
Specifically, we only consider zero-shot prompts aimed at revising each code snippet within a single forward pass.
The aim of our study is to improve the intrinsic confidence calibration of the studied models, which can be difficult to interpret when conflated with the effects of advanced prompting mechanisms.
For each benchmark, we use the same context provided in the original study with the addition of instructions to only generate the code revision.
This controlled design allows us to examine the models' confidence in its own code revision conditioned solely on the input prompt.
The exact prompt templates are shown in Figure~\ref{fig:prompt_templates}.
We use greedy decoding and deterministically select the single most probable candidate.
This represents the models' strongest belief and is the sample most trusted by developers~\cite{copilot}.
To faithfully represent the models, we deploy them in their native bfloat-16 precision.
For Platt-scaling, we use the canonical implementation of logistic regression, which optimises via the limited-memory BFGS algorithm with a L2 penalty term for regularisation.
For local Platt-scaling, we optimise HDBSCAN's hyperparameters via grid search.
For \textit{minimum cluster size} we search 50-150 (step=25). 
The lower bound ensures adequate support for logistic regression~\cite{peduzzi1996simulation}, whilst the upper bound restricts cluster growth from approximating global Platt-scaling. 
For \textit{minimum samples}, we search 5-80 (step=15) to regulate the trade-off between local manifold sensitivity and density smoothing~\cite{campello2013density}.
The final results are presented based on the most optimal settings according to the three calibration metrics.

\begin{figure}[htbp]
\centering
\begin{promptbox}
{\footnotesize
\textbf{Prompt Template: DCF-Bug and DCF-Vul}
\begin{verbatim}
Fix {rule} {rule_description}.
[BEGIN_BUGGY_CODE] {code} [END_BUGGY_CODE]
Please return only the fixed code wrapped in the 
special markers. 
[BEGIN_FIXED_CODE]
\end{verbatim}
\textbf{Prompt Template: CR-Trans}
\begin{verbatim}
The code snippet has received a code review.
[BEGIN_BUGGY_CODE] {code} [END_BUGGY_CODE]
[BEGIN_CODE_REVIEW] {review} [END_CODE_REVIEW]
Generate a revised version of the code snippet 
according to the code review. Please return only 
the fixed code wrapped in the special markers. 
[BEGIN_FIXED_CODE]
\end{verbatim}
}
\end{promptbox}
\caption{Zero-Shot Prompt Templates for ACR Tasks}
\label{fig:prompt_templates}
\end{figure}

\subsection{Model Performance on ACR Tasks}
We preliminarily assess the models' competencies in the ACR tasks. Table~\ref{Tab:correctness} shows their performance in terms of each correctness metric.
Across the tasks, models generally struggle with DCF-Vul (EM: 3.7-12.0, EP: 15.7-31.2) more than DCF-Bug (EM: 13.6-40.5, EP: 35.1-66.8) and CR-Trans (EM: 17.2-51.8, EP: 35.0-72.4), when using $EM$ and $EP$ as correctness metrics.
However, when considering the more lenient $CP$ metric in DCF-Vul, all models yielded far stronger performances (CP: 40.2-72.9).
For DCF-Vul, we omit exact match as a calibration target because models typically achieve less than 10\% accuracy, resulting in the available positive signal being too sparse for effective calibration.
The remaining correctness metrics are retained as models attain comparatively reliable performances by their measurements, rendering them more suitable for decision-making or thresholding.
As expected, we find that larger model variants generally outperform their lightweight counterparts across all correctness metrics and tasks.
Overall, CodeLlama-70B achieves the best performances across both DCF-Bug and CR-Trans, whilst Llama-3.1-70B achieves the best performance for DCF-Vul.

\begin{table}[!htbp]
\caption{Correctness (\%) Achieved by LLMs in ACR Tasks}
\centering
\resizebox{\columnwidth}{!}{%
\begin{tabular}{!{\vrule width 1.2pt}
>{\columncolor[HTML]{F0C88C}}l |
>{\columncolor[HTML]{FFE7C8}}r !{\vrule width 1.2pt}
>{\columncolor[HTML]{FFF5E3}}r 
>{\columncolor[HTML]{FFF5E3}}r |
>{\columncolor[HTML]{FFF5E3}}r 
>{\columncolor[HTML]{FFF5E3}}r 
>{\columncolor[HTML]{FFF5E3}}r |
>{\columncolor[HTML]{FFF5E3}}r 
>{\columncolor[HTML]{FFF5E3}}r !{\vrule width 1.2pt}}
\Xhline{2\arrayrulewidth}
\cellcolor[HTML]{F0C88C} & \cellcolor[HTML]{F0C88C} & \multicolumn{2}{c|}{\cellcolor[HTML]{F0C88C}\textbf{DCF-Bug}} & \multicolumn{3}{c|}{\cellcolor[HTML]{F0C88C}\textbf{DCF-Vul}} & \multicolumn{2}{c!{\vrule width 1.2pt}}{\cellcolor[HTML]{F0C88C}\textbf{CR-Trans}} \\
\cellcolor[HTML]{F0C88C} & \cellcolor[HTML]{F0C88C} & \multicolumn{2}{c|}{\cellcolor[HTML]{F0C88C}\textit{n=777}} & \multicolumn{3}{c|}{\cellcolor[HTML]{F0C88C}\textit{n=1,041}} & \multicolumn{2}{c!{\vrule width 1.2pt}}{\cellcolor[HTML]{F0C88C}\textit{n=900}} \\ \cline{3-9} 
\multirow{-3}{*}{\cellcolor[HTML]{F0C88C}\textbf{Model}} & \multirow{-3}{*}{\cellcolor[HTML]{F0C88C}\textbf{Size}} & \multicolumn{1}{c}{\cellcolor[HTML]{FFE7C8}\textbf{EM}} & \multicolumn{1}{c|}{\cellcolor[HTML]{FFE7C8}\textbf{EP}} & \multicolumn{1}{c}{\cellcolor[HTML]{FFE7C8}\textbf{EM}} & \multicolumn{1}{c}{\cellcolor[HTML]{FFE7C8}\textbf{EP}} & \multicolumn{1}{l|}{\cellcolor[HTML]{FFE7C8}\textbf{CP}} & \multicolumn{1}{c}{\cellcolor[HTML]{FFE7C8}\textbf{EM}} & \multicolumn{1}{c!{\vrule width 1.2pt}}{\cellcolor[HTML]{FFE7C8}\textbf{EP}} \\ \Xhline{2\arrayrulewidth}
\cellcolor[HTML]{F0C88C} & \textbf{8B} & 32.9 & 59.3 & {\color[HTML]{000000} 7.7} & {\color[HTML]{000000} 29.2} & {\color[HTML]{000000} 59.6} & 36.3 & 59.3 \\ \cline{2-9} 
\multirow{-2}{*}{\cellcolor[HTML]{F0C88C}\textbf{Llama-3.1}} & \textbf{70B} & 34.5 & 58.0 & {\color[HTML]{000000} 9.7} & {\color[HTML]{000000} \textbf{31.2}} & {\color[HTML]{000000} \textbf{72.9}} & 45.9 & 67.1 \\ \hline
\cellcolor[HTML]{F0C88C} & \textbf{7B} & 32.3 & 55.6 & {\color[HTML]{000000} 8.8} & {\color[HTML]{000000} 26.1} & {\color[HTML]{000000} 40.2} & 40.1 & 60.4 \\ \cline{2-9} 
\multirow{-2}{*}{\cellcolor[HTML]{F0C88C}\textbf{CodeLlama}} & \textbf{70B} & \textbf{40.5} & \textbf{66.8} & {\color[HTML]{000000} \textbf{12.0}} & {\color[HTML]{000000} 29.0} & {\color[HTML]{000000} 44.9} & \textbf{51.8} & \textbf{72.4} \\ \hline
\cellcolor[HTML]{F0C88C} & \textbf{7B} & 26.6 & 51.1 & {\color[HTML]{000000} 5.7} & {\color[HTML]{000000} 18.6} & {\color[HTML]{000000} 62.6} & 28.2 & 53.8 \\ \cline{2-9} 
\multirow{-2}{*}{\cellcolor[HTML]{F0C88C}\textbf{Qwen2.5}} & \textbf{72B} & 27.9 & 52.1 & {\color[HTML]{000000} 3.7} & {\color[HTML]{000000} 16.0} & {\color[HTML]{000000} 70.2} & 46.0 & 71.0 \\ \hline
\cellcolor[HTML]{F0C88C} & \textbf{7B} & 28.8 & 55.3 & {\color[HTML]{000000} 5.6} & {\color[HTML]{000000} 20.2} & {\color[HTML]{000000} 66.0} & 27.9 & 49.9 \\ \cline{2-9} 
\multirow{-2}{*}{\cellcolor[HTML]{F0C88C}\textbf{Qwen2.5-Coder}} & \textbf{32B} & 32.8 & 57.4 & {\color[HTML]{000000} 5.8} & {\color[HTML]{000000} 20.3} & {\color[HTML]{000000} 70.3} & 37.9 & 64.0 \\ \hline
\cellcolor[HTML]{F0C88C} & \textbf{7B} & 13.6 & 35.1 & {\color[HTML]{000000} 4.0} & {\color[HTML]{000000} 15.7} & {\color[HTML]{000000} 46.0} & 17.7 & 40.1 \\ \cline{2-9} 
\multirow{-2}{*}{\cellcolor[HTML]{F0C88C}\textbf{DS-LLM}} & \textbf{67B} & 35.3 & 62.7 & {\color[HTML]{000000} 9.1} & {\color[HTML]{000000} 28.9} & {\color[HTML]{000000} 49.9} & 38.4 & 62.1 \\ \hline
\cellcolor[HTML]{F0C88C} & \textbf{7B} & 30.5 & 58.0 & {\color[HTML]{000000} 8.6} & {\color[HTML]{000000} 27.1} & {\color[HTML]{000000} 62.4} & 35.6 & 58.9 \\ \cline{2-9} 
\multirow{-2}{*}{\cellcolor[HTML]{F0C88C}\textbf{DS-Coder}} & \textbf{33B} & 38.0 & 63.6 & {\color[HTML]{000000} 8.6} & {\color[HTML]{000000} 30.3} & {\color[HTML]{000000} 66.3} & 42.1 & 67.9 \\ \hline
\textbf{DS-V2} & \textbf{16B} & 18.9 & 39.0 & {\color[HTML]{000000} 4.5} & {\color[HTML]{000000} 19.2} & {\color[HTML]{000000} 47.8} & 17.2 & 35.0 \\ \hline
\textbf{DS-Coder-V2} & \textbf{16B} & 32.9 & 56.9 & {\color[HTML]{000000} 6.3} & {\color[HTML]{000000} 23.9} & {\color[HTML]{000000} 57.7} & 33.1 & 55.1 \\ \Xhline{2\arrayrulewidth}

\multicolumn{9}{l}{\footnotesize\textbf{Bold Number:} Best Result for that Task and Correctness Measure} \\

\multicolumn{9}{l}{\footnotesize\textbf{EM:} Exact Match, \textbf{EP:} Edit Progress, \textbf{CP:} Checks Passed} \\

\end{tabular}
}
\label{Tab:correctness}
\end{table}

\section{Results}
\label{sec:results}

\begin{figure}[h]
    \centering
    \includegraphics[width=0.80\columnwidth]{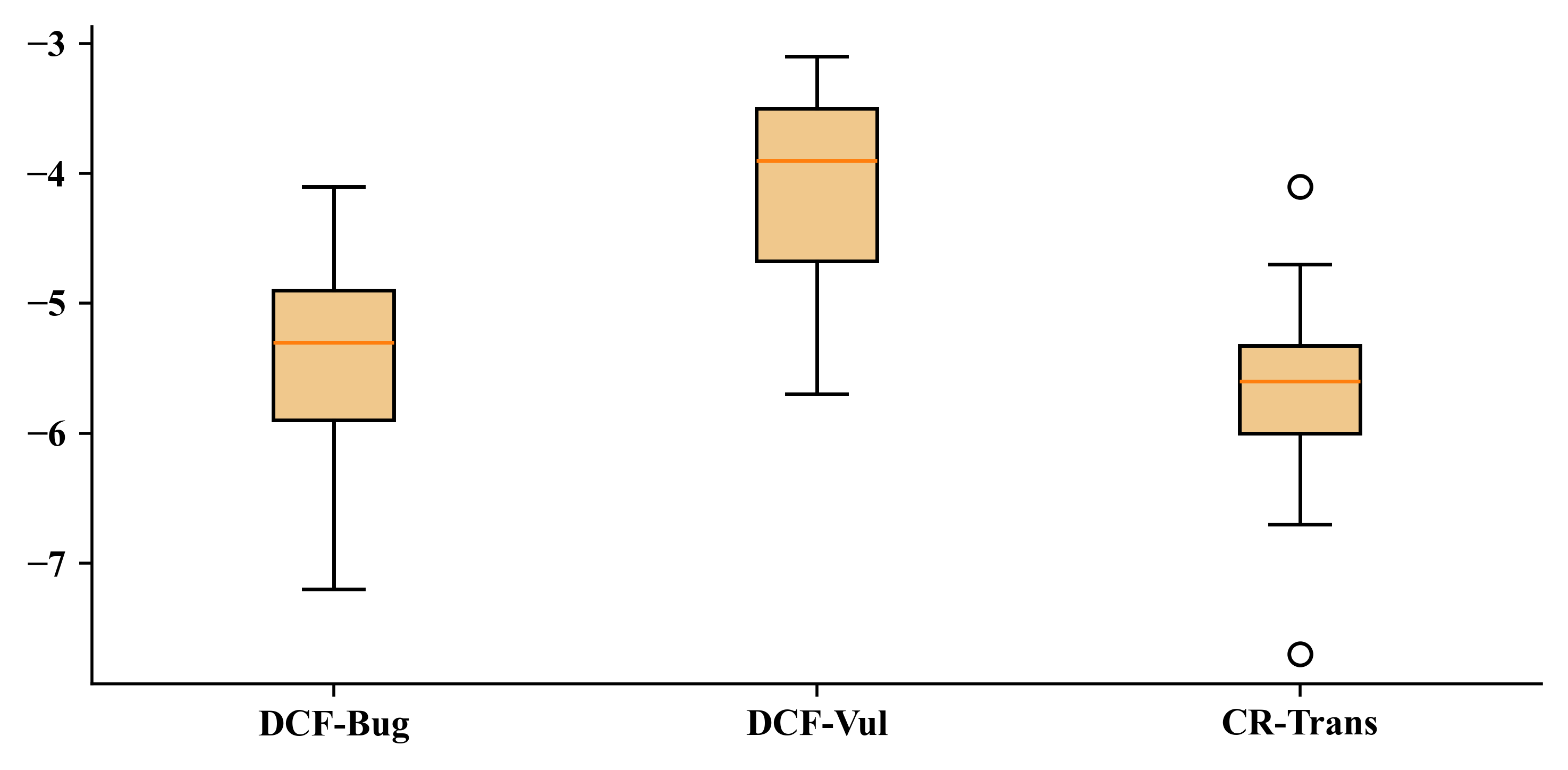} 
    \captionsetup{justification=centering} 
    \caption{Median Skewness of Token-Level Softmax Probabilities for LLM-Generated Code Revisions ($\tilde{\gamma}_1$)}
    \label{fig:median_skew}
\end{figure}

\textbf{RQ1. How well can fine-grained confidence scores separate correct from incorrect code revisions in ACR tasks?}
This forms the preliminary analysis on  token-level softmax probabilities and confidence scores before conducting calibration.
Figure~\ref{fig:median_skew} shows $\tilde{\gamma}_1$ values for each model and ACR task.~\footnote{\label{fn:myfoot}All p-values are statistically significant and are therefore omitted.}
Confirming our rationale, the distributions of softmax probabilities consistently exhibit strong negative skewness for all models and tasks.
This indicates that low softmax probabilities are consistently centred around a few outlier tokens in the generated code revisions.
The $\tilde{\gamma}_1$ values range from -7.2 to -4.1 for DCF-Bug, -5.7 to -3.1 for DCF-Vul, and -7.7 to -4.1 for CR-Trans.
In the case of sequence-level confidence scores, signals from these outlier tokens will be averaged away by the cluster of higher probabilities, either by the geometric mean in normalised sequence likelihood or by the arithmetic mean in the average token probability.
In contrast, the fine-grained confidence scores are designed to identify and focus on these particular outliers, whilst discarding the overpowering effects from the cluster of tokens in the high probability region.

We also compare the distributions of correct and incorrect code revisions under each confidence score to assess their inherent ability to separate and rank them.
We leverage two statistical measures, $W_1$ and $\tau_b$~\cite{kendall}, to facilitate the preliminary analysis. 
These two statistical measures capture the extent to which each confidence score exhibits a monotonic, calibratable relationship with the correctness label, providing an initial indication of their amenability to Platt-scaling.

\begin{figure}[htbp]
    \centering
    \includegraphics[width=\columnwidth]{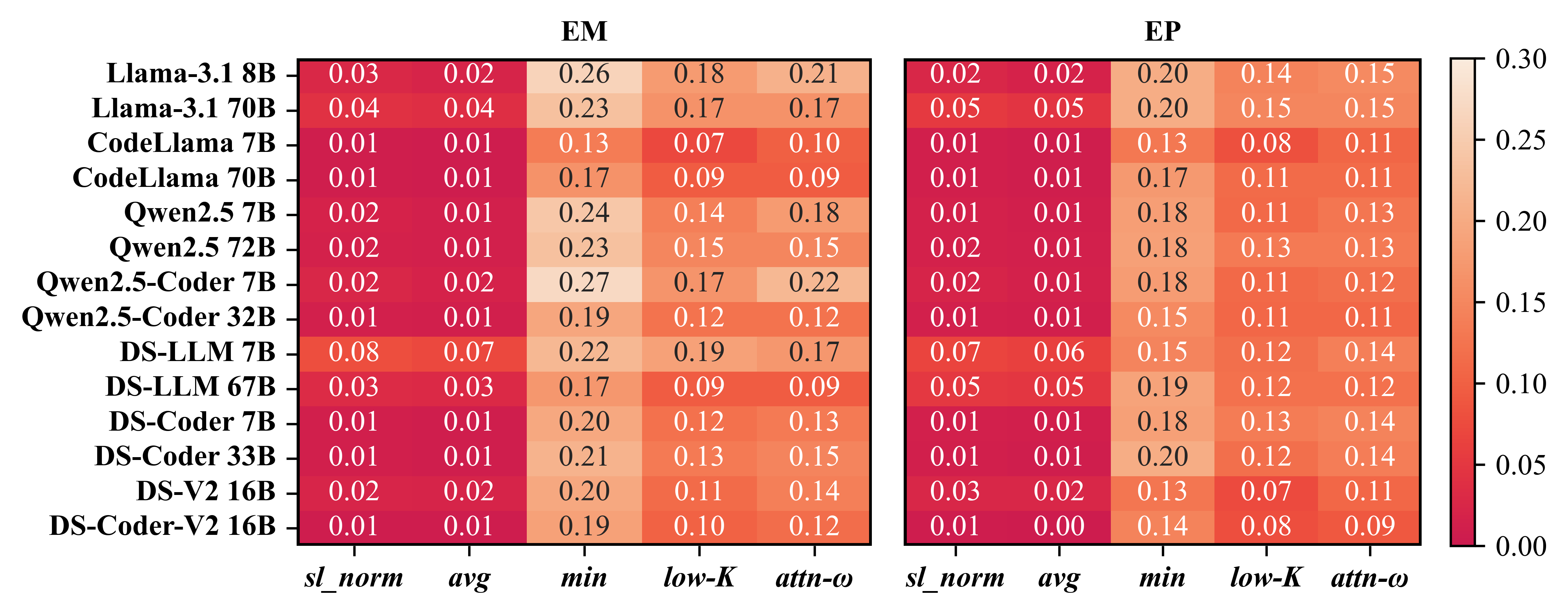} 
    \captionsetup{justification=centering} 
    \caption{DCF-Bug (Wasserstein Distance $W_1$)}
    \label{fig:DCF_Bug_ws}
\end{figure}

\begin{figure}[htbp]
    \centering
    \includegraphics[width=\columnwidth]{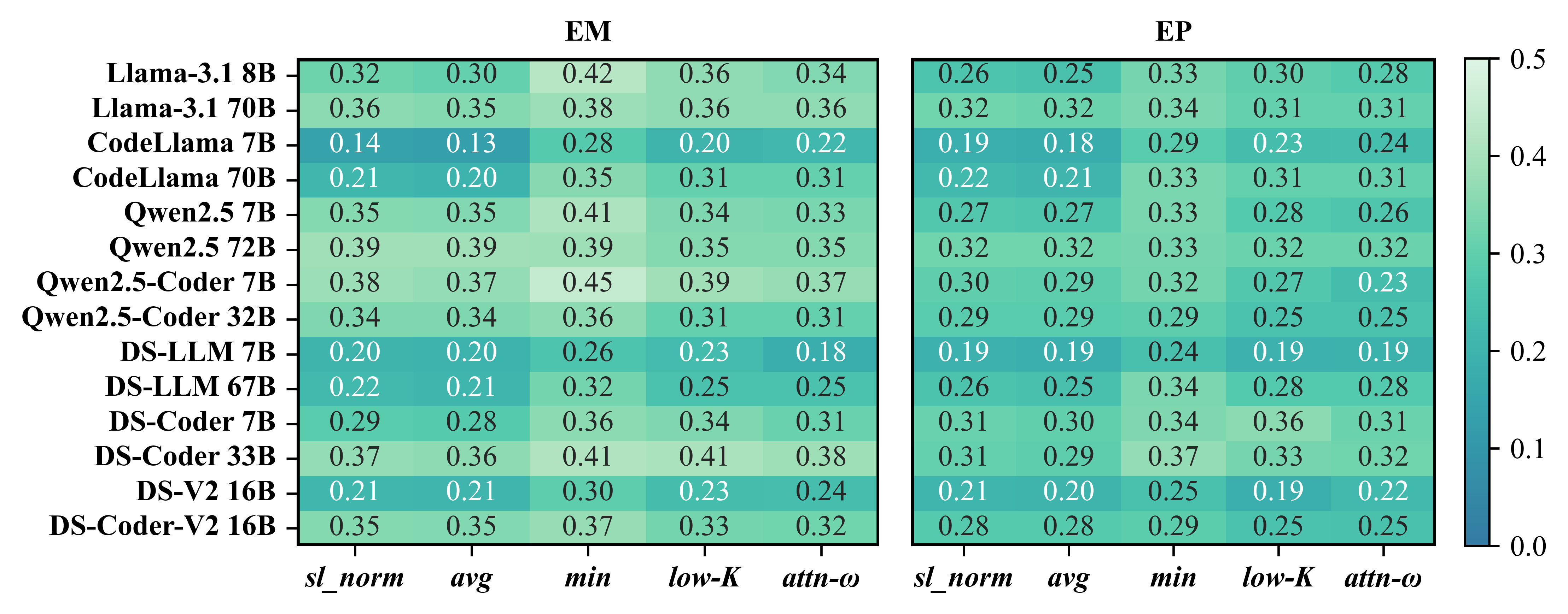} 
    \captionsetup{justification=centering} 
    \caption{DCF-Bug (Kendall's $\tau_b$ Coefficient)}
    \label{fig:DCF_Bug_Tau}
\end{figure}

Figures~\ref{fig:DCF_Bug_ws} and~\ref{fig:DCF_Bug_Tau} show the $W_1$ and $\tau_b$ for each confidence score and model in DCF-Bug.~\footnotemark[\value{footnote}]
For all models, we find that the fine-grained confidence scores are far better separated than sequence-level confidence scores.
Specifically, minimum token probability yields the widest separation (EM-$W_1$: 0.13-0.26, EP-$W_1$: 0.13-0.20), whilst sequence-level confidence scores have a separation of 0.01-0.03 for the majority of models against both correctness metrics.
In terms of rank correlation for both correctness metrics, we find that minimum token probability is consistently the top performer (EM-$\tau_b$: 0.26-0.45, EP-$\tau_b$: 0.24-0.37), whilst lowest-$K$ token probability and attention-weighted uncertainty can equal or outperform the sequence-level confidence scores for 10 out of 14 models.
The sequence-level confidence scores achieve near identical rank correlation (EM-$\tau_b$: 0.13-0.39, EP-$\tau_b$: 0.18-0.32), which are on average the worst performers.
These results suggest that fine-grained confidence scores are on average more amenable to Platt-scaling for DCF-Bug. 
As such, we expect more accurate confidence calibration when using them.

\begin{figure}[htbp]
    \centering
    \includegraphics[width=\columnwidth]{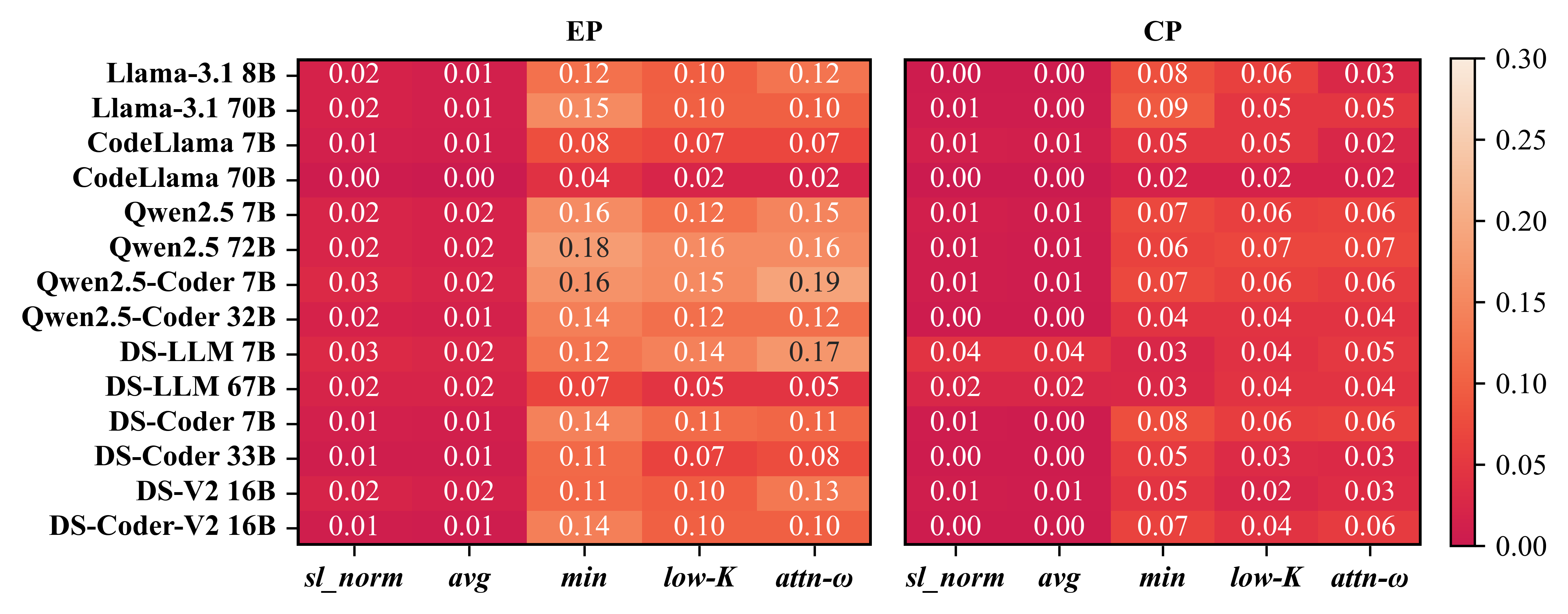} 
    \captionsetup{justification=centering} 
    \caption{DCF-Vul (Wasserstein Distance $W_1$)}
    \label{fig:DCF_Vul_ws}
\end{figure}

\begin{figure}[htbp]
    \centering
    \includegraphics[width=\columnwidth]{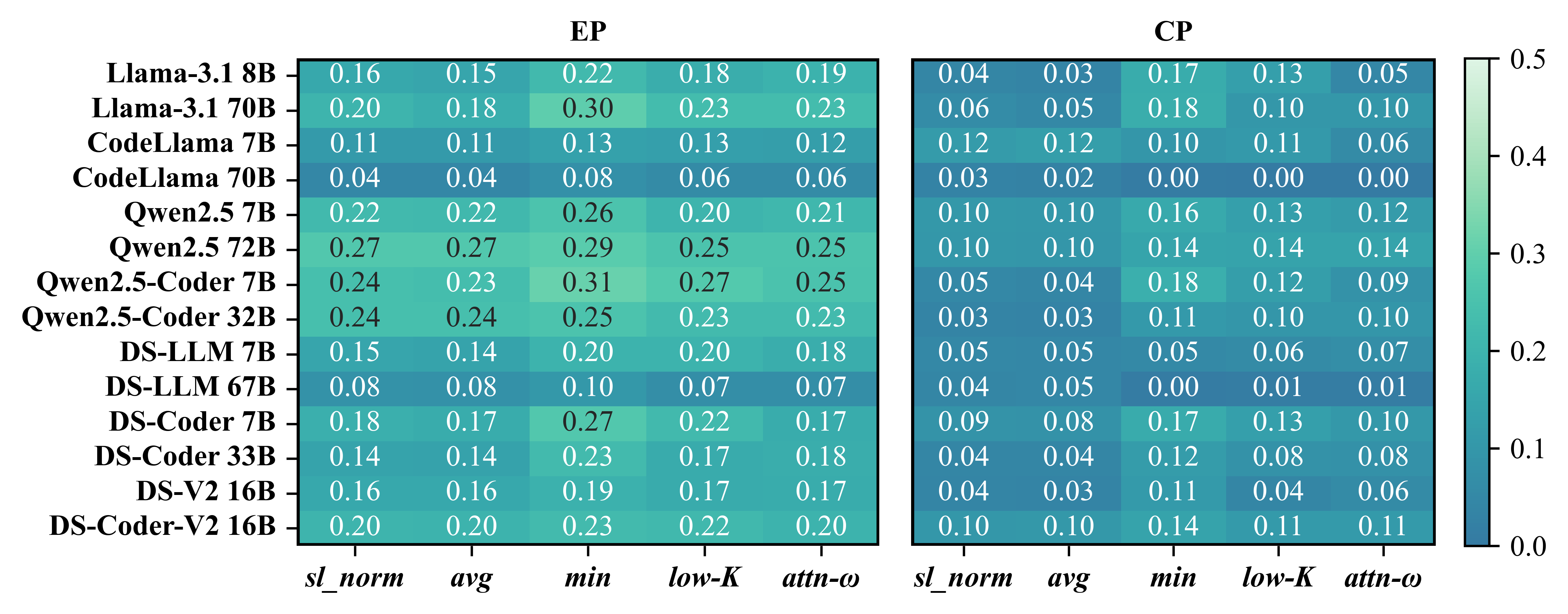} 
    \captionsetup{justification=centering} 
    \caption{DCF-Vul (Kendall's $\tau_b$  Coefficient)}
    \label{fig:DCF_Vul_Tau}
\end{figure}

Figures~\ref{fig:DCF_Vul_ws} and~\ref{fig:DCF_Vul_Tau} show the $W_1$ and $\tau_b$ for each confidence score and model in DCF-Vul.~\footnotemark[\value{footnote}]
Similar to the DCF-Bug, we find that fine-grained confidence scores are better separated than sequence-level confidence scores, with minimum token probability as the top performer (EP-$W_1$: 0.04-0.18, CP-$W_1$: 0.02-0.08).
The sequence-level confidence scores have a separation of 0.00-0.02 for the majority of models against both correctness metrics.
In terms of rank correlation for both correctness metrics, we still find that minimum token probability is consistently the top performer (EP-$\tau_b$: 0.08-0.31, CP-$\tau_b$: 0.00-0.18), whilst lowest-$K$ token probability and attention-weighted uncertainty can equal or outperform the sequence-level confidence scores for 10 out of 14 models.
Again, the sequence-level confidence scores achieve near identical rank correlation (EP-$\tau_b$: 0.04-0.27, CP-$\tau_b$: 0.03-0.12), which are on average the worst performers.
Whilst these results suggest that fine-grained confidence scores are more amenable to Platt-scaling for DCF-Vul, all $W_1$ and $\tau_b$ results are much weaker than found in DCF-Bug.
Therefore, it is expected that this task induces weaker calibration results for all confidence scores.

\begin{figure}[htbp]
    \centering
    \includegraphics[width=\columnwidth]{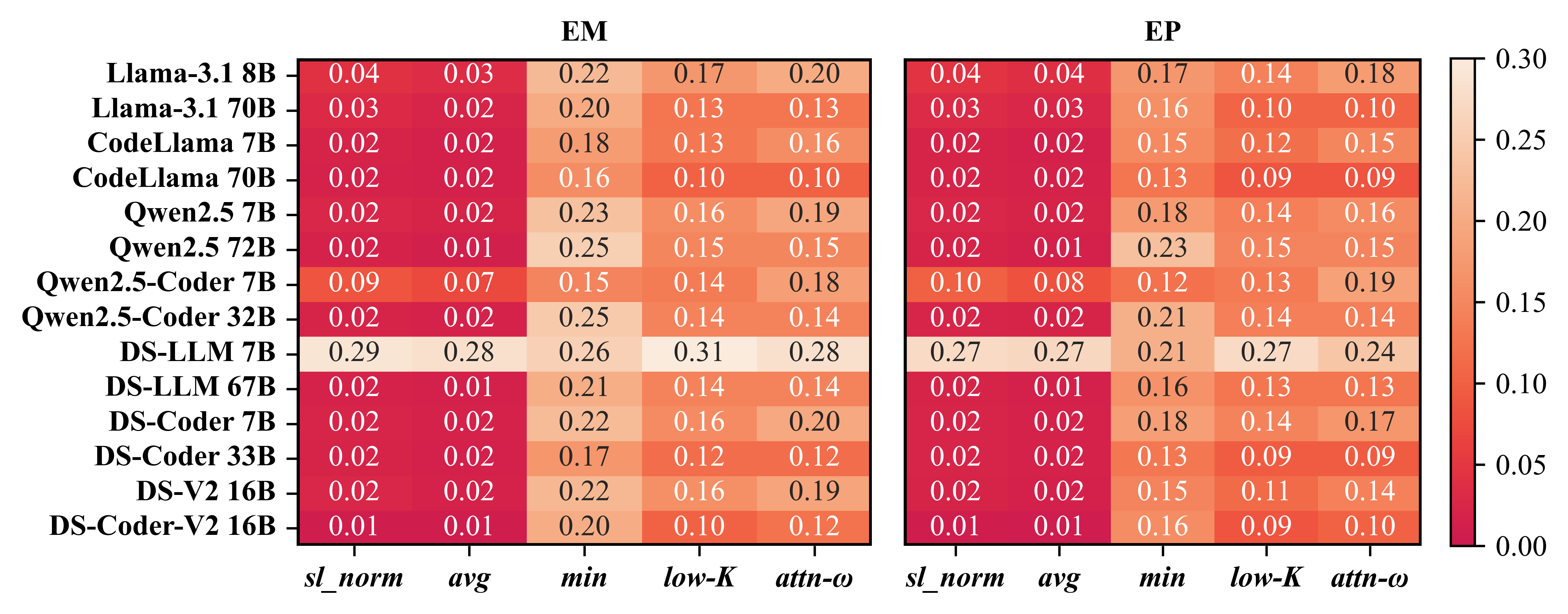} 
    \captionsetup{justification=centering} 
    \caption{CR-Trans (Wasserstein Distance $W_1$)}
    \label{fig:CR_Trans_ws}
\end{figure}

\begin{figure}[htbp]
    \centering
    \includegraphics[width=\columnwidth]{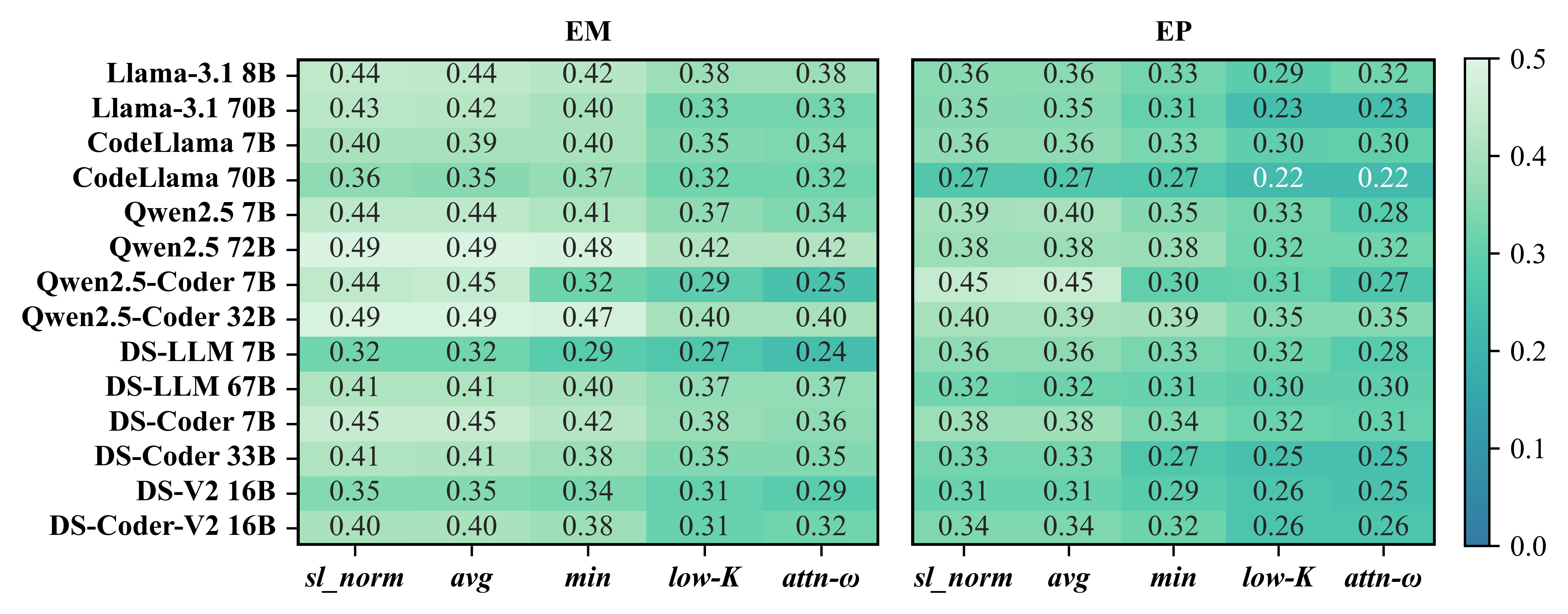} 
    \captionsetup{justification=centering} 
    \caption{CR-Trans (Kendall's $\tau_b$  Coefficient)}
    \label{fig:CR_Trans_Tau}
\end{figure}

Figures~\ref{fig:CR_Trans_ws} and~\ref{fig:CR_Trans_Tau} show the $W_1$ and $\tau_b$ for each confidence score and model in CR-Trans.~\footnotemark[\value{footnote}]
Similar to the prior tasks, we find that fine-grained confidence scores are far better separated than sequence-level confidence scores.
In general, we find minimum token probability yields the widest separation (EM-$W_1$: 0.15-0.26, EP-$W_1$: 0.12-0.23), whilst sequence-level confidence scores again have a separation of only 0.01-0.03 for the majority of models against both correctness metrics.
In terms of rank correlation, we find an interesting phenomenon that contrasts behaviours found in DCF-Bug and DCF-Vul, where sequence-level confidence scores are consistently the top performers (EM-$\tau_b$: 0.32-0.49, EP-$\tau_b$: 0.31-0.45).
On the other hand, fine-grained confidence scores yield weaker correlations for this task (EM-$\tau_b$: 0.25-0.48, EP-$\tau_b$: 0.22-0.39).
This behaviour can be explained by the fundamental difference in solution space between CR-Trans and the two prior tasks.
In short, DCF-Bug and DCF-Vul are both expecting code revisions to address explicit static analysis warnings that have narrow solutions spaces (e.g., \textit{Cookie has the Secure attribute set to false. Set it to true to protect the cookie from man-in-the-middle attacks.}), whilst CR-Trans expects code revisions to address human-written reviews that allow for wider solution spaces (e.g., \textit{We'll have to think about a unified system for integrating regularization as part of the loss}).
Therefore, lower token probabilities in CR-Trans can be a result of multiple correct alternative code edits sharing the model's solution space (aleatoric uncertainty), rather than the model truly being unaware of the correct code edit (epistemic uncertainty)~\cite{KuhnGF23}.
This weakens lower token probabilities' correlations with correctness.
Nevertheless, we do not expect sequence-level confidence scores to be more amenable to Platt-scaling for CR-Trans.
Since $W_1$ for sequence-level measures are still extremely low, the calibrator will be required to learn an unstable $w$ that conflicts with standard regularisation.
Ultimately, given the substantially stronger $W_1$ results for fine-grained confidence scores, alongside $\tau_b$ results that remain comparable, we still expect these methods to yield more accurate confidence calibration overall.

\summary{The empirical evidence shows that generated code revisions exhibit strong negative skewness ($\tilde{\gamma}_1$) in token probabilities. 
As such, fine-grained confidence scores preserve low-probability outlier signals that are otherwise suppressed by sequence-level averaging.
This directly induces stronger separation of correct from incorrect code revisions (as evidenced by higher $W_1$ and $\tau_b$), rendering them more amenable to Platt-scaling for ACR tasks.}

\begin{table*}[!t]  
\caption{Global Platt-Scaling Results for {DCF-Bug} in Terms of Exact Match and Edit Progress}
\centering
\resizebox{\textwidth}{!}{%
\begin{tabular}{!{\vrule width 1.2pt}
>{\columncolor[HTML]{F9B83C}}l |
>{\columncolor[HTML]{F9B83C}}r !{\vrule width 1.2pt}
>{\columncolor[HTML]{FFE7C8}}c 
>{\columncolor[HTML]{FFE7C8}}c 
>{\columncolor[HTML]{FFE7C8}}c 
>{\columncolor[HTML]{FFE7C8}}c 
>{\columncolor[HTML]{FFE7C8}}c 
>{\columncolor[HTML]{FFE7C8}}c 
>{\columncolor[HTML]{FFE7C8}}c 
>{\columncolor[HTML]{FFE7C8}}c 
>{\columncolor[HTML]{FFE7C8}}c 
>{\columncolor[HTML]{FFE7C8}}c 
>{\columncolor[HTML]{FFE7C8}}c 
>{\columncolor[HTML]{FFE7C8}}c 
>{\columncolor[HTML]{FFE7C8}}c 
>{\columncolor[HTML]{FFE7C8}}c 
>{\columncolor[HTML]{FFE7C8}}c 
>{\columncolor[HTML]{FFF5E3}}c 
>{\columncolor[HTML]{FFF5E3}}c 
>{\columncolor[HTML]{FFF5E3}}c 
>{\columncolor[HTML]{FFF5E3}}c 
>{\columncolor[HTML]{FFF5E3}}c 
>{\columncolor[HTML]{FFF5E3}}c 
>{\columncolor[HTML]{FFF5E3}}c 
>{\columncolor[HTML]{FFF5E3}}c 
>{\columncolor[HTML]{FFF5E3}}c 
>{\columncolor[HTML]{FFF5E3}}c 
>{\columncolor[HTML]{FFF5E3}}c 
>{\columncolor[HTML]{FFF5E3}}c 
>{\columncolor[HTML]{FFF5E3}}c 
>{\columncolor[HTML]{FFF5E3}}c 
>{\columncolor[HTML]{FFF5E3}}c !{\vrule width 1.2pt}}
\Xhline{2\arrayrulewidth}

\rowcolor[HTML]{F0C88C} 
\cellcolor[HTML]{F0C88C} & \multicolumn{1}{l!{\vrule width 1.2pt}}{\cellcolor[HTML]{F0C88C}} & \multicolumn{15}{c!{\vrule width 1.2pt}}{\cellcolor[HTML]{F0C88C}\textbf{EM}} & \multicolumn{15}{c!{\vrule width 1.2pt}}{\cellcolor[HTML]{F0C88C}\textbf{EP}} \\
\end{tabular}
}
\label{Tab:dcf_bug}
\end{table*}

\begin{table*}[!t]  
\caption{Global Platt-Scaling Results for {DCF-Vul} in Terms of Edit Progress and Checks Passed}
\centering
\resizebox{\textwidth}{!}{%
\begin{tabular}{!{\vrule width 1.2pt}
>{\columncolor[HTML]{F9B83C}}l |
>{\columncolor[HTML]{F9B83C}}r !{\vrule width 1.2pt}
>{\columncolor[HTML]{FFE7C8}}c 
>{\columncolor[HTML]{FFE7C8}}c 
>{\columncolor[HTML]{FFE7C8}}c 
>{\columncolor[HTML]{FFE7C8}}c 
>{\columncolor[HTML]{FFE7C8}}c 
>{\columncolor[HTML]{FFE7C8}}c 
>{\columncolor[HTML]{FFE7C8}}c 
>{\columncolor[HTML]{FFE7C8}}c 
>{\columncolor[HTML]{FFE7C8}}c 
>{\columncolor[HTML]{FFE7C8}}c 
>{\columncolor[HTML]{FFE7C8}}c 
>{\columncolor[HTML]{FFE7C8}}c 
>{\columncolor[HTML]{FFE7C8}}c 
>{\columncolor[HTML]{FFE7C8}}c 
>{\columncolor[HTML]{FFE7C8}}c 
>{\columncolor[HTML]{FFF5E3}}c 
>{\columncolor[HTML]{FFF5E3}}c 
>{\columncolor[HTML]{FFF5E3}}c 
>{\columncolor[HTML]{FFF5E3}}c 
>{\columncolor[HTML]{FFF5E3}}c 
>{\columncolor[HTML]{FFF5E3}}c 
>{\columncolor[HTML]{FFF5E3}}c 
>{\columncolor[HTML]{FFF5E3}}c 
>{\columncolor[HTML]{FFF5E3}}c 
>{\columncolor[HTML]{FFF5E3}}c 
>{\columncolor[HTML]{FFF5E3}}c 
>{\columncolor[HTML]{FFF5E3}}c 
>{\columncolor[HTML]{FFF5E3}}c 
>{\columncolor[HTML]{FFF5E3}}c 
>{\columncolor[HTML]{FFF5E3}}c !{\vrule width 1.2pt}}
\Xhline{2\arrayrulewidth}

\rowcolor[HTML]{F0C88C} 
\cellcolor[HTML]{F0C88C} & \multicolumn{1}{l!{\vrule width 1.2pt}}{\cellcolor[HTML]{F0C88C}} & \multicolumn{15}{c!{\vrule width 1.2pt}}{\cellcolor[HTML]{F0C88C}\textbf{EP}} & \multicolumn{15}{c!{\vrule width 1.2pt}}{\cellcolor[HTML]{F0C88C}\textbf{CP}} \\
\end{tabular}
}
\label{Tab:dcf_vul}
\end{table*}

\begin{table*}[!t]  
\caption{Global Platt-Scaling Results for {CR-Trans} in Terms of Exact Match and Edit Progress}
\centering
\resizebox{\textwidth}{!}{%
\begin{tabular}{!{\vrule width 1.2pt}
>{\columncolor[HTML]{F9B83C}}l |
>{\columncolor[HTML]{F9B83C}}r !{\vrule width 1.2pt}
>{\columncolor[HTML]{FFE7C8}}c 
>{\columncolor[HTML]{FFE7C8}}c 
>{\columncolor[HTML]{FFE7C8}}c 
>{\columncolor[HTML]{FFE7C8}}c 
>{\columncolor[HTML]{FFE7C8}}c 
>{\columncolor[HTML]{FFE7C8}}c 
>{\columncolor[HTML]{FFE7C8}}c 
>{\columncolor[HTML]{FFE7C8}}c 
>{\columncolor[HTML]{FFE7C8}}c 
>{\columncolor[HTML]{FFE7C8}}c 
>{\columncolor[HTML]{FFE7C8}}c 
>{\columncolor[HTML]{FFE7C8}}c 
>{\columncolor[HTML]{FFE7C8}}c 
>{\columncolor[HTML]{FFE7C8}}c 
>{\columncolor[HTML]{FFE7C8}}c 
>{\columncolor[HTML]{FFF5E3}}c 
>{\columncolor[HTML]{FFF5E3}}c 
>{\columncolor[HTML]{FFF5E3}}c 
>{\columncolor[HTML]{FFF5E3}}c 
>{\columncolor[HTML]{FFF5E3}}c 
>{\columncolor[HTML]{FFF5E3}}c 
>{\columncolor[HTML]{FFF5E3}}c 
>{\columncolor[HTML]{FFF5E3}}c 
>{\columncolor[HTML]{FFF5E3}}c 
>{\columncolor[HTML]{FFF5E3}}c 
>{\columncolor[HTML]{FFF5E3}}c 
>{\columncolor[HTML]{FFF5E3}}c 
>{\columncolor[HTML]{FFF5E3}}c 
>{\columncolor[HTML]{FFF5E3}}c 
>{\columncolor[HTML]{FFF5E3}}c !{\vrule width 1.2pt}}
\Xhline{2\arrayrulewidth}

\rowcolor[HTML]{F0C88C} 
\cellcolor[HTML]{F0C88C} & \multicolumn{1}{l!{\vrule width 1.2pt}}{\cellcolor[HTML]{F0C88C}} & \multicolumn{15}{c!{\vrule width 1.2pt}}{\cellcolor[HTML]{F0C88C}\textbf{EM}} & \multicolumn{15}{c!{\vrule width 1.2pt}}{\cellcolor[HTML]{F0C88C}\textbf{EP}} \\
\end{tabular}
}
\label{Tab:cr_trans}
\end{table*}

\textbf{RQ2. How effective is Platt-scaling for ACR tasks when using fine-grained confidence scores?}
The main calibration metrics are used to facilitate this analysis.
Specifically, we compare fine-grained and sequence-level confidence scores in terms of $ECE$, $\mathcal{B}$, and $BC$ after global Platt-scaling for each ACR task.
To platt-scale a confidence score, we train a single global calibrator with each task's training set and selected measure of correctness.
This produces a separate calibrator for each combination of model, task, confidence score, and correctness metric.
The calibrators are then deployed for inference on their respective test sets for evaluation.
Overall, the consistent improvements from fine-grained confidence scores across all models, correctness metrics, and ACR tasks demonstrate that our rationale holds in practice.

Table~\ref{Tab:dcf_bug} shows the $ECE$, $\mathcal{B}$, and $BC$ results after global Platt-scaling for DCF-Bug.
We find that sequence-level confidence scores exhibit the same phenomenon as found in past research~\cite{Spiess0PPRAJDA25}, where they are susceptible to single bin collapse.
In these cases, the calibrator was unable to learn an effective $w$ and defaulted to a global average for the task.
For normalised sequence likelihood, we find that single bin collapse occurred for three models against $EM$, and two models against $EP$.
For average token probability, we find that single bin collapse occurred for five models against both $EM$ and $EP$.
Whilst each sequence-level confidence score appear to attain the lowest $ECE$ for five models, they were achieved at the expense of extremely low bin coverage ($BC$: 2-3).
Thus, indicating inability to support granular decision-making for DCF-Bug.
Sequence-level confidence scores also consistently exhibit the highest instance-level miscalibrations ($\mathcal{B}$: 0.12-0.24).
In comparison, lowest-$K$ token probability and attention-weighted uncertainty can achieve the lowest $ECE$ for five models against $EM$ and four models against $EP$, with reasonable bin coverage ($BC$: 4-8).
We find that attention-weighted uncertainty can outperform lowest-$K$ token probability for six models against $EM$ and four models against $EP$, but does not produce a meaningful difference for the rest of the cases.
Overall, we find that minimum token probability achieves the lowest calibration errors on average ($ECE$: 0.04-0.08, $\mathcal{B}$: 0.11-0.22), whilst achieving the highest bin coverage ($BC$: 5-9).
Against both $EM$ and $EP$, minimum token probability can achieve the lowest $ECE$ for five models and the lowest $\mathcal{B}$ for all models.
These results align with our analysis in RQ1, where the measure achieved the highest $W_1$ and $\tau_b$.
Thus, minimum token probability is the most optimal choice when considering $EM$ and $EP$ in DCF-Bug, as it can provide the most accurate prediction for likelihood of correctness across the widest range of probability intervals.

Table~\ref{Tab:dcf_vul} shows the $ECE$, $\mathcal{B}$, and $BC$ results after global Platt-scaling for DCF-Vul.
In line with the low $W_1$ and $\tau_b$ results from RQ1, all confidence scores exhibit worse calibration performance compared to DCF-Bug.
For sequence-level confidence scores, the single bin collapse issue is more severe in DCF-Vul. 
Normalised sequence likelihood is completely degenerate for five models against $EP$ and eight models against $CP$, whilst average token probability is completely degenerate for eight models against $EP$, and 11 models against $CP$.
For the remaining cases, sequence-level confidence scores still achieve extremely low bin coverage ($BC$: 2-3), indicating an inability to support granular decision-making for DCF-Vul.
For $EP$, lowest-$K$ token probability and attention-weighted uncertainty produce similar calibration errors ($ECE$: 0.02-0.05, $\mathcal{B}$: 0.12-0.20), achieving the lowest $ECE$ for 11 and nine models, respectively.
However, their bin coverage is generally low ($BC$: 2-5).
Against $EP$, minimum token probability achieves slightly lower calibration error than other fine-grained confidence scores ($ECE$: 0.01-0.05, $\mathcal{B}$: 0.12-0.20) with higher bin coverage ($BC$: 2-7).
This results in the lowest $ECE$ for seven models and lowest $\mathcal{B}$ for all models.
Against $CP$, minimum token probability, lowest-$K$ token probability, and attention-weighted uncertainty all achieve similar calibration error ($ECE$: 0.01-0.08, $\mathcal{B}$: 0.19-0.25), producing the lowest $ECE$ for four, eight and six models, respectively.
In general, the bin coverage against $CP$ was lower ($BC$: 2-5).
Against $CP$, single bin collapse also occurred with fine-grained confidence scores, albeit to a far lesser extent, affecting two, one, and five models, respectively.
Similar to prior results, minimum token probability is the most optimal choice when considering $EP$ and $CP$ in DCF-Vul, since it still achieves the lowest calibration error for the widest bin coverage.

Table~\ref{Tab:cr_trans} shows the $ECE$, $\mathcal{B}$, and $BC$ results after global Platt-scaling for CR-Trans.
Whilst fine-grained confidence scores can provide improved calibration results over sequence-level confidence scores, they struggled to achieve below 0.14 $ECE$. 
This indicates that global Platt-scaling is insufficient for supporting decision-making in CR-Trans.
Against $EM$, single bin collapse occurred for six models using normalised sequence likelihood and nine models using average token probability.
For the remaining cases, they exhibited the highest calibration error ($ECE$: 0.11-0.31, $\mathcal{B}$: 0.20-0.34), whilst covering the lowest amount of bins ($BC$: 2-3).
These results reflect our $W_1$ findings in RQ1.
We find that attention-weighted uncertainty can outperform lowest-$K$ token probability for three models, but achieve identical results for the rest of the cases ($ECE$: 0.05-0.25, $\mathcal{B}$: 0.09-0.29).
Similar to prior tasks, minimum token probability generally yields the lowest calibration errors ($ECE$: 0.06-0.21, $\mathcal{B}$: 0.09-0.25) with the highest bin coverage ($BC$: 2-7).
Against $EP$, single bin collapse occurred for two models using normalised sequence likelihood and three models using average token probability.
For the remaining cases, they still exhibited the highest calibration error ($ECE$: 0.09-0.32, $\mathcal{B}$: 0.17-0.31), whilst covering the lowest amount of bins ($BC$: 2-3).
Against $EP$, lowest-$K$ token probability and attention-weighted uncertainty also achieved near identical results ($ECE$: 0.09-0.28, $\mathcal{B}$: 0.16-0.29).
Similar to prior results, minimum token probability generally achieves the lowest calibration error ($ECE$: 0.09-0.26, $\mathcal{B}$: 0.16-0.27), with the highest bin coverage ($BC$: 4-7).
Although fine-grained confidence scores can improve global Platt-scaling for CR-Trans, the extent of miscalibration is still unfit for facilitating accurate decision-making.

\summary{The empirical results demonstrate that Platt-scaling is highly effective for ACR tasks when using fine-grained confidence scores. 
They consistently produce lower calibration errors over a wider range of probability intervals compared to sequence-level scores. 
Specifically, minimum token probability yields the strongest performance across $ECE$, $\mathcal{B}$, and $BC$ for all ACR tasks.}

\textbf{RQ3. To what extent can local Platt-scaling improve over conventional Platt-scaling in ACR tasks?}
The main calibration metrics are used to facilitate this analysis.
Based on prior results, we find that global Platt-scaling is generally sufficient for achieving low calibration error in DCF-Bug and DCF-Vul (best calibration results $\leq$ 0.08 $ECE$), yet falls short for CR-Trans (best calibration results generally $\geq$ 0.14 $ECE$).
This discrepancy suggests that CR-Trans exhibits higher error heterogeneity, presenting a significant opportunity for local Platt-scaling to further improve confidence calibration.
In contrast, the two former tasks offer only marginal opportunities for further calibration gains.
We apply local Platt-scaling to each scenario, training a separate set of clusters and calibrators for each combination of model, confidence score, correctness metric, and ACR task, before evaluating on their respective test sets. 
Overall, the consistent improvements from local Platt-scaling across all models, correctness metrics, and ACR tasks demonstrate that our rationale holds in practice.

\begin{table*}[!t]  
\caption{Local Platt-Scaling Results for {DCF-Bug} in Terms of Exact Match and Edit Progress}
\centering
\resizebox{\textwidth}{!}{%
\begin{tabular}{!{\vrule width 1.2pt}
>{\columncolor[HTML]{F9B83C}}l |
>{\columncolor[HTML]{F9B83C}}r !{\vrule width 1.2pt}
>{\columncolor[HTML]{FFE7C8}}c 
>{\columncolor[HTML]{FFE7C8}}c 
>{\columncolor[HTML]{FFE7C8}}c 
>{\columncolor[HTML]{FFE7C8}}c 
>{\columncolor[HTML]{FFE7C8}}c 
>{\columncolor[HTML]{FFE7C8}}c 
>{\columncolor[HTML]{FFE7C8}}c 
>{\columncolor[HTML]{FFE7C8}}c 
>{\columncolor[HTML]{FFE7C8}}c 
>{\columncolor[HTML]{FFF5E3}}c 
>{\columncolor[HTML]{FFF5E3}}c 
>{\columncolor[HTML]{FFF5E3}}c 
>{\columncolor[HTML]{FFF5E3}}c 
>{\columncolor[HTML]{FFF5E3}}c 
>{\columncolor[HTML]{FFF5E3}}c 
>{\columncolor[HTML]{FFF5E3}}c 
>{\columncolor[HTML]{FFF5E3}}c
>{\columncolor[HTML]{FFF5E3}}c !{\vrule width 1.2pt}}
\Xhline{2\arrayrulewidth}

\rowcolor[HTML]{F0C88C} 
\cellcolor[HTML]{F0C88C} & \multicolumn{1}{l!{\vrule width 1.2pt}}{\cellcolor[HTML]{F0C88C}} & \multicolumn{9}{c!{\vrule width 1.2pt}}{\cellcolor[HTML]{F0C88C}\textbf{EM}} & \multicolumn{9}{c!{\vrule width 1.2pt}}{\cellcolor[HTML]{F0C88C}\textbf{EP}} \\ \cline{3-20}

\rowcolor[HTML]{F0C88C} 
\cellcolor[HTML]{F0C88C} & \multicolumn{1}{l!{\vrule width 1.2pt}}{\cellcolor[HTML]{F0C88C}} & \multicolumn{3}{c|}{\cellcolor[HTML]{F0C88C}\textit{\textbf{min}}} & \multicolumn{3}{c|}{\cellcolor[HTML]{F0C88C}\textit{\textbf{low-K}}} & \multicolumn{3}{c!{\vrule width 1.2pt}}{\cellcolor[HTML]{F0C88C}\textit{\textbf{attn-$\omega$}}} & \multicolumn{3}{c|}{\cellcolor[HTML]{F0C88C}\textit{\textbf{min}}} & \multicolumn{3}{c|}{\cellcolor[HTML]{F0C88C}\textit{\textbf{low-K}}} & \multicolumn{3}{c!{\vrule width 1.2pt}}{\cellcolor[HTML]{F0C88C}\textit{\textbf{attn-$\omega$}}} \\ \cline{3-20}

\rowcolor[HTML]{FFE7C8} 
\multirow{-3}{*}{\cellcolor[HTML]{F0C88C}\textbf{Model}} & \multicolumn{1}{l!{\vrule width 1.2pt}}{\multirow{-3}{*}{\cellcolor[HTML]{F0C88C}\textbf{Size}}} & \textit{\textbf{ECE}} & \textit{\textbf{$\boldsymbol{\mathcal{B}}$}} & \multicolumn{1}{c|}{\cellcolor[HTML]{FFE7C8}\textit{\textbf{BC}}} & \textit{\textbf{ECE}} & \textit{\textbf{$\boldsymbol{\mathcal{B}}$}} & \multicolumn{1}{c|}{\cellcolor[HTML]{FFE7C8}\textit{\textbf{BC}}} & \textit{\textbf{ECE}} & \textit{\textbf{$\boldsymbol{\mathcal{B}}$}} & \multicolumn{1}{c!{\vrule width 1.2pt}}{\cellcolor[HTML]{FFE7C8}\textit{\textbf{BC}}} & \textit{\textbf{ECE}} & \textit{\textbf{$\boldsymbol{\mathcal{B}}$}} & \multicolumn{1}{c|}{\cellcolor[HTML]{FFE7C8}\textit{\textbf{BC}}} & \textit{\textbf{ECE}} & \textit{\textbf{$\boldsymbol{\mathcal{B}}$}} & \multicolumn{1}{c|}{\cellcolor[HTML]{FFE7C8}\textit{\textbf{BC}}} & \textit{\textbf{ECE}} & \textit{\textbf{$\boldsymbol{\mathcal{B}}$}} & \textit{\textbf{BC}} \\ \Xhline{2\arrayrulewidth}

\rowcolor[HTML]{FFF5E3} 
\cellcolor[HTML]{F0C88C} & \cellcolor[HTML]{FFE7C8}\textbf{8B} & {\color[HTML]{38761D} \textbf{0.03 (-0.02)}} & {\color[HTML]{38761D} \textbf{0.15 (-0.01)}} & \multicolumn{1}{c|}{\cellcolor[HTML]{FFF5E3}{\color[HTML]{000000} \textbf{9 (=0)}}} & {\color[HTML]{38761D} \textbf{0.03 (-0.03)}} & {\color[HTML]{000000} 0.18 (=0)} & \multicolumn{1}{c|}{\cellcolor[HTML]{FFF5E3}{\color[HTML]{000000} 7 (=0)}} & {\color[HTML]{38761D} \textbf{0.03 (-0.01)}} & {\color[HTML]{000000} 0.18 (=0)} & \multicolumn{1}{c!{\vrule width 1.2pt}}{\color[HTML]{000000} 7 (=0)} & {\color[HTML]{38761D} 0.06 (-0.02)} & {\color[HTML]{38761D} \textbf{0.20 (-0.01)}} & \multicolumn{1}{c|}{\cellcolor[HTML]{FFF5E3}{\color[HTML]{38761D} \textbf{9 (+1)}}} & {\color[HTML]{38761D} \textbf{0.05 (-0.03)}} & {\color[HTML]{000000} 0.21 (=0)} & \multicolumn{1}{c|}{\cellcolor[HTML]{FFF5E3}{\color[HTML]{38761D} 8 (+1)}} & {\color[HTML]{38761D} 0.06 (-0.02)} & {\color[HTML]{000000} 0.22 (=0)} & {\color[HTML]{38761D} \textbf{9 (+2)}} \\ \cline{2-20}

\rowcolor[HTML]{FFF5E3} 
\multirow{-2}{*}{\cellcolor[HTML]{F0C88C}\textbf{Llama-3.1}} & \cellcolor[HTML]{FFE7C8}\textbf{70B} & {\color[HTML]{38761D} \textbf{0.03 (-0.01)}} & {\color[HTML]{38761D} \textbf{0.17 (-0.01)}} & \multicolumn{1}{c|}{\cellcolor[HTML]{FFF5E3}{\color[HTML]{000000} \textbf{8 (=0)}}} & {\color[HTML]{000000} 0.04 (=0)} & {\color[HTML]{000000} 0.18 (=0)} & \multicolumn{1}{c|}{\cellcolor[HTML]{FFF5E3}{\color[HTML]{000000} 7 (=0)}} & {\color[HTML]{000000} 0.04 (=0)} & {\color[HTML]{000000} 0.18 (=0)} & \multicolumn{1}{c!{\vrule width 1.2pt}}{\color[HTML]{000000} 7 (=0)} & {\color[HTML]{38761D} \textbf{0.03 (-0.01)}} & {\color[HTML]{000000} \textbf{0.20 (=0)}} & \multicolumn{1}{c|}{\cellcolor[HTML]{FFF5E3}{\color[HTML]{38761D} \textbf{8 (+1)}}} & {\color[HTML]{000000} 0.04 (=0)} & {\color[HTML]{000000} 0.21 (=0)} & \multicolumn{1}{c|}{\cellcolor[HTML]{FFF5E3}{\color[HTML]{000000} 6 (=0)}} & {\color[HTML]{000000} 0.04 (=0)} & {\color[HTML]{000000} 0.21 (=0)} & {\color[HTML]{000000} 6 (=0)} \\ \hline

\rowcolor[HTML]{FFF5E3} 
\cellcolor[HTML]{F0C88C} & \cellcolor[HTML]{FFE7C8}\textbf{7B} & {\color[HTML]{38761D} \textbf{0.02 (-0.02)}} & {\color[HTML]{000000} \textbf{0.19 (=0)}} & \multicolumn{1}{c|}{\cellcolor[HTML]{FFF5E3}{\color[HTML]{000000} \textbf{8 (=0)}}} & {\color[HTML]{38761D} 0.03 (-0.03)} & {\color[HTML]{000000} 0.21 (=0)} & \multicolumn{1}{c|}{\cellcolor[HTML]{FFF5E3}{\color[HTML]{000000} 5 (=0)}} & {\color[HTML]{38761D} \textbf{0.02 (-0.02)}} & {\color[HTML]{000000} 0.20 (=0)} & \multicolumn{1}{c!{\vrule width 1.2pt}}{\color[HTML]{000000} 6 (=0)} & {\color[HTML]{38761D} \textbf{0.03 (-0.01)}} & {\color[HTML]{000000} \textbf{0.22 (=0)}} & \multicolumn{1}{c|}{\cellcolor[HTML]{FFF5E3}{\color[HTML]{000000} \textbf{7 (=0)}}} & {\color[HTML]{000000} \textbf{0.03 (=0)}} & {\color[HTML]{000000} 0.23 (=0)} & \multicolumn{1}{c|}{\cellcolor[HTML]{FFF5E3}{\color[HTML]{000000} 6 (=0)}} & {\color[HTML]{000000} \textbf{0.03 (=0)}} & {\color[HTML]{000000} 0.23 (=0)} & {\color[HTML]{000000} 6 (=0)} \\ \cline{2-20}

\rowcolor[HTML]{FFF5E3} 
\multirow{-2}{*}{\cellcolor[HTML]{F0C88C}\textbf{CodeLlama}} & \cellcolor[HTML]{FFE7C8}\textbf{70B} & {\color[HTML]{000000} \textbf{0.04 (=0)}} & {\color[HTML]{000000} \textbf{0.20 (=0)}} & \multicolumn{1}{c|}{\cellcolor[HTML]{FFF5E3}{\color[HTML]{000000} \textbf{8 (=0)}}} & {\color[HTML]{38761D} 0.07 (-0.01)} & {\color[HTML]{38761D} \textbf{0.20 (-0.01)}} & \multicolumn{1}{c|}{\cellcolor[HTML]{FFF5E3}{\color[HTML]{000000} 7 (=0)}} & {\color[HTML]{38761D} 0.07 (-0.01)} & {\color[HTML]{38761D} \textbf{0.20 (-0.01)}} & \multicolumn{1}{c!{\vrule width 1.2pt}}{\color[HTML]{000000} 7 (=0)} & {\color[HTML]{38761D} \textbf{0.04 (-0.01)}} & {\color[HTML]{000000} \textbf{0.18 (=0)}} & \multicolumn{1}{c|}{\cellcolor[HTML]{FFF5E3}{\color[HTML]{38761D} \textbf{7 (+1)}}} & {\color[HTML]{000000} 0.06 (=0)} & {\color[HTML]{000000} 0.19 (=0)} & \multicolumn{1}{c|}{\cellcolor[HTML]{FFF5E3}{\color[HTML]{000000} 5 (=0)}} & {\color[HTML]{000000} 0.06 (=0)} & {\color[HTML]{000000} 0.19 (=0)} & {\color[HTML]{000000} 5 (=0)} \\ \hline

\rowcolor[HTML]{FFF5E3} 
\cellcolor[HTML]{F0C88C} & \cellcolor[HTML]{FFE7C8}\textbf{7B} & {\color[HTML]{38761D} \textbf{0.04 (-0.02)}} & {\color[HTML]{000000} \textbf{0.15 (=0)}} & \multicolumn{1}{c|}{\cellcolor[HTML]{FFF5E3}{\color[HTML]{000000} \textbf{7 (=0)}}} & {\color[HTML]{38761D} 0.05 (-0.01)} & {\color[HTML]{000000} 0.16 (=0)} & \multicolumn{1}{c|}{\cellcolor[HTML]{FFF5E3}{\color[HTML]{000000} 6 (=0)}} & {\color[HTML]{38761D} \textbf{0.04 (-0.01)}} & {\color[HTML]{000000} 0.16 (=0)} & \multicolumn{1}{c!{\vrule width 1.2pt}}{\color[HTML]{000000} 6 (=0)} & {\color[HTML]{990000} 0.05 (+0.01)} & {\color[HTML]{000000} \textbf{0.21 (=0)}} & \multicolumn{1}{c|}{\cellcolor[HTML]{FFF5E3}{\color[HTML]{38761D} \textbf{9 (+2)}}} & {\color[HTML]{38761D} 0.05 (-0.04)} & {\color[HTML]{000000} 0.22 (=0)} & \multicolumn{1}{c|}{\cellcolor[HTML]{FFF5E3}{\color[HTML]{000000} 6 (=0)}} & {\color[HTML]{38761D} \textbf{0.04 (-0.03)}} & {\color[HTML]{000000} 0.23 (=0)} & {\color[HTML]{38761D} 7 (+1)} \\ \cline{2-20}

\rowcolor[HTML]{FFF5E3} 
\multirow{-2}{*}{\cellcolor[HTML]{F0C88C}\textbf{Qwen2.5}} & \cellcolor[HTML]{FFE7C8}\textbf{72B} & {\color[HTML]{38761D} \textbf{0.04 (-0.02)}} & {\color[HTML]{000000} \textbf{0.16 (=0)}} & \multicolumn{1}{c|}{\cellcolor[HTML]{FFF5E3}{\color[HTML]{000000} \textbf{7 (=0)}}} & {\color[HTML]{000000} 0.06 (=0)} & {\color[HTML]{38761D} \textbf{0.16 (-0.01)}} & \multicolumn{1}{c|}{\cellcolor[HTML]{FFF5E3}{\color[HTML]{000000} 6 (=0)}} & {\color[HTML]{000000} 0.06 (=0)} & {\color[HTML]{38761D} \textbf{0.16 (-0.01)}} & \multicolumn{1}{c!{\vrule width 1.2pt}}{\color[HTML]{000000} 6 (=0)} & {\color[HTML]{38761D} \textbf{0.03 (-0.05)}} & {\color[HTML]{000000} \textbf{0.21 (=0)}} & \multicolumn{1}{c|}{\cellcolor[HTML]{FFF5E3}{\color[HTML]{000000} \textbf{8 (=0)}}} & {\color[HTML]{38761D} 0.05 (-0.01)} & {\color[HTML]{000000} \textbf{0.21 (=0)}} & \multicolumn{1}{c|}{\cellcolor[HTML]{FFF5E3}{\color[HTML]{38761D} \textbf{8 (+1)}}} & {\color[HTML]{38761D} 0.05 (-0.01)} & {\color[HTML]{000000} \textbf{0.21 (=0)}} & {\color[HTML]{38761D} \textbf{8 (+1)}} \\ \hline

\rowcolor[HTML]{FFF5E3} 
\cellcolor[HTML]{F0C88C} & \cellcolor[HTML]{FFE7C8}\textbf{7B} & {\color[HTML]{38761D} 0.04 (-0.02)} & {\color[HTML]{000000} \textbf{0.14 (=0)}} & \multicolumn{1}{c|}{\cellcolor[HTML]{FFF5E3}{\color[HTML]{000000} \textbf{9 (=0)}}} & {\color[HTML]{38761D} 0.05 (-0.01)} & {\color[HTML]{38761D} 0.15 (-0.01)} & \multicolumn{1}{c|}{\cellcolor[HTML]{FFF5E3}{\color[HTML]{000000} 7 (=0)}} & {\color[HTML]{000000} \textbf{0.03 (=0)}} & {\color[HTML]{000000} 0.16 (=0)} & \multicolumn{1}{c!{\vrule width 1.2pt}}{\color[HTML]{000000} 7 (=0)} & {\color[HTML]{38761D} \textbf{0.05 (-0.02)}} & {\color[HTML]{000000} \textbf{0.21 (=0)}} & \multicolumn{1}{c|}{\cellcolor[HTML]{FFF5E3}{\color[HTML]{38761D} \textbf{9 (+1)}}} & {\color[HTML]{38761D} 0.08 (-0.01)} & {\color[HTML]{38761D} 0.22 (-0.01)} & \multicolumn{1}{c|}{\cellcolor[HTML]{FFF5E3}{\color[HTML]{000000} 8 (=0)}} & {\color[HTML]{000000} 0.10 (=0)} & {\color[HTML]{000000} 0.24 (=0)} & {\color[HTML]{000000} 7 (=0)} \\ \cline{2-20}

\rowcolor[HTML]{FFF5E3} 
\multirow{-2}{*}{\cellcolor[HTML]{F0C88C}\textbf{Qwen2.5-Coder}} & \cellcolor[HTML]{FFE7C8}\textbf{32B} & {\color[HTML]{38761D} \textbf{0.03 (-0.02)}} & {\color[HTML]{38761D} \textbf{0.17 (-0.01)}} & \multicolumn{1}{c|}{\cellcolor[HTML]{FFF5E3}{\color[HTML]{000000} \textbf{8 (=0)}}} & {\color[HTML]{000000} 0.04 (=0)} & {\color[HTML]{38761D} \textbf{0.17 (-0.02)}} & \multicolumn{1}{c|}{\cellcolor[HTML]{FFF5E3}{\color[HTML]{38761D} \textbf{8 (+2)}}} & {\color[HTML]{000000} 0.04 (=0)} & {\color[HTML]{38761D} \textbf{0.17 (-0.02)}} & \multicolumn{1}{c!{\vrule width 1.2pt}}{\color[HTML]{38761D} \textbf{8 (+2)}} & {\color[HTML]{38761D} \textbf{0.03 (-0.03)}} & {\color[HTML]{38761D} \textbf{0.20 (-0.02)}} & \multicolumn{1}{c|}{\cellcolor[HTML]{FFF5E3}{\color[HTML]{38761D} \textbf{8 (+1)}}} & {\color[HTML]{000000} 0.04 (=0)} & {\color[HTML]{38761D} 0.21 (-0.01)} & \multicolumn{1}{c|}{\cellcolor[HTML]{FFF5E3}{\color[HTML]{000000} 7 (=0)}} & {\color[HTML]{000000} 0.04 (=0)} & {\color[HTML]{38761D} 0.21 (-0.01)} & {\color[HTML]{38761D} \textbf{8 (+1)}} \\ \hline

\rowcolor[HTML]{FFF5E3} 
\cellcolor[HTML]{F0C88C} & \cellcolor[HTML]{FFE7C8}\textbf{7B} & {\color[HTML]{38761D} \textbf{0.02 (-0.02)}} & {\color[HTML]{38761D} \textbf{0.10 (-0.01)}} & \multicolumn{1}{c|}{\cellcolor[HTML]{FFF5E3}{\color[HTML]{000000} \textbf{5 (=0)}}} & {\color[HTML]{38761D} \textbf{0.02 (-0.03)}} & {\color[HTML]{000000} 0.11 (=0)} & \multicolumn{1}{c|}{\cellcolor[HTML]{FFF5E3}{\color[HTML]{38761D} 4 (+1)}} & {\color[HTML]{38761D} \textbf{0.02 (-0.02)}} & {\color[HTML]{000000} 0.11 (=0)} & \multicolumn{1}{c!{\vrule width 1.2pt}}{\color[HTML]{000000} 4 (=0)} & {\color[HTML]{38761D} \textbf{0.04 (-0.02)}} & {\color[HTML]{000000} \textbf{0.21 (=0)}} & \multicolumn{1}{c|}{\cellcolor[HTML]{FFF5E3}{\color[HTML]{000000} \textbf{8 (=0)}}} & {\color[HTML]{38761D} 0.05 (-0.03)} & {\color[HTML]{000000} 0.22 (=0)} & \multicolumn{1}{c|}{\cellcolor[HTML]{FFF5E3}{\color[HTML]{000000} 7 (=0)}} & {\color[HTML]{000000} 0.06 (=0)} & {\color[HTML]{000000} 0.22 (=0)} & {\color[HTML]{000000} 6 (=0)} \\ \cline{2-20}

\rowcolor[HTML]{FFF5E3} 
\multirow{-2}{*}{\cellcolor[HTML]{F0C88C}\textbf{DS-LLM}} & \cellcolor[HTML]{FFE7C8}\textbf{67B} & {\color[HTML]{38761D} \textbf{0.05 (-0.02)}} & {\color[HTML]{000000} \textbf{0.19 (=0)}} & \multicolumn{1}{c|}{\cellcolor[HTML]{FFF5E3}{\color[HTML]{000000} \textbf{8 (=0)}}} & {\color[HTML]{000000} 0.07 (=0)} & {\color[HTML]{38761D} 0.20 (-0.01)} & \multicolumn{1}{c|}{\cellcolor[HTML]{FFF5E3}{\color[HTML]{000000} 6 (=0)}} & {\color[HTML]{000000} 0.07 (=0)} & {\color[HTML]{38761D} 0.20 (-0.01)} & \multicolumn{1}{c!{\vrule width 1.2pt}}{\color[HTML]{000000} 6 (=0)} & {\color[HTML]{38761D} \textbf{0.03 (-0.04)}} & {\color[HTML]{38761D} \textbf{0.19 (-0.01)}} & \multicolumn{1}{c|}{\cellcolor[HTML]{FFF5E3}{\color[HTML]{000000} \textbf{7 (=0)}}} & {\color[HTML]{38761D} 0.06 (-0.02)} & {\color[HTML]{000000} 0.21 (=0)} & \multicolumn{1}{c|}{\cellcolor[HTML]{FFF5E3}{\color[HTML]{38761D} 6 (+1)}} & {\color[HTML]{38761D} 0.06 (-0.02)} & {\color[HTML]{000000} 0.21 (=0)} & {\color[HTML]{38761D} 6 (+1)} \\ \hline

\rowcolor[HTML]{FFF5E3} 
\cellcolor[HTML]{F0C88C} & \cellcolor[HTML]{FFE7C8}\textbf{7B} & {\color[HTML]{38761D} \textbf{0.03 (-0.04)}} & {\color[HTML]{000000} \textbf{0.17 (=0)}} & \multicolumn{1}{c|}{\cellcolor[HTML]{FFF5E3}{\color[HTML]{000000} \textbf{8 (=0)}}} & {\color[HTML]{38761D} 0.06 (-0.01)} & {\color[HTML]{38761D} \textbf{0.17 (-0.01)}} & \multicolumn{1}{c|}{\cellcolor[HTML]{FFF5E3}{\color[HTML]{000000} 6 (=0)}} & {\color[HTML]{38761D} 0.06 (-0.01)} & {\color[HTML]{38761D} \textbf{0.17 (-0.01)}} & \multicolumn{1}{c!{\vrule width 1.2pt}}{\color[HTML]{38761D} 7 (+1)} & {\color[HTML]{38761D} \textbf{0.04 (-0.02)}} & {\color[HTML]{000000} \textbf{0.20 (=0)}} & \multicolumn{1}{c|}{\cellcolor[HTML]{FFF5E3}{\color[HTML]{000000} \textbf{9 (=0)}}} & {\color[HTML]{000000} 0.06 (=0)} & {\color[HTML]{000000} \textbf{0.20 (=0)}} & \multicolumn{1}{c|}{\cellcolor[HTML]{FFF5E3}{\color[HTML]{000000} 7 (=0)}} & {\color[HTML]{000000} 0.07 (=0)} & {\color[HTML]{000000} 0.21 (=0)} & {\color[HTML]{000000} 7 (=0)} \\ \cline{2-20}

\rowcolor[HTML]{FFF5E3} 
\multirow{-2}{*}{\cellcolor[HTML]{F0C88C}\textbf{DS-Coder}} & \cellcolor[HTML]{FFE7C8}\textbf{33B} & {\color[HTML]{000000} \textbf{0.04 (=0)}} & {\color[HTML]{38761D} \textbf{0.17 (-0.01)}} & \multicolumn{1}{c|}{\cellcolor[HTML]{FFF5E3}{\color[HTML]{000000} \textbf{8 (=0)}}} & {\color[HTML]{38761D} 0.06 (-0.02)} & {\color[HTML]{000000} 0.18 (=0)} & \multicolumn{1}{c|}{\cellcolor[HTML]{FFF5E3}{\color[HTML]{000000} 7 (=0)}} & {\color[HTML]{38761D} 0.05 (-0.03)} & {\color[HTML]{38761D} 0.18 (-0.01)} & \multicolumn{1}{c!{\vrule width 1.2pt}}{\color[HTML]{000000} 7 (=0)} & {\color[HTML]{38761D} \textbf{0.03 (-0.02)}} & {\color[HTML]{000000} \textbf{0.18 (=0)}} & \multicolumn{1}{c|}{\cellcolor[HTML]{FFF5E3}{\color[HTML]{38761D} \textbf{8 (+1)}}} & {\color[HTML]{38761D} 0.04 (-0.02)} & {\color[HTML]{000000} 0.20 (=0)} & \multicolumn{1}{c|}{\cellcolor[HTML]{FFF5E3}{\color[HTML]{38761D} 7 (+1)}} & {\color[HTML]{38761D} 0.04 (-0.03)} & {\color[HTML]{000000} 0.20 (=0)} & {\color[HTML]{38761D} 7 (+1)} \\ \hline

\rowcolor[HTML]{FFF5E3} 
\cellcolor[HTML]{F0C88C}\textbf{DS-V2} & \cellcolor[HTML]{FFE7C8}\textbf{16B} & {\color[HTML]{38761D} \textbf{0.03 (-0.04)}} & {\color[HTML]{38761D} \textbf{0.13 (-0.01)}} & \multicolumn{1}{c|}{\cellcolor[HTML]{FFF5E3}{\color[HTML]{990000} \textbf{6 (-1)}}} & {\color[HTML]{38761D} 0.05 (-0.02)} & {\color[HTML]{38761D} \textbf{0.13 (-0.01)}} & \multicolumn{1}{c|}{\cellcolor[HTML]{FFF5E3}{\color[HTML]{000000} 5 (=0)}} & {\color[HTML]{000000} 0.06 (=0)} & {\color[HTML]{38761D} \textbf{0.13 (-0.02)}} & \multicolumn{1}{c!{\vrule width 1.2pt}}{\color[HTML]{38761D} 5 (+1)} & {\color[HTML]{38761D} 0.05 (-0.03)} & {\color[HTML]{000000} \textbf{0.22 (=0)}} & \multicolumn{1}{c|}{\cellcolor[HTML]{FFF5E3}{\color[HTML]{000000} \textbf{8 (=0)}}} & {\color[HTML]{38761D} \textbf{0.04 (-0.03)}} & {\color[HTML]{38761D} \textbf{0.22 (-0.01)}} & \multicolumn{1}{c|}{\cellcolor[HTML]{FFF5E3}{\color[HTML]{000000} 6 (=0)}} & {\color[HTML]{38761D} \textbf{0.04 (-0.01)}} & {\color[HTML]{000000} \textbf{0.22 (=0)}} & {\color[HTML]{38761D} 6 (+1)} \\ \hline

\rowcolor[HTML]{FFF5E3} 
\cellcolor[HTML]{F0C88C}\textbf{DS-Coder-V2} & \cellcolor[HTML]{FFE7C8}\textbf{16B} & {\color[HTML]{38761D} \textbf{0.03 (-0.01)}} & {\color[HTML]{38761D} \textbf{0.17 (-0.01)}} & \multicolumn{1}{c|}{\cellcolor[HTML]{FFF5E3}{\color[HTML]{000000} 7 (=0)}} & {\color[HTML]{38761D} 0.04 (-0.02)} & {\color[HTML]{000000} 0.19 (=0)} & \multicolumn{1}{c|}{\cellcolor[HTML]{FFF5E3}{\color[HTML]{000000} 6 (=0)}} & {\color[HTML]{38761D} 0.04 (-0.04)} & {\color[HTML]{38761D} 0.18 (-0.01)} & \multicolumn{1}{c!{\vrule width 1.2pt}}{\color[HTML]{38761D} \textbf{8 (+2)}} & {\color[HTML]{000000} 0.04 (=0)} & {\color[HTML]{000000} \textbf{0.22 (=0)}} & \multicolumn{1}{c|}{\cellcolor[HTML]{FFF5E3}{\color[HTML]{000000} \textbf{6 (=0)}}} & {\color[HTML]{38761D} \textbf{0.03 (-0.07)}} & {\color[HTML]{38761D} \textbf{0.22 (-0.01)}} & \multicolumn{1}{c|}{\cellcolor[HTML]{FFF5E3}{\color[HTML]{000000} \textbf{6 (=0)}}} & {\color[HTML]{38761D} 0.04 (-0.05)} & {\color[HTML]{38761D} \textbf{0.22 (-0.01)}} & {\color[HTML]{000000} \textbf{6 (=0)}} \\ 
 \Xhline{2\arrayrulewidth}

\multicolumn{20}{l}{\textbf{\textit{min}:} Minimum Token Probability, \textbf{\textit{low-K}:} Lowest-$K$ Token Probability \textbf{\textit{attn-$\omega$}:} Attention-Weighted Uncertainty, \textbf{EM:} Exact Match, \textbf{EP:} Edit Progress, \textbf{ECE:} Expected Calibration Error (↓), $\boldsymbol{\mathcal{B}}$\textbf{:} Brier Score (↓), \textbf{BC:} Bin Coverage (↑)} \\

\multicolumn{20}{l}{\footnotesize\textbf{Bold Number:} Best Result for that Model and Correctness Measure, \textbf{(+/-/=):} Difference from Global Platt-Scaling} \\

\end{tabular}
}
\label{Tab:dcf_bug_local}
\end{table*}

\begin{table*}[!t]  
\caption{Local Platt-Scaling Results for {DCF-Vul} in Terms of Edit Progress and Checks Passed}
\centering
\resizebox{\textwidth}{!}{%
\begin{tabular}{!{\vrule width 1.2pt}
>{\columncolor[HTML]{F9B83C}}l |
>{\columncolor[HTML]{F9B83C}}r !{\vrule width 1.2pt}
>{\columncolor[HTML]{FFE7C8}}c 
>{\columncolor[HTML]{FFE7C8}}c 
>{\columncolor[HTML]{FFE7C8}}c 
>{\columncolor[HTML]{FFE7C8}}c 
>{\columncolor[HTML]{FFE7C8}}c 
>{\columncolor[HTML]{FFE7C8}}c 
>{\columncolor[HTML]{FFE7C8}}c 
>{\columncolor[HTML]{FFE7C8}}c 
>{\columncolor[HTML]{FFE7C8}}c 
>{\columncolor[HTML]{FFF5E3}}c 
>{\columncolor[HTML]{FFF5E3}}c 
>{\columncolor[HTML]{FFF5E3}}c 
>{\columncolor[HTML]{FFF5E3}}c 
>{\columncolor[HTML]{FFF5E3}}c 
>{\columncolor[HTML]{FFF5E3}}c 
>{\columncolor[HTML]{FFF5E3}}c 
>{\columncolor[HTML]{FFF5E3}}c
>{\columncolor[HTML]{FFF5E3}}c !{\vrule width 1.2pt}}
\Xhline{2\arrayrulewidth}

\rowcolor[HTML]{F0C88C} 
\cellcolor[HTML]{F0C88C} & \multicolumn{1}{l!{\vrule width 1.2pt}}{\cellcolor[HTML]{F0C88C}} & \multicolumn{9}{c!{\vrule width 1.2pt}}{\cellcolor[HTML]{F0C88C}\textbf{EP}} & \multicolumn{9}{c!{\vrule width 1.2pt}}{\cellcolor[HTML]{F0C88C}\textbf{CP}} \\ \cline{3-20}

\rowcolor[HTML]{F0C88C} 
\cellcolor[HTML]{F0C88C} & \multicolumn{1}{l!{\vrule width 1.2pt}}{\cellcolor[HTML]{F0C88C}} & \multicolumn{3}{c|}{\cellcolor[HTML]{F0C88C}\textit{\textbf{min}}} & \multicolumn{3}{c|}{\cellcolor[HTML]{F0C88C}\textit{\textbf{low-K}}} & \multicolumn{3}{c!{\vrule width 1.2pt}}{\cellcolor[HTML]{F0C88C}\textit{\textbf{attn-$\omega$}}} & \multicolumn{3}{c|}{\cellcolor[HTML]{F0C88C}\textit{\textbf{min}}} & \multicolumn{3}{c|}{\cellcolor[HTML]{F0C88C}\textit{\textbf{low-K}}} & \multicolumn{3}{c!{\vrule width 1.2pt}}{\cellcolor[HTML]{F0C88C}\textit{\textbf{attn-$\omega$}}} \\ \cline{3-20}

\rowcolor[HTML]{FFE7C8} 
\multirow{-3}{*}{\cellcolor[HTML]{F0C88C}\textbf{Model}} & \multicolumn{1}{l!{\vrule width 1.2pt}}{\multirow{-3}{*}{\cellcolor[HTML]{F0C88C}\textbf{Size}}} & \textit{\textbf{ECE}} & \textit{\textbf{$\boldsymbol{\mathcal{B}}$}} & \multicolumn{1}{c|}{\cellcolor[HTML]{FFE7C8}\textit{\textbf{BC}}} & \textit{\textbf{ECE}} & \textit{\textbf{$\boldsymbol{\mathcal{B}}$}} & \multicolumn{1}{c|}{\cellcolor[HTML]{FFE7C8}\textit{\textbf{BC}}} & \textit{\textbf{ECE}} & \textit{\textbf{$\boldsymbol{\mathcal{B}}$}} & \multicolumn{1}{c!{\vrule width 1.2pt}}{\cellcolor[HTML]{FFE7C8}\textit{\textbf{BC}}} & \textit{\textbf{ECE}} & \textit{\textbf{$\boldsymbol{\mathcal{B}}$}} & \multicolumn{1}{c|}{\cellcolor[HTML]{FFE7C8}\textit{\textbf{BC}}} & \textit{\textbf{ECE}} & \textit{\textbf{$\boldsymbol{\mathcal{B}}$}} & \multicolumn{1}{c|}{\cellcolor[HTML]{FFE7C8}\textit{\textbf{BC}}} & \textit{\textbf{ECE}} & \textit{\textbf{$\boldsymbol{\mathcal{B}}$}} & \textit{\textbf{BC}} \\ \Xhline{2\arrayrulewidth}

\rowcolor[HTML]{FFF5E3} 
\cellcolor[HTML]{F0C88C} & \cellcolor[HTML]{FFE7C8}\textbf{8B} & {\color[HTML]{000000} \textbf{0.02 (=0)}} & {\color[HTML]{000000} \textbf{0.19 (=0)}} & \multicolumn{1}{c|}{\cellcolor[HTML]{FFF5E3}{\color[HTML]{000000} \textbf{7 (=0)}}} & {\color[HTML]{000000} \textbf{0.02 (=0)}} & {\color[HTML]{000000} 0.20 (=0)} & \multicolumn{1}{c|}{\cellcolor[HTML]{FFF5E3}{\color[HTML]{000000} 4 (=0)}} & {\color[HTML]{000000} \textbf{0.02 (=0)}} & {\color[HTML]{000000} 0.20 (=0)} & \multicolumn{1}{c!{\vrule width 1.2pt}}{\color[HTML]{38761D} 5 (+1)} & {\color[HTML]{38761D} 0.03 (-0.03)} & {\color[HTML]{000000} \textbf{0.23 (=0)}} & \multicolumn{1}{c|}{\cellcolor[HTML]{FFF5E3}{\color[HTML]{000000} \textbf{5 (=0)}}} & {\color[HTML]{38761D} 0.03 (-0.01)} & {\color[HTML]{000000} 0.24 (=0)} & \multicolumn{1}{c|}{\cellcolor[HTML]{FFF5E3}{\color[HTML]{000000} 4 (=0)}} & {\color[HTML]{38761D} \textbf{0.02}} & {\color[HTML]{38761D} 0.24} & {\color[HTML]{38761D} 3 (+2)} \\ \cline{2-20}

\rowcolor[HTML]{FFF5E3} 
\multirow{-2}{*}{\cellcolor[HTML]{F0C88C}\textbf{Llama-3.1}} & \cellcolor[HTML]{FFE7C8}\textbf{70B} & {\color[HTML]{000000} \textbf{0.05 (=0)}} & {\color[HTML]{000000} \textbf{0.19 (=0)}} & \multicolumn{1}{c|}{\cellcolor[HTML]{FFF5E3}{\color[HTML]{000000} \textbf{6 (=0)}}} & {\color[HTML]{000000} \textbf{0.05 (=0)}} & {\color[HTML]{000000} 0.20 (=0)} & \multicolumn{1}{c|}{\cellcolor[HTML]{FFF5E3}{\color[HTML]{000000} 5 (=0)}} & {\color[HTML]{000000} \textbf{0.05 (=0)}} & {\color[HTML]{000000} 0.20 (=0)} & \multicolumn{1}{c!{\vrule width 1.2pt}}{\color[HTML]{000000} 5 (=0)} & {\color[HTML]{38761D} \textbf{0.04 (-0.01)}} & {\color[HTML]{000000} \textbf{0.19 (=0)}} & \multicolumn{1}{c|}{\cellcolor[HTML]{FFF5E3}{\color[HTML]{000000} \textbf{3 (=0)}}} & {\color[HTML]{38761D} 0.05} & {\color[HTML]{38761D} 0.20} & \multicolumn{1}{c|}{\cellcolor[HTML]{FFF5E3}{\color[HTML]{38761D} \textbf{3 (+2)}}} & {\color[HTML]{38761D} 0.05} & {\color[HTML]{38761D} 0.20} & {\color[HTML]{38761D} \textbf{3 (+2)}} \\ \hline

\rowcolor[HTML]{FFF5E3} 
\cellcolor[HTML]{F0C88C} & \cellcolor[HTML]{FFE7C8}\textbf{7B} & {\color[HTML]{000000} 0.03 (=0)} & {\color[HTML]{000000} \textbf{0.19 (=0)}} & \multicolumn{1}{c|}{\cellcolor[HTML]{FFF5E3}{\color[HTML]{000000} \textbf{3 (=0)}}} & {\color[HTML]{000000} 0.03 (=0)} & {\color[HTML]{000000} \textbf{0.19 (=0)}} & \multicolumn{1}{c|}{\cellcolor[HTML]{FFF5E3}{\color[HTML]{000000} \textbf{3 (=0)}}} & {\color[HTML]{000000} \textbf{0.01 (=0)}} & {\color[HTML]{000000} \textbf{0.19 (=0)}} & \multicolumn{1}{c!{\vrule width 1.2pt}}{\color[HTML]{000000} \textbf{3 (=0)}} & {\color[HTML]{38761D} 0.03 (-0.01)} & {\color[HTML]{000000} \textbf{0.24 (=0)}} & \multicolumn{1}{c|}{\cellcolor[HTML]{FFF5E3}{\color[HTML]{000000} 3 (=0)}} & {\color[HTML]{000000} \textbf{0.02 (=0)}} & {\color[HTML]{000000} \textbf{0.24 (=0)}} & \multicolumn{1}{c|}{\cellcolor[HTML]{FFF5E3}{\color[HTML]{000000} \textbf{4 (=0)}}} & {\color[HTML]{000000} 0.03 (=0)} & {\color[HTML]{000000} \textbf{0.24 (=0)}} & {\color[HTML]{000000} 2 (=0)} \\ \cline{2-20}

\rowcolor[HTML]{FFF5E3} 
\multirow{-2}{*}{\cellcolor[HTML]{F0C88C}\textbf{CodeLlama}} & \cellcolor[HTML]{FFE7C8}\textbf{70B} & {\color[HTML]{38761D} \textbf{0.02 (-0.01)}} & {\color[HTML]{000000} \textbf{0.20 (=0)}} & \multicolumn{1}{c|}{\cellcolor[HTML]{FFF5E3}{\color[HTML]{38761D} 3 (+1)}} & {\color[HTML]{000000} 0.03 (=0)} & {\color[HTML]{000000} \textbf{0.20 (=0)}} & \multicolumn{1}{c|}{\cellcolor[HTML]{FFF5E3}{\color[HTML]{38761D} \textbf{4 (+1)}}} & {\color[HTML]{000000} 0.03 (=0)} & {\color[HTML]{000000} \textbf{0.20 (=0)}} & \multicolumn{1}{c!{\vrule width 1.2pt}}{\color[HTML]{38761D} \textbf{4 (+1)}} & {\color[HTML]{38761D} \textbf{0.05}} & {\color[HTML]{38761D} \textbf{0.23}} & \multicolumn{1}{c|}{\cellcolor[HTML]{FFF5E3}{\color[HTML]{38761D} \textbf{6 (+5)}}} & {\color[HTML]{38761D} \textbf{0.05 (-0.03)}} & {\color[HTML]{38761D} 0.24 (-0.01)} & \multicolumn{1}{c|}{\cellcolor[HTML]{FFF5E3}{\color[HTML]{000000} 2 (=0)}} & {\color[HTML]{38761D} \textbf{0.05 (-0.03)}} & {\color[HTML]{38761D} 0.24 (-0.01)} & {\color[HTML]{000000} 2 (=0)} \\ \hline

\rowcolor[HTML]{FFF5E3} 
\cellcolor[HTML]{F0C88C} & \cellcolor[HTML]{FFE7C8}\textbf{7B} & {\color[HTML]{000000} 0.03 (=0)} & {\color[HTML]{000000} \textbf{0.13 (=0)}} & \multicolumn{1}{c|}{\cellcolor[HTML]{FFF5E3}{\color[HTML]{000000} \textbf{4 (=0)}}} & {\color[HTML]{38761D} \textbf{0.02 (-0.01)}} & {\color[HTML]{000000} 0.14 (=0)} & \multicolumn{1}{c|}{\cellcolor[HTML]{FFF5E3}{\color[HTML]{000000} \textbf{4 (=0)}}} & {\color[HTML]{000000} 0.03 (=0)} & {\color[HTML]{000000} 0.14 (=0)} & \multicolumn{1}{c!{\vrule width 1.2pt}}{\color[HTML]{000000} \textbf{4 (=0)}} & {\color[HTML]{000000} 0.04 (=0)} & {\color[HTML]{38761D} \textbf{0.22 (-0.01)}} & \multicolumn{1}{c|}{\cellcolor[HTML]{FFF5E3}{\color[HTML]{38761D} \textbf{6 (+3)}}} & {\color[HTML]{38761D} 0.04 (-0.01)} & {\color[HTML]{38761D} \textbf{0.22 (-0.01)}} & \multicolumn{1}{c|}{\cellcolor[HTML]{FFF5E3}{\color[HTML]{38761D} \textbf{6 (+3)}}} & {\color[HTML]{000000} \textbf{0.03 (=0)}} & {\color[HTML]{38761D} \textbf{0.22 (-0.01)}} & {\color[HTML]{38761D} \textbf{6 (+3)}} \\ \cline{2-20}

\rowcolor[HTML]{FFF5E3} 
\multirow{-2}{*}{\cellcolor[HTML]{F0C88C}\textbf{Qwen2.5}} & \cellcolor[HTML]{FFE7C8}\textbf{72B} & {\color[HTML]{000000} \textbf{0.02 (=0)}} & {\color[HTML]{000000} \textbf{0.12 (=0)}} & \multicolumn{1}{c|}{\cellcolor[HTML]{FFF5E3}{\color[HTML]{38761D} \textbf{5 (+1)}}} & {\color[HTML]{000000} \textbf{0.02 (=0)}} & {\color[HTML]{000000} \textbf{0.12 (=0)}} & \multicolumn{1}{c|}{\cellcolor[HTML]{FFF5E3}{\color[HTML]{000000} \textbf{5 (=0)}}} & {\color[HTML]{000000} \textbf{0.02 (=0)}} & {\color[HTML]{000000} \textbf{0.12 (=0)}} & \multicolumn{1}{c!{\vrule width 1.2pt}}{\color[HTML]{000000} \textbf{5 (=0)}} & {\color[HTML]{38761D} \textbf{0.03 (-0.02)}} & {\color[HTML]{38761D} \textbf{0.20 (-0.01)}} & \multicolumn{1}{c|}{\cellcolor[HTML]{FFF5E3}{\color[HTML]{38761D} \textbf{5 (+2)}}} & {\color[HTML]{38761D} \textbf{0.03 (-0.01)}} & {\color[HTML]{38761D} \textbf{0.20 (-0.01)}} & \multicolumn{1}{c|}{\cellcolor[HTML]{FFF5E3}{\color[HTML]{38761D} \textbf{5 (+3)}}} & {\color[HTML]{38761D} \textbf{0.03 (-0.01)}} & {\color[HTML]{38761D} \textbf{0.20 (-0.01)}} & {\color[HTML]{38761D} \textbf{5 (+3)}} \\ \hline

\rowcolor[HTML]{FFF5E3} 
\cellcolor[HTML]{F0C88C} & \cellcolor[HTML]{FFE7C8}\textbf{7B} & {\color[HTML]{38761D} \textbf{0.04 (-0.01)}} & {\color[HTML]{000000} \textbf{0.13 (=0)}} & \multicolumn{1}{c|}{\cellcolor[HTML]{FFF5E3}{\color[HTML]{000000} \textbf{5 (=0)}}} & {\color[HTML]{38761D} \textbf{0.04 (-0.01)}} & {\color[HTML]{000000} 0.15 (=0)} & \multicolumn{1}{c|}{\cellcolor[HTML]{FFF5E3}{\color[HTML]{000000} 4 (=0)}} & {\color[HTML]{000000} \textbf{0.04 (=0)}} & {\color[HTML]{000000} 0.15 (=0)} & \multicolumn{1}{c!{\vrule width 1.2pt}}{\color[HTML]{000000} 4 (=0)} & {\color[HTML]{000000} 0.07 (=0)} & {\color[HTML]{38761D} \textbf{0.21 (-0.01)}} & \multicolumn{1}{c|}{\cellcolor[HTML]{FFF5E3}{\color[HTML]{38761D} 4 (+2)}} & {\color[HTML]{000000} \textbf{0.05 (=0)}} & {\color[HTML]{38761D} \textbf{0.21 (-0.01)}} & \multicolumn{1}{c|}{\cellcolor[HTML]{FFF5E3}{\color[HTML]{38761D} 4 (+2)}} & {\color[HTML]{38761D} 0.06} & {\color[HTML]{38761D} 0.22} & {\color[HTML]{38761D} \textbf{7 (+6)}} \\ \cline{2-20}

\rowcolor[HTML]{FFF5E3} 
\multirow{-2}{*}{\cellcolor[HTML]{F0C88C}\textbf{Qwen2.5-Coder}} & \cellcolor[HTML]{FFE7C8}\textbf{32B} & {\color[HTML]{000000} \textbf{0.02 (=0)}} & {\color[HTML]{000000} \textbf{0.14 (=0)}} & \multicolumn{1}{c|}{\cellcolor[HTML]{FFF5E3}{\color[HTML]{000000} \textbf{5 (=0)}}} & {\color[HTML]{000000} \textbf{0.02 (=0)}} & {\color[HTML]{000000} 0.15 (=0)} & \multicolumn{1}{c|}{\cellcolor[HTML]{FFF5E3}{\color[HTML]{000000} \textbf{5 (=0)}}} & {\color[HTML]{000000} \textbf{0.02 (=0)}} & {\color[HTML]{000000} 0.15 (=0)} & \multicolumn{1}{c!{\vrule width 1.2pt}}{\color[HTML]{000000} \textbf{5 (=0)}} & {\color[HTML]{38761D} 0.03 (-0.02)} & {\color[HTML]{38761D} \textbf{0.20 (-0.02)}} & \multicolumn{1}{c|}{\cellcolor[HTML]{FFF5E3}{\color[HTML]{38761D} \textbf{6 (+4)}}} & {\color[HTML]{38761D} 0.03 (-0.02)} & {\color[HTML]{38761D} 0.21 (-0.01)} & \multicolumn{1}{c|}{\cellcolor[HTML]{FFF5E3}{\color[HTML]{38761D} 3 (+1)}} & {\color[HTML]{38761D} \textbf{0.02}} & {\color[HTML]{38761D} 0.21} & {\color[HTML]{38761D} 3 (+2)} \\ \hline

\rowcolor[HTML]{FFF5E3} 
\cellcolor[HTML]{F0C88C} & \cellcolor[HTML]{FFE7C8}\textbf{7B} & {\color[HTML]{000000} \textbf{0.01 (=0)}} & {\color[HTML]{000000} \textbf{0.12 (=0)}} & \multicolumn{1}{c|}{\cellcolor[HTML]{FFF5E3}{\color[HTML]{000000} \textbf{4 (=0)}}} & {\color[HTML]{000000} \textbf{0.01 (=0)}} & {\color[HTML]{000000} \textbf{0.12 (=0)}} & \multicolumn{1}{c|}{\cellcolor[HTML]{FFF5E3}{\color[HTML]{38761D} \textbf{4 (+1)}}} & {\color[HTML]{000000} 0.02 (=0)} & {\color[HTML]{38761D} \textbf{0.12 (-0.01)}} & \multicolumn{1}{c!{\vrule width 1.2pt}}{\color[HTML]{000000} 3 (=0)} & {\color[HTML]{38761D} 0.03} & {\color[HTML]{38761D} \textbf{0.24}} & \multicolumn{1}{c|}{\cellcolor[HTML]{FFF5E3}{\color[HTML]{38761D} \textbf{5 (+4)}}} & {\color[HTML]{000000} 0.03 (=0)} & {\color[HTML]{38761D} \textbf{0.24 (-0.01)}} & \multicolumn{1}{c|}{\cellcolor[HTML]{FFF5E3}{\color[HTML]{38761D} \textbf{5 (+2)}}} & {\color[HTML]{38761D} \textbf{0.02}} & {\color[HTML]{38761D} \textbf{0.24}} & {\color[HTML]{38761D} \textbf{5 (+4)}} \\ \cline{2-20}

\rowcolor[HTML]{FFF5E3} 
\multirow{-2}{*}{\cellcolor[HTML]{F0C88C}\textbf{DS-LLM}} & \cellcolor[HTML]{FFE7C8}\textbf{67B} & {\color[HTML]{38761D} 0.03 (-0.01)} & {\color[HTML]{000000} \textbf{0.20 (=0)}} & \multicolumn{1}{c|}{\cellcolor[HTML]{FFF5E3}{\color[HTML]{000000} \textbf{4 (=0)}}} & {\color[HTML]{000000} 0.04 (=0)} & {\color[HTML]{000000} \textbf{0.20 (=0)}} & \multicolumn{1}{c|}{\cellcolor[HTML]{FFF5E3}{\color[HTML]{000000} 3 (=0)}} & {\color[HTML]{000000} \textbf{0.02 (=0)}} & {\color[HTML]{000000} \textbf{0.20 (=0)}} & \multicolumn{1}{c!{\vrule width 1.2pt}}{\color[HTML]{000000} 3 (=0)} & {\color[HTML]{38761D} \textbf{0.02 (-0.02)}} & {\color[HTML]{000000} \textbf{0.25 (=0)}} & \multicolumn{1}{c|}{\cellcolor[HTML]{FFF5E3}{\color[HTML]{000000} \textbf{2 (=0)}}} & {\color[HTML]{38761D} \textbf{0.02 (-0.04)}} & {\color[HTML]{000000} \textbf{0.25 (=0)}} & \multicolumn{1}{c|}{\cellcolor[HTML]{FFF5E3}{\color[HTML]{000000} \textbf{2 (=0)}}} & {\color[HTML]{38761D} \textbf{0.02 (-0.04)}} & {\color[HTML]{000000} \textbf{0.25 (=0)}} & {\color[HTML]{000000} \textbf{2 (=0)}} \\ \hline

\rowcolor[HTML]{FFF5E3} 
\cellcolor[HTML]{F0C88C} & \cellcolor[HTML]{FFE7C8}\textbf{7B} & {\color[HTML]{000000} 0.04 (=0)} & {\color[HTML]{000000} \textbf{0.17 (=0)}} & \multicolumn{1}{c|}{\cellcolor[HTML]{FFF5E3}{\color[HTML]{000000} \textbf{7 (=0)}}} & {\color[HTML]{38761D} \textbf{0.03 (-0.01)}} & {\color[HTML]{990000} 0.19 (+0.01)} & \multicolumn{1}{c|}{\cellcolor[HTML]{FFF5E3}{\color[HTML]{38761D} 5 (+1)}} & {\color[HTML]{000000} \textbf{0.03 (=0)}} & {\color[HTML]{000000} 0.19 (=0)} & \multicolumn{1}{c!{\vrule width 1.2pt}}{\color[HTML]{000000} 5 (=0)} & {\color[HTML]{000000} \textbf{0.03 (=0)}} & {\color[HTML]{000000} \textbf{0.23 (=0)}} & \multicolumn{1}{c|}{\cellcolor[HTML]{FFF5E3}{\color[HTML]{000000} 4 (=0)}} & {\color[HTML]{000000} \textbf{0.03 (=0)}} & {\color[HTML]{000000} \textbf{0.23 (=0)}} & \multicolumn{1}{c|}{\cellcolor[HTML]{FFF5E3}{\color[HTML]{38761D} \textbf{5 (+3)}}} & {\color[HTML]{990000} \textbf{0.03 (+0.02)}} & {\color[HTML]{000000} \textbf{0.23 (=0)}} & {\color[HTML]{38761D} \textbf{5 (+3)}} \\ \cline{2-20}

\rowcolor[HTML]{FFF5E3} 
\multirow{-2}{*}{\cellcolor[HTML]{F0C88C}\textbf{DS-Coder}} & \cellcolor[HTML]{FFE7C8}\textbf{33B} & {\color[HTML]{38761D} \textbf{0.05 (-0.01)}} & {\color[HTML]{000000} \textbf{0.19 (=0)}} & \multicolumn{1}{c|}{\cellcolor[HTML]{FFF5E3}{\color[HTML]{000000} \textbf{6 (=0)}}} & {\color[HTML]{38761D} \textbf{0.05 (-0.01)}} & {\color[HTML]{000000} 0.20 (=0)} & \multicolumn{1}{c|}{\cellcolor[HTML]{FFF5E3}{\color[HTML]{000000} 4 (=0)}} & {\color[HTML]{000000} \textbf{0.05 (=0)}} & {\color[HTML]{000000} 0.20 (=0)} & \multicolumn{1}{c!{\vrule width 1.2pt}}{\color[HTML]{000000} 4 (=0)} & {\color[HTML]{000000} 0.03 (=0)} & {\color[HTML]{38761D} \textbf{0.22 (-0.01)}} & \multicolumn{1}{c|}{\cellcolor[HTML]{FFF5E3}{\color[HTML]{38761D} \textbf{5 (+2)}}} & {\color[HTML]{38761D} \textbf{0.01 (-0.02)}} & {\color[HTML]{38761D} \textbf{0.22 (-0.01)}} & \multicolumn{1}{c|}{\cellcolor[HTML]{FFF5E3}{\color[HTML]{38761D} \textbf{5 (+2)}}} & {\color[HTML]{000000} 0.03 (=0)} & {\color[HTML]{38761D} \textbf{0.22 (-0.01)}} & {\color[HTML]{38761D} \textbf{5 (+2)}} \\ \hline

\rowcolor[HTML]{FFF5E3} 
\cellcolor[HTML]{F0C88C}\textbf{DS-V2} & \cellcolor[HTML]{FFE7C8}\textbf{16B} & {\color[HTML]{38761D} \textbf{0.02 (-0.01)}} & {\color[HTML]{000000} \textbf{0.15 (=0)}} & \multicolumn{1}{c|}{\cellcolor[HTML]{FFF5E3}{\color[HTML]{000000} \textbf{4 (=0)}}} & {\color[HTML]{000000} 0.03 (=0)} & {\color[HTML]{000000} \textbf{0.15 (=0)}} & \multicolumn{1}{c|}{\cellcolor[HTML]{FFF5E3}{\color[HTML]{38761D} \textbf{4 (+1)}}} & {\color[HTML]{000000} 0.03 (=0)} & {\color[HTML]{000000} \textbf{0.15 (=0)}} & \multicolumn{1}{c!{\vrule width 1.2pt}}{\color[HTML]{000000} 3 (=0)} & {\color[HTML]{000000} 0.03 (=0)} & {\color[HTML]{38761D} \textbf{0.24 (-0.01)}} & \multicolumn{1}{c|}{\cellcolor[HTML]{FFF5E3}{\color[HTML]{38761D} 3 (+1)}} & {\color[HTML]{000000} \textbf{0.02 (=0)}} & {\color[HTML]{38761D} \textbf{0.24 (-0.01)}} & \multicolumn{1}{c|}{\cellcolor[HTML]{FFF5E3}{\color[HTML]{38761D} 3 (+1)}} & {\color[HTML]{990000} 0.03 (+0.02)} & {\color[HTML]{38761D} \textbf{0.24 (-0.01)}} & {\color[HTML]{38761D} \textbf{4 (+1)}} \\ \hline

\rowcolor[HTML]{FFF5E3} 
\cellcolor[HTML]{F0C88C}\textbf{DS-Coder-V2} & \cellcolor[HTML]{FFE7C8}\textbf{16B} & {\color[HTML]{000000} \textbf{0.04 (=0)}} & {\color[HTML]{000000} \textbf{0.16 (=0)}} & \multicolumn{1}{c|}{\cellcolor[HTML]{FFF5E3}{\color[HTML]{000000} \textbf{5 (=0)}}} & {\color[HTML]{000000} \textbf{0.04 (=0)}} & {\color[HTML]{000000} 0.17 (=0)} & \multicolumn{1}{c|}{\cellcolor[HTML]{FFF5E3}{\color[HTML]{38761D} \textbf{5 (+1)}}} & {\color[HTML]{000000} 0.05 (=0)} & {\color[HTML]{000000} 0.17 (=0)} & \multicolumn{1}{c!{\vrule width 1.2pt}}{\color[HTML]{38761D} \textbf{5 (+1)}} & {\color[HTML]{38761D} \textbf{0.02 (-0.02)}} & {\color[HTML]{38761D} \textbf{0.23 (-0.01)}} & \multicolumn{1}{c|}{\cellcolor[HTML]{FFF5E3}{\color[HTML]{38761D} \textbf{5 (+1)}}} & {\color[HTML]{38761D} \textbf{0.02 (-0.01)}} & {\color[HTML]{000000} 0.24 (=0)} & \multicolumn{1}{c|}{\cellcolor[HTML]{FFF5E3}{\color[HTML]{38761D} 4 (+1)}} & {\color[HTML]{38761D} 0.03 (-0.02)} & {\color[HTML]{38761D} 0.24 (-0.01)} & {\color[HTML]{38761D} \textbf{5 (+3)}} \\ 
\Xhline{2\arrayrulewidth}

\multicolumn{20}{l}{\textbf{\textit{min}:} Minimum Token Probability, \textbf{\textit{low-K}:} Lowest-$K$ Token Probability \textbf{\textit{attn-$\omega$}:} Attention-Weighted Uncertainty, \textbf{EM:} Exact Match, \textbf{EP:} Edit Progress, \textbf{ECE:} Expected Calibration Error (↓), $\boldsymbol{\mathcal{B}}$\textbf{:} Brier Score (↓), \textbf{BC:} Bin Coverage (↑)} \\

\multicolumn{20}{l}{\footnotesize\textbf{Bold Number:} Best Result for that Model and Correctness Measure, \textbf{(+/-/=):} Difference from Global Platt-Scaling} \\

\end{tabular}
}
\label{Tab:dcf_vul_local}
\end{table*}

\begin{table*}[!t]  
\caption{Local Platt-Scaling Results for {CR-Trans} in Terms of Exact Match and Edit Progress}
\centering
\resizebox{\textwidth}{!}{%
\begin{tabular}{!{\vrule width 1.2pt}
>{\columncolor[HTML]{F9B83C}}l |
>{\columncolor[HTML]{F9B83C}}r !{\vrule width 1.2pt}
>{\columncolor[HTML]{FFE7C8}}c 
>{\columncolor[HTML]{FFE7C8}}c 
>{\columncolor[HTML]{FFE7C8}}c 
>{\columncolor[HTML]{FFE7C8}}c 
>{\columncolor[HTML]{FFE7C8}}c 
>{\columncolor[HTML]{FFE7C8}}c 
>{\columncolor[HTML]{FFE7C8}}c 
>{\columncolor[HTML]{FFE7C8}}c 
>{\columncolor[HTML]{FFE7C8}}c 
>{\columncolor[HTML]{FFF5E3}}c 
>{\columncolor[HTML]{FFF5E3}}c 
>{\columncolor[HTML]{FFF5E3}}c 
>{\columncolor[HTML]{FFF5E3}}c 
>{\columncolor[HTML]{FFF5E3}}c 
>{\columncolor[HTML]{FFF5E3}}c 
>{\columncolor[HTML]{FFF5E3}}c 
>{\columncolor[HTML]{FFF5E3}}c
>{\columncolor[HTML]{FFF5E3}}c !{\vrule width 1.2pt}}
\Xhline{2\arrayrulewidth}

\rowcolor[HTML]{F0C88C} 
\cellcolor[HTML]{F0C88C} & \multicolumn{1}{l!{\vrule width 1.2pt}}{\cellcolor[HTML]{F0C88C}} & \multicolumn{9}{c!{\vrule width 1.2pt}}{\cellcolor[HTML]{F0C88C}\textbf{EM}} & \multicolumn{9}{c!{\vrule width 1.2pt}}{\cellcolor[HTML]{F0C88C}\textbf{EP}} \\ \cline{3-20}

\rowcolor[HTML]{F0C88C} 
\cellcolor[HTML]{F0C88C} & \multicolumn{1}{l!{\vrule width 1.2pt}}{\cellcolor[HTML]{F0C88C}} & \multicolumn{3}{c|}{\cellcolor[HTML]{F0C88C}\textit{\textbf{min}}} & \multicolumn{3}{c|}{\cellcolor[HTML]{F0C88C}\textit{\textbf{low-K}}} & \multicolumn{3}{c!{\vrule width 1.2pt}}{\cellcolor[HTML]{F0C88C}\textit{\textbf{attn-$\omega$}}} & \multicolumn{3}{c|}{\cellcolor[HTML]{F0C88C}\textit{\textbf{min}}} & \multicolumn{3}{c|}{\cellcolor[HTML]{F0C88C}\textit{\textbf{low-K}}} & \multicolumn{3}{c!{\vrule width 1.2pt}}{\cellcolor[HTML]{F0C88C}\textit{\textbf{attn-$\omega$}}} \\ \cline{3-20}

\rowcolor[HTML]{FFE7C8} 
\multirow{-3}{*}{\cellcolor[HTML]{F0C88C}\textbf{Model}} & \multicolumn{1}{l!{\vrule width 1.2pt}}{\multirow{-3}{*}{\cellcolor[HTML]{F0C88C}\textbf{Size}}} & \textit{\textbf{ECE}} & \textit{\textbf{$\boldsymbol{\mathcal{B}}$}} & \multicolumn{1}{c|}{\cellcolor[HTML]{FFE7C8}\textit{\textbf{BC}}} & \textit{\textbf{ECE}} & \textit{\textbf{$\boldsymbol{\mathcal{B}}$}} & \multicolumn{1}{c|}{\cellcolor[HTML]{FFE7C8}\textit{\textbf{BC}}} & \textit{\textbf{ECE}} & \textit{\textbf{$\boldsymbol{\mathcal{B}}$}} & \multicolumn{1}{c!{\vrule width 1.2pt}}{\cellcolor[HTML]{FFE7C8}\textit{\textbf{BC}}} & \textit{\textbf{ECE}} & \textit{\textbf{$\boldsymbol{\mathcal{B}}$}} & \multicolumn{1}{c|}{\cellcolor[HTML]{FFE7C8}\textit{\textbf{BC}}} & \textit{\textbf{ECE}} & \textit{\textbf{$\boldsymbol{\mathcal{B}}$}} & \multicolumn{1}{c|}{\cellcolor[HTML]{FFE7C8}\textit{\textbf{BC}}} & \textit{\textbf{ECE}} & \textit{\textbf{$\boldsymbol{\mathcal{B}}$}} & \textit{\textbf{BC}} \\ \Xhline{2\arrayrulewidth}

\rowcolor[HTML]{FFF5E3} 
\cellcolor[HTML]{F0C88C} & \cellcolor[HTML]{FFE7C8}\textbf{8B} & {\color[HTML]{38761D} \textbf{0.13 (-0.03)}} & {\color[HTML]{000000} \textbf{0.20 (=0)}} & \multicolumn{1}{c|}{\cellcolor[HTML]{FFF5E3}{\color[HTML]{38761D} \textbf{9 (+3)}}} & {\color[HTML]{38761D} 0.18 (-0.01)} & {\color[HTML]{000000} 0.22 (=0)} & \multicolumn{1}{c|}{\cellcolor[HTML]{FFF5E3}{\color[HTML]{000000} 4 (=0)}} & {\color[HTML]{000000} 0.17 (=0)} & {\color[HTML]{000000} 0.23 (=0)} & \multicolumn{1}{c!{\vrule width 1.2pt}}{\color[HTML]{000000} 4 (=0)} & {\color[HTML]{38761D} \textbf{0.07 (-0.10)}} & {\color[HTML]{38761D} \textbf{0.22 (-0.01)}} & \multicolumn{1}{c|}{\cellcolor[HTML]{FFF5E3}{\color[HTML]{38761D} \textbf{9 (+2)}}} & {\color[HTML]{38761D} 0.11 (-0.07)} & {\color[HTML]{38761D} 0.23 (-0.01)} & \multicolumn{1}{c|}{\cellcolor[HTML]{FFF5E3}{\color[HTML]{38761D} 8 (+2)}} & {\color[HTML]{000000} 0.18 (=0)} & {\color[HTML]{000000} 0.24 (=0)} & {\color[HTML]{000000} 5 (=0)} \\ \cline{2-20}

\rowcolor[HTML]{FFF5E3} 
\multirow{-2}{*}{\cellcolor[HTML]{F0C88C}\textbf{Llama-3.1}} & \cellcolor[HTML]{FFE7C8}\textbf{70B} & {\color[HTML]{38761D} \textbf{0.08 (-0.09)}} & {\color[HTML]{000000} \textbf{0.22 (=0)}} & \multicolumn{1}{c|}{\cellcolor[HTML]{FFF5E3}{\color[HTML]{38761D} \textbf{9 (+2)}}} & {\color[HTML]{000000} 0.19 (=0)} & {\color[HTML]{000000} 0.25 (=0)} & \multicolumn{1}{c|}{\cellcolor[HTML]{FFF5E3}{\color[HTML]{000000} 5 (=0)}} & {\color[HTML]{000000} 0.19 (=0)} & {\color[HTML]{000000} 0.25 (=0)} & \multicolumn{1}{c!{\vrule width 1.2pt}}{\color[HTML]{000000} 5 (=0)} & {\color[HTML]{38761D} 0.09 (-0.05)} & {\color[HTML]{38761D} \textbf{0.20 (-0.01)}} & \multicolumn{1}{c|}{\cellcolor[HTML]{FFF5E3}{\color[HTML]{38761D} \textbf{8 (+1)}}} & {\color[HTML]{38761D} \textbf{0.08 (-0.08)}} & {\color[HTML]{38761D} 0.21 (-0.02)} & \multicolumn{1}{c|}{\cellcolor[HTML]{FFF5E3}{\color[HTML]{38761D} 7 (+2)}} & {\color[HTML]{000000} 0.16 (=0)} & {\color[HTML]{000000} 0.23 (=0)} & {\color[HTML]{000000} 5 (=0)} \\ \hline

\rowcolor[HTML]{FFF5E3} 
\cellcolor[HTML]{F0C88C} & \cellcolor[HTML]{FFE7C8}\textbf{7B} & {\color[HTML]{38761D} \textbf{0.12 (-0.07)}} & {\color[HTML]{38761D} \textbf{0.22 (-0.01)}} & \multicolumn{1}{c|}{\cellcolor[HTML]{FFF5E3}{\color[HTML]{38761D} \textbf{8 (+2)}}} & {\color[HTML]{000000} 0.21 (=0)} & {\color[HTML]{000000} 0.25 (=0)} & \multicolumn{1}{c|}{\cellcolor[HTML]{FFF5E3}{\color[HTML]{000000} 4 (=0)}} & {\color[HTML]{000000} 0.21 (=0)} & {\color[HTML]{000000} 0.25 (=0)} & \multicolumn{1}{c!{\vrule width 1.2pt}}{\color[HTML]{000000} 4 (=0)} & {\color[HTML]{38761D} \textbf{0.10 (-0.15)}} & {\color[HTML]{38761D} \textbf{0.22 (-0.05)}} & \multicolumn{1}{c|}{\cellcolor[HTML]{FFF5E3}{\color[HTML]{38761D} \textbf{7 (+2)}}} & {\color[HTML]{38761D} 0.18 (-0.08)} & {\color[HTML]{38761D} 0.25 (-0.04)} & \multicolumn{1}{c|}{\cellcolor[HTML]{FFF5E3}{\color[HTML]{38761D} \textbf{7 (+3)}}} & {\color[HTML]{000000} 0.26 (=0)} & {\color[HTML]{000000} 0.29 (=0)} & {\color[HTML]{000000} 4 (=0)} \\ \cline{2-20}

\rowcolor[HTML]{FFF5E3} 
\multirow{-2}{*}{\cellcolor[HTML]{F0C88C}\textbf{CodeLlama}} & \cellcolor[HTML]{FFE7C8}\textbf{70B} & {\color[HTML]{38761D} \textbf{0.10 (-0.11)}} & {\color[HTML]{38761D} \textbf{0.22 (-0.03)}} & \multicolumn{1}{c|}{\cellcolor[HTML]{FFF5E3}{\color[HTML]{38761D} \textbf{8 (+1)}}} & {\color[HTML]{990000} 0.23 (+0.02)} & {\color[HTML]{38761D} 0.28 (-0.01)} & \multicolumn{1}{c|}{\cellcolor[HTML]{FFF5E3}{\color[HTML]{38761D} \textbf{8 (+3)}}} & {\color[HTML]{000000} 0.25 (=0)} & {\color[HTML]{000000} 0.29 (=0)} & \multicolumn{1}{c!{\vrule width 1.2pt}}{\color[HTML]{000000} 5 (=0)} & {\color[HTML]{38761D} 0.18 (-0.08)} & {\color[HTML]{38761D} 0.22 (-0.03)} & \multicolumn{1}{c|}{\cellcolor[HTML]{FFF5E3}{\color[HTML]{38761D} 7 (+2)}} & {\color[HTML]{38761D} \textbf{0.12 (-0.16)}} & {\color[HTML]{38761D} \textbf{0.21 (-0.06)}} & \multicolumn{1}{c|}{\cellcolor[HTML]{FFF5E3}{\color[HTML]{38761D} \textbf{8 (+4)}}} & {\color[HTML]{000000} 0.28 (=0)} & {\color[HTML]{000000} 0.27 (=0)} & {\color[HTML]{000000} 4 (=0)} \\ \hline

\rowcolor[HTML]{FFF5E3} 
\cellcolor[HTML]{F0C88C} & \cellcolor[HTML]{FFE7C8}\textbf{7B} & {\color[HTML]{38761D} \textbf{0.13 (-0.01)}} & {\color[HTML]{000000} \textbf{0.18 (=0)}} & \multicolumn{1}{c|}{\cellcolor[HTML]{FFF5E3}{\color[HTML]{38761D} \textbf{8 (+3)}}} & {\color[HTML]{000000} 0.14 (=0)} & {\color[HTML]{000000} 0.20 (=0)} & \multicolumn{1}{c|}{\cellcolor[HTML]{FFF5E3}{\color[HTML]{000000} 3 (=0)}} & {\color[HTML]{000000} 0.14 (=0)} & {\color[HTML]{000000} 0.20 (=0)} & \multicolumn{1}{c!{\vrule width 1.2pt}}{\color[HTML]{000000} 3 (=0)} & {\color[HTML]{38761D} \textbf{0.07 (-0.10)}} & {\color[HTML]{38761D} \textbf{0.23 (-0.01)}} & \multicolumn{1}{c|}{\cellcolor[HTML]{FFF5E3}{\color[HTML]{38761D} \textbf{8 (+2)}}} & {\color[HTML]{38761D} 0.16 (-0.01)} & {\color[HTML]{000000} 0.24 (=0)} & \multicolumn{1}{c|}{\cellcolor[HTML]{FFF5E3}{\color[HTML]{38761D} 7 (+2)}} & {\color[HTML]{000000} 0.17 (=0)} & {\color[HTML]{000000} 0.25 (=0)} & {\color[HTML]{000000} 5 (=0)} \\ \cline{2-20}

\rowcolor[HTML]{FFF5E3} 
\multirow{-2}{*}{\cellcolor[HTML]{F0C88C}\textbf{Qwen2.5}} & \cellcolor[HTML]{FFE7C8}\textbf{72B} & {\color[HTML]{38761D} \textbf{0.12 (-0.07)}} & {\color[HTML]{000000} \textbf{0.21 (=0)}} & \multicolumn{1}{c|}{\cellcolor[HTML]{FFF5E3}{\color[HTML]{38761D} \textbf{10 (+4)}}} & {\color[HTML]{000000} 0.20 (=0)} & {\color[HTML]{000000} 0.24 (=0)} & \multicolumn{1}{c|}{\cellcolor[HTML]{FFF5E3}{\color[HTML]{000000} 5 (=0)}} & {\color[HTML]{000000} 0.20 (=0)} & {\color[HTML]{000000} 0.24 (=0)} & \multicolumn{1}{c!{\vrule width 1.2pt}}{\color[HTML]{000000} 5 (=0)} & {\color[HTML]{38761D} \textbf{0.08 (-0.12)}} & {\color[HTML]{38761D} \textbf{0.17 (-0.03)}} & \multicolumn{1}{c|}{\cellcolor[HTML]{FFF5E3}{\color[HTML]{38761D} \textbf{8 (+1)}}} & {\color[HTML]{38761D} 0.14 (-0.07)} & {\color[HTML]{38761D} 0.20 (-0.02)} & \multicolumn{1}{c|}{\cellcolor[HTML]{FFF5E3}{\color[HTML]{000000} 6 (=0)}} & {\color[HTML]{000000} 0.21 (=0)} & {\color[HTML]{000000} 0.22 (=0)} & {\color[HTML]{000000} 6 (=0)} \\ \hline

\rowcolor[HTML]{FFF5E3} 
\cellcolor[HTML]{F0C88C} & \cellcolor[HTML]{FFE7C8}\textbf{7B} & {\color[HTML]{38761D} \textbf{0.06 (-0.07)}} & {\color[HTML]{38761D} \textbf{0.18 (-0.01)}} & \multicolumn{1}{c|}{\cellcolor[HTML]{FFF5E3}{\color[HTML]{38761D} \textbf{6 (+2)}}} & {\color[HTML]{38761D} 0.10 (-0.02)} & {\color[HTML]{000000} 0.20 (=0)} & \multicolumn{1}{c|}{\cellcolor[HTML]{FFF5E3}{\color[HTML]{000000} 3 (=0)}} & {\color[HTML]{38761D} 0.10 (-0.01)} & {\color[HTML]{000000} 0.20 (=0)} & \multicolumn{1}{c!{\vrule width 1.2pt}}{\color[HTML]{000000} 3 (=0)} & {\color[HTML]{38761D} 0.11 (-0.06)} & {\color[HTML]{38761D} \textbf{0.24 (-0.01)}} & \multicolumn{1}{c|}{\cellcolor[HTML]{FFF5E3}{\color[HTML]{38761D} 6 (+2)}} & {\color[HTML]{38761D} \textbf{0.09 (-0.07)}} & {\color[HTML]{38761D} \textbf{0.24 (-0.01)}} & \multicolumn{1}{c|}{\cellcolor[HTML]{FFF5E3}{\color[HTML]{38761D} \textbf{7 (+4)}}} & {\color[HTML]{38761D} 0.14 (-0.01)} & {\color[HTML]{38761D} 0.25 (-0.01)} & {\color[HTML]{000000} 3 (=0)} \\ \cline{2-20}

\rowcolor[HTML]{FFF5E3} 
\multirow{-2}{*}{\cellcolor[HTML]{F0C88C}\textbf{Qwen2.5-Coder}} & \cellcolor[HTML]{FFE7C8}\textbf{32B} & {\color[HTML]{38761D} \textbf{0.11 (-0.03)}} & {\color[HTML]{38761D} \textbf{0.19 (-0.01)}} & \multicolumn{1}{c|}{\cellcolor[HTML]{FFF5E3}{\color[HTML]{38761D} \textbf{10 (+4)}}} & {\color[HTML]{000000} 0.15 (=0)} & {\color[HTML]{000000} 0.22 (=0)} & \multicolumn{1}{c|}{\cellcolor[HTML]{FFF5E3}{\color[HTML]{000000} 4 (=0)}} & {\color[HTML]{000000} 0.15 (=0)} & {\color[HTML]{000000} 0.22 (=0)} & \multicolumn{1}{c!{\vrule width 1.2pt}}{\color[HTML]{000000} 4 (=0)} & {\color[HTML]{38761D} \textbf{0.09 (-0.07)}} & {\color[HTML]{38761D} \textbf{0.19 (-0.01)}} & \multicolumn{1}{c|}{\cellcolor[HTML]{FFF5E3}{\color[HTML]{38761D} \textbf{8 (+1)}}} & {\color[HTML]{38761D} 0.14 (-0.03)} & {\color[HTML]{000000} 0.22 (=0)} & \multicolumn{1}{c|}{\cellcolor[HTML]{FFF5E3}{\color[HTML]{38761D} \textbf{8 (+3)}}} & {\color[HTML]{000000} 0.17 (=0)} & {\color[HTML]{000000} 0.22 (=0)} & {\color[HTML]{000000} 5 (=0)} \\ \hline

\rowcolor[HTML]{FFF5E3} 
\cellcolor[HTML]{F0C88C} & \cellcolor[HTML]{FFE7C8}\textbf{7B} & {\color[HTML]{990000} 0.10 (+0.04)} & {\color[HTML]{990000} 0.10 (+0.01)} & \multicolumn{1}{c|}{\cellcolor[HTML]{FFF5E3}{\color[HTML]{38761D} \textbf{6 (+4)}}} & {\color[HTML]{000000} 0.06 (=0)} & {\color[HTML]{000000} \textbf{0.09 (=0)}} & \multicolumn{1}{c|}{\cellcolor[HTML]{FFF5E3}{\color[HTML]{000000} 2 (=0)}} & {\color[HTML]{000000} \textbf{0.05 (=0)}} & {\color[HTML]{000000} \textbf{0.09 (=0)}} & \multicolumn{1}{c!{\vrule width 1.2pt}}{\color[HTML]{000000} 2 (=0)} & {\color[HTML]{38761D} \textbf{0.06 (-0.03)}} & {\color[HTML]{000000} \textbf{0.16 (=0)}} & \multicolumn{1}{c|}{\cellcolor[HTML]{FFF5E3}{\color[HTML]{38761D} \textbf{7 (+3)}}} & {\color[HTML]{000000} 0.09 (=0)} & {\color[HTML]{000000} \textbf{0.16 (=0)}} & \multicolumn{1}{c|}{\cellcolor[HTML]{FFF5E3}{\color[HTML]{38761D} 4 (+1)}} & {\color[HTML]{38761D} 0.08 (-0.02)} & {\color[HTML]{000000} 0.17 (=0)} & {\color[HTML]{000000} 3 (=0)} \\ \cline{2-20}

\rowcolor[HTML]{FFF5E3} 
\multirow{-2}{*}{\cellcolor[HTML]{F0C88C}\textbf{DS-LLM}} & \cellcolor[HTML]{FFE7C8}\textbf{67B} & {\color[HTML]{38761D} \textbf{0.15 (-0.02)}} & {\color[HTML]{000000} \textbf{0.22 (=0)}} & \multicolumn{1}{c|}{\cellcolor[HTML]{FFF5E3}{\color[HTML]{38761D} \textbf{10 (+5)}}} & {\color[HTML]{38761D} 0.18(-0.01)} & {\color[HTML]{000000} 0.24 (=0)} & \multicolumn{1}{c|}{\cellcolor[HTML]{FFF5E3}{\color[HTML]{000000} 4 (=0)}} & {\color[HTML]{38761D} 0.18 (-0.01)} & {\color[HTML]{000000} 0.24 (=0)} & \multicolumn{1}{c!{\vrule width 1.2pt}}{\color[HTML]{000000} 4 (=0)} & {\color[HTML]{38761D} \textbf{0.08 (-0.16)}} & {\color[HTML]{38761D} \textbf{0.21 (-0.05)}} & \multicolumn{1}{c|}{\cellcolor[HTML]{FFF5E3}{\color[HTML]{38761D} \textbf{8 (+3)}}} & {\color[HTML]{38761D} 0.15 (-0.10)} & {\color[HTML]{38761D} 0.24 (-0.03)} & \multicolumn{1}{c|}{\cellcolor[HTML]{FFF5E3}{\color[HTML]{38761D} 7 (+4)}} & {\color[HTML]{000000} 0.25 (=0)} & {\color[HTML]{000000} 0.27 (=0)} & {\color[HTML]{000000} 3 (=0)} \\ \hline

\rowcolor[HTML]{FFF5E3} 
\cellcolor[HTML]{F0C88C} & \cellcolor[HTML]{FFE7C8}\textbf{7B} & {\color[HTML]{38761D} \textbf{0.16 (-0.03)}} & {\color[HTML]{000000} \textbf{0.21 (=0)}} & \multicolumn{1}{c|}{\cellcolor[HTML]{FFF5E3}{\color[HTML]{38761D} \textbf{7 (+1)}}} & {\color[HTML]{000000} 0.19 (=0)} & {\color[HTML]{000000} 0.23 (=0)} & \multicolumn{1}{c|}{\cellcolor[HTML]{FFF5E3}{\color[HTML]{000000} 4 (=0)}} & {\color[HTML]{000000} 0.19 (=0)} & {\color[HTML]{000000} 0.23 (=0)} & \multicolumn{1}{c!{\vrule width 1.2pt}}{\color[HTML]{000000} 4 (=0)} & {\color[HTML]{38761D} \textbf{0.08 (-0.13)}} & {\color[HTML]{38761D} \textbf{0.21 (-0.03)}} & \multicolumn{1}{c|}{\cellcolor[HTML]{FFF5E3}{\color[HTML]{38761D} \textbf{9 (+2)}}} & {\color[HTML]{38761D} 0.14 (-0.07)} & {\color[HTML]{38761D} 0.24 (-0.01)} & \multicolumn{1}{c|}{\cellcolor[HTML]{FFF5E3}{\color[HTML]{38761D} 8 (+3)}} & {\color[HTML]{000000} 0.21 (=0)} & {\color[HTML]{000000} 0.25 (=0)} & {\color[HTML]{000000} 5 (=0)} \\ \cline{2-20}

\rowcolor[HTML]{FFF5E3} 
\multirow{-2}{*}{\cellcolor[HTML]{F0C88C}\textbf{DS-Coder}} & \cellcolor[HTML]{FFE7C8}\textbf{33B} & {\color[HTML]{38761D} \textbf{0.11 (-0.06)}} & {\color[HTML]{000000} \textbf{0.22 (=0)}} & \multicolumn{1}{c|}{\cellcolor[HTML]{FFF5E3}{\color[HTML]{38761D} \textbf{10 (+3)}}} & {\color[HTML]{000000} 0.19 (=0)} & {\color[HTML]{38761D} 0.25 (-0.02)} & \multicolumn{1}{c|}{\cellcolor[HTML]{FFF5E3}{\color[HTML]{000000} 5 (=0)}} & {\color[HTML]{000000} 0.19 (=0)} & {\color[HTML]{000000} 0.25 (=0)} & \multicolumn{1}{c!{\vrule width 1.2pt}}{\color[HTML]{000000} 5 (=0)} & {\color[HTML]{38761D} 0.13 (-0.09)} & {\color[HTML]{38761D} \textbf{0.21 (-0.03)}} & \multicolumn{1}{c|}{\cellcolor[HTML]{FFF5E3}{\color[HTML]{38761D} \textbf{8 (+2)}}} & {\color[HTML]{38761D} \textbf{0.11 (-0.12)}} & {\color[HTML]{38761D} \textbf{0.21 (-0.04)}} & \multicolumn{1}{c|}{\cellcolor[HTML]{FFF5E3}{\color[HTML]{38761D} 7 (+3)}} & {\color[HTML]{38761D} 0.24 (-0.01)} & {\color[HTML]{000000} 0.25 (=0)} & {\color[HTML]{000000} 4 (=0)} \\ \hline

\rowcolor[HTML]{FFF5E3} 
\cellcolor[HTML]{F0C88C}\textbf{DS-V2} & \cellcolor[HTML]{FFE7C8}\textbf{16B} & {\color[HTML]{000000} \textbf{0.10 (=0)}} & {\color[HTML]{000000} \textbf{0.13 (=0)}} & \multicolumn{1}{c|}{\cellcolor[HTML]{FFF5E3}{\color[HTML]{000000} \textbf{3 (=0)}}} & {\color[HTML]{000000} \textbf{0.10 (=0)}} & {\color[HTML]{000000} 0.14 (=0)} & \multicolumn{1}{c|}{\cellcolor[HTML]{FFF5E3}{\color[HTML]{000000} 2 (=0)}} & {\color[HTML]{000000} \textbf{0.10 (=0)}} & {\color[HTML]{000000} 0.14 (=0)} & \multicolumn{1}{c!{\vrule width 1.2pt}}{\color[HTML]{000000} 2 (=0)} & {\color[HTML]{000000} \textbf{0.13 (=0)}} & {\color[HTML]{000000} \textbf{0.22 (=0)}} & \multicolumn{1}{c|}{\cellcolor[HTML]{FFF5E3}{\color[HTML]{000000} \textbf{5 (=0)}}} & {\color[HTML]{000000} \textbf{0.13 (=0)}} & {\color[HTML]{000000} 0.23 (=0)} & \multicolumn{1}{c|}{\cellcolor[HTML]{FFF5E3}{\color[HTML]{000000} 4 (=0)}} & {\color[HTML]{000000} \textbf{0.13 (=0)}} & {\color[HTML]{000000} 0.23 (=0)} & {\color[HTML]{000000} 2 (=0)} \\ \hline

\rowcolor[HTML]{FFF5E3} 
\cellcolor[HTML]{F0C88C}\textbf{DS-Coder-V2} & \cellcolor[HTML]{FFE7C8}\textbf{16B} & {\color[HTML]{000000} \textbf{0.14 (=0)}} & {\color[HTML]{000000} \textbf{0.20 (=0)}} & \multicolumn{1}{c|}{\cellcolor[HTML]{FFF5E3}{\color[HTML]{000000} \textbf{5 (=0)}}} & {\color[HTML]{000000} 0.16 (=0)} & {\color[HTML]{000000} 0.22 (=0)} & \multicolumn{1}{c|}{\cellcolor[HTML]{FFF5E3}{\color[HTML]{000000} 3 (=0)}} & {\color[HTML]{000000} 0.15 (=0)} & {\color[HTML]{000000} 0.22 (=0)} & \multicolumn{1}{c!{\vrule width 1.2pt}}{\color[HTML]{000000} 3 (=0)} & {\color[HTML]{000000} \textbf{0.18 (=0)}} & {\color[HTML]{000000} \textbf{0.25 (=0)}} & \multicolumn{1}{c|}{\cellcolor[HTML]{FFF5E3}{\color[HTML]{000000} \textbf{5 (=0)}}} & {\color[HTML]{000000} 0.19 (=0)} & {\color[HTML]{000000} 0.26 (=0)} & \multicolumn{1}{c|}{\cellcolor[HTML]{FFF5E3}{\color[HTML]{000000} 4 (=0)}} & {\color[HTML]{000000} 0.19 (=0)} & {\color[HTML]{000000} 0.27 (=0)} & {\color[HTML]{38761D} 4 (+1)} \\ \hline
 \Xhline{2\arrayrulewidth}

\multicolumn{20}{l}{\textbf{\textit{min}:} Minimum Token Probability, \textbf{\textit{low-K}:} Lowest-$K$ Token Probability \textbf{\textit{attn-$\omega$}:} Attention-Weighted Uncertainty, \textbf{EM:} Exact Match, \textbf{EP:} Edit Progress, \textbf{ECE:} Expected Calibration Error (↓), $\boldsymbol{\mathcal{B}}$\textbf{:} Brier Score (↓), \textbf{BC:} Bin Coverage (↑)} \\

\multicolumn{20}{l}{\footnotesize\textbf{Bold Number:} Best Result for that Model and Correctness Measure, \textbf{(+/-/=):} Difference from Global Platt-Scaling} \\

\end{tabular}
}
\label{Tab:cr_trans_local}
\end{table*}

Table~\ref{Tab:dcf_bug_local} shows the $ECE$, $\mathcal{B}$, and $BC$ results after local Platt-scaling for DCF-Bug.
In general, we find that confidence calibration can be improved regardless of model, confidence score, or correctness metric.
Specifically, minimum token probability is the most amenable to local Platt-scaling, yielding reductions in $ECE$ and $\mathcal{B}$ across the highest number of models.
This further solidifies its position as the best performing confidence score for both correctness metrics considered in DCF-Bug.
In terms of ($ECE$, $\mathcal{B}$) against $EM$, (12, 7), (10, 7), and (8, 8) models experienced improvements for minimum token probability, lowest-$K$ token probability, and attention-weighted uncertainty.
In terms of ($ECE$, $\mathcal{B}$) against $EP$, (12, 3), (9, 4), and (7, 2) models experienced improvements, respectively.
In terms of bin coverage, the results show that local Platt-scaling can also induce higher $BC$ for certain models.
These findings are most prevalent for $EP$, where seven, four, and seven models experienced increases in $BC$ for minimum token probability, lowest-$K$ token probability, and attention-weighted uncertainty, respectively.
Despite the favourable findings, we note that all calibration improvements are still marginal ($ECE$: $\Delta$0.01-0.07, $\mathcal{B}$: $\Delta$0.01-0.02, $BC$: $\Delta$1-2).
These results suggest that while error heterogeneity exists for automated program repair, it is relatively modest.

Table~\ref{Tab:dcf_vul} shows the $ECE$, $\mathcal{B}$, and $BC$ results after local Platt-scaling for DCF-Vul.
In general, we find that confidence calibration can be improved for the majority of models against $CP$, but is relatively ineffective against $EP$.
As in DCF-Bug, minimum token probability remains the most effective confidence score, owing to its strong baseline performance and amenability to local Platt-scaling.
In terms of ($ECE$, $\mathcal{B}$) against $EP$, only (5, 0), (4, 0), and (0, 1) models experienced improvements for minimum token probability, lowest-$K$ token probability, and attention-weighted uncertainty.
In contrast, for ($ECE$, $\mathcal{B}$) against $CP$, (9, 9), (9, 9), and (9, 11) models experienced improvements, respectively.
These results suggest that error heterogeneity is largely confined against $EP$.
Interestingly, we find that the low bin coverage issue against $CP$ can be greatly improved, where nine, 10, and 11 models experienced increases in $BC$ for minimum token probability, lowest-$K$ token probability, and attention-weighted uncertainty, respectively.
Despite persistent cases of bin collapse under global Platt-scaling for these fine-grained confidence scores, local Platt-scaling entirely mitigates the issue.
Overall, while the reduction in calibration error against $CP$ is comparable to that of DCF-Bug ($ECE$: $\Delta$0.01-0.04, $\mathcal{B}$: $\Delta$0.01-0.02), local Platt-scaling for DCF-Vul achieves a significantly greater increase in bin coverage ($BC$: $\Delta$1-6).
This suggests that errors in the task of vulnerability repair, as governed by security rules, exhibit greater variation.

Table~\ref{Tab:cr_trans} shows the $ECE$, $\mathcal{B}$, and $BC$ results after local Platt-scaling for CR-Trans.
In general, this task yields the largest improvement in confidence calibration.
Specifically, local Platt-scaling is highly effective for minimum token probability against both $EM$ and $EP$.
In terms of ($ECE$, $\mathcal{B}$) against $EM$, (11, 4) models experienced improvements for minimum token probability, whilst only (3, 2) and (2, 0) models experienced improvements for lowest-$K$ token probability and attention-weighted uncertainty.
In contrast, for ($ECE$, $\mathcal{B}$) against $EP$, (12, 11) and (11, 9) models experienced improvements for minimum token probability and lowest-$K$ token probability, whilst only (3, 2) models experienced improvements for attention-weighted uncertainty.
Under local Platt-scaling, $EP$ is easier to calibrate than $EM$.
Using minimum token probability, nine models can achieve $\leq$ 0.10 $ECE$ against $EP$, compared to only five against $EM$.
Additionally, using this confidence score can increase bin coverage for 12 models against both correctness metrics.
Compared to the prior tasks, CR-Trans exhibits higher error heterogeneity, much of which can be captured by local Platt-scaling ($ECE$: $\Delta$0.01-0.16, $\mathcal{B}$: $\Delta$0.01-0.06, $BC$: $\Delta$1-5).
Overall, the results suggest that using minimum token probability with local Platt-scaling is necessary for facilitating accurate decision-making with generated code revisions in automated code refinement.

\summary{
The empirical results show that local Platt-scaling yields substantial improvements over conventional Platt-scaling in CR-Trans, with moderate improvements in DCF-Bug and DCF-Vul. 
Its combination with minimum token probability consistently achieves the strongest performance across $ECE$, $\mathcal{B}$, and $BC$ for all ACR tasks.
}

\section{Discussion}
\label{sec:discussion}

Although the results demonstrate the effectiveness of our fine-grained calibration approach, a more principled understanding of its behaviour is still needed. 
This section provides such an analysis by examining the role of Platt-scaling within the end-to-end approach, the varying impact of local Platt-scaling across the different tasks, its underlying clustering dynamics, sensitivity to hyperparameters, validation data requirements, and considerations for practical deployment.

\textbf{To what extent is fine-grained confidence calibration dependent on Platt-scaling?}
As fine-grained confidence scores and local clustering are scaler-agnostic, we further evaluate their pairing with other traditional post-hoc scalers, in addition to Platt-scaling, that require only the original inference step.
Temperature-scaling~\cite{neuralnetwork_calibration} is a restricted form of Platt-scaling that omits the bias term, learning only a single scaling parameter.
Histogram binning~\cite{zadrozny2001obtaining} is a model-free alternative that calibrates confidence estimates by grouping them into bins and assigning each bin the empirical likelihood of correctness observed within it.
Isotonic regression~\cite{zadrozny2002transforming} is a more flexible generalisation of histogram binning that learns bin boundaries adaptively from training examples, rather than fixing them in advance, by fitting the best non-decreasing step function to the empirical likelihood of correctness.
To facilitate this analysis, we focus on the two most performant models for each task.

Table~\ref{Tab:dcf_bug_ablation} demonstrates the performance variation when applying different scalers to fine-grained confidence calibration using minimum token probability on DCF-Bug.
Calibration errors vary minimally when replacing Platt-scaling with histogram binning or isotonic regression ($ECE$: $\Delta$0.01-0.03, $\mathcal{B}$: $\Delta$0.01); however, $BC$ can decrease by $\Delta$1-3 depending on the setting.
In contrast, temperature-scaling can substantially increase $ECE$ and $\mathcal{B}$ by up to $\Delta$0.18 and $\Delta$0.06, respectively. 
It also consistently reduces $BC$ by $\Delta$1–4, limiting coverage.

Table~\ref{Tab:dcf_vul_ablation} demonstrates the performance variation when applying different scalers to fine-grained confidence calibration using minimum token probability on DCF-Vul.
Utilising either histogram binning or isotonic regression yields similar results to Platt-scaling across all calibration metrics ($ECE$: $\Delta$0.01-0.02, $\mathcal{B}$: $\Delta$0.01, $BC$: $\Delta$1).
In contrast, utilising temperature-scaling can substantially degrade calibration, increasing $ECE$ and $\mathcal{B}$ by up to $\Delta$ 0.19 and $\Delta$0.06, respectively.

Table~\ref{Tab:cr_trans_ablation} demonstrates the performance variation when applying different scalers to fine-grained confidence calibration using minimum token probability on CR-Trans.
Aligning with prior findings, histogram binning and isotonic regression are both comparable to Platt-scaling ($ECE$: $\Delta$0.01-0.03, $\mathcal{B}$: $\Delta$0.01, $BC$: $\Delta$1-2).
Although temperature-scaling attains lower calibration error under certain settings ($ECE$: $\Delta$0.01-0.12, $\mathcal{B}$: $\Delta$0.01-0.04), these gains are obtained at the expense of a substantial reduction in coverage ($BC$: $\Delta$1-6), thereby diminishing the practical utility in downstream application.

Overall, Platt-scaling is consistently competitive across all settings and calibration metrics, typically achieving either the lowest calibration error or the highest coverage without sacrificing the other.
The benefits of fine-grained calibration extend to histogram binning and isotonic regression given the larger training set sizes in our study, whereas temperature-scaling is generally ineffective.
This indicates the importance of the additional bias term. 
Since Platt-scaling fully subsumes temperature-scaling as a special case ($\beta = 0$), restricting calibration to a single parameter scaler is typically unnecessary.

\begin{table}[!htbp]
\caption{Comparing Scalers for Fine-Grained Confidence Calibration using Minimum Token Probability on DCF-Bug}
\centering
\resizebox{\columnwidth}{!}{%
\begin{tabular}{!{\vrule width 1.2pt}l!{\vrule width 1.2pt}l|ccc|ccc!{\vrule width 1.2pt}}
\Xhline{2\arrayrulewidth}
\rowcolor[HTML]{F0C88C} 
\cellcolor[HTML]{F0C88C} & \cellcolor[HTML]{F0C88C} & \multicolumn{3}{c|}{\cellcolor[HTML]{F0C88C}\textbf{EM}} & \multicolumn{3}{c!{\vrule width 1.2pt}}{\cellcolor[HTML]{F0C88C}\textbf{EP}} \\ \cline{3-8} 
\rowcolor[HTML]{FFE7C8} 
\multirow{-2}{*}{\cellcolor[HTML]{F0C88C}\textbf{Model}} & \multirow{-2}{*}{\cellcolor[HTML]{F0C88C}\textbf{Scaler}} & \textit{\textbf{ECE}} & \textit{\textbf{$\mathcal{B}$}} & \textit{\textbf{BC}} & \textit{\textbf{ECE}} & \textit{\textbf{$\mathcal{B}$}} & \textit{\textbf{BC}} \\ \Xhline{2\arrayrulewidth}
\rowcolor[HTML]{FFF5E3} 
\cellcolor[HTML]{F0C88C} & \cellcolor[HTML]{FFE7C8}\textbf{Platt-scaling} & {\color[HTML]{000000} \textbf{0.04 (0.04)}} & {\color[HTML]{000000} \textbf{0.20 (0.20)}} & {\color[HTML]{000000} \textbf{8 (8)}} & {\color[HTML]{000000} 0.05 (0.04)} & {\color[HTML]{000000} \textbf{0.18 (0.18)}} & {\color[HTML]{000000} \textbf{6 (7)}} \\ \cline{2-8} 
\rowcolor[HTML]{FFF5E3} 
\cellcolor[HTML]{F0C88C} & \cellcolor[HTML]{FFE7C8}\textbf{Temperature-scaling} & {\color[HTML]{000000} 0.18 (0.18)} & {\color[HTML]{000000} 0.24 (0.24)} & {\color[HTML]{000000} 4 (4)} & {\color[HTML]{000000} \textbf{0.02 (0.02)}} & {\color[HTML]{000000} \textbf{0.18 (0.18)}} & {\color[HTML]{000000} 4 (6)} \\ \cline{2-8} 
\rowcolor[HTML]{FFF5E3} 
\cellcolor[HTML]{F0C88C} & \cellcolor[HTML]{FFE7C8}\textbf{Histogram Binning} & {\color[HTML]{000000} 0.04 (0.05)} & {\color[HTML]{000000} \textbf{0.20 (0.20)}} & {\color[HTML]{000000} 5 (7)} & {\color[HTML]{000000} 0.05 (0.05)} & {\color[HTML]{000000} \textbf{0.18 (0.18)}} & {\color[HTML]{000000} 6 (6)} \\ \cline{2-8} 
\rowcolor[HTML]{FFF5E3} 
\multirow{-4}{*}{\cellcolor[HTML]{F0C88C}\textbf{CodeLlama-70b}} & \cellcolor[HTML]{FFE7C8}\textbf{Isotonic Regression} & {\color[HTML]{000000} 0.05 (0.04)} & {\color[HTML]{000000} \textbf{0.20 (0.20)}} & {\color[HTML]{000000} 8 (6)} & {\color[HTML]{000000} 0.05 (0.05)} & {\color[HTML]{000000} 0.18 (0.19)} & {\color[HTML]{000000} 5 (6)} \\ \hline
\rowcolor[HTML]{FFF5E3} 
\cellcolor[HTML]{F0C88C} & \cellcolor[HTML]{FFE7C8}\textbf{Platt-scaling} & 0.04 (0.04) & \textbf{0.18 (0.17)} & 8 (8) & \textbf{0.05 (0.03)} & \textbf{0.18 (0.18)} & \textbf{7 (8)} \\ \cline{2-8} 
\rowcolor[HTML]{FFF5E3} 
\cellcolor[HTML]{F0C88C} & \cellcolor[HTML]{FFE7C8}\textbf{Temperature-scaling} & 0.22 (0.21) & 0.24 (0.23) & 4 (6) & 0.04 (0.04) & \textbf{0.18 (0.18)} & 4 (6) \\ \cline{2-8} 
\rowcolor[HTML]{FFF5E3} 
\cellcolor[HTML]{F0C88C} & \cellcolor[HTML]{FFE7C8}\textbf{Histogram Binning} & \textbf{0.04 (0.02)} & 0.18 (0.18) & 6 (6) & 0.04 (0.06) & 0.19 (0.19) & 8 (7) \\ \cline{2-8} 
\rowcolor[HTML]{FFF5E3} 
\multirow{-4}{*}{\cellcolor[HTML]{F0C88C}\textbf{DS-Coder-33B}} & \cellcolor[HTML]{FFE7C8}\textbf{Isotonic Regression} & 0.03 (0.03) & \textbf{0.18 (0.17)} & \textbf{7 (9)} & 0.04 (0.05) & 0.19 (0.19) & 6 (8) \\ \Xhline{2\arrayrulewidth}
\multicolumn{8}{l}{\footnotesize \textbf{Bold Number:} Best Result for that Model and Correctness Measure} \\
\multicolumn{8}{l}{\footnotesize Global-Scaling (Local-Scaling)} \\
\end{tabular}
}
\label{Tab:dcf_bug_ablation}
\end{table}
\begin{table}[!htbp]
\caption{Comparing Scalers for Fine-Grained Confidence Calibration using Minimum Token Probability on DCF-Vul}
\centering
\resizebox{\columnwidth}{!}{%
\begin{tabular}{!{\vrule width 1.2pt}l!{\vrule width 1.2pt}l|ccc|ccc!{\vrule width 1.2pt}}
\Xhline{2\arrayrulewidth}
\rowcolor[HTML]{F0C88C} 
\cellcolor[HTML]{F0C88C} & \cellcolor[HTML]{F0C88C} & \multicolumn{3}{c|}{\cellcolor[HTML]{F0C88C}\textbf{EP}} & \multicolumn{3}{c!{\vrule width 1.2pt}}{\cellcolor[HTML]{F0C88C}\textbf{CP}} \\ \cline{3-8} 
\rowcolor[HTML]{FFE7C8} 
\multirow{-2}{*}{\cellcolor[HTML]{F0C88C}\textbf{Model}} & \multirow{-2}{*}{\cellcolor[HTML]{F0C88C}\textbf{Scaler}} & \textit{\textbf{ECE}} & \textit{\textbf{$\mathcal{B}$}} & \textit{\textbf{BC}} & \textit{\textbf{ECE}} & \textit{\textbf{$\mathcal{B}$}} & \textit{\textbf{BC}} \\ \Xhline{2\arrayrulewidth}
\cellcolor[HTML]{F0C88C} & \cellcolor[HTML]{FFE7C8}\textbf{Platt-scaling} & \cellcolor[HTML]{FFF5E3}{\color[HTML]{000000} 0.05 (0.05)} & \cellcolor[HTML]{FFF5E3}{\color[HTML]{000000} \textbf{0.19 (0.19)}} & \cellcolor[HTML]{FFF5E3}{\color[HTML]{000000} 6 (6)} & \cellcolor[HTML]{FFF5E3}{\color[HTML]{000000} \textbf{0.05 (0.04)}} & \cellcolor[HTML]{FFF5E3}{\color[HTML]{000000} \textbf{0.19 (0.19)}} & \cellcolor[HTML]{FFF5E3}{\color[HTML]{000000} \textbf{3 (3)}} \\ \cline{2-8} 
\rowcolor[HTML]{FFF5E3} 
\cellcolor[HTML]{F0C88C} & \cellcolor[HTML]{FFE7C8}\textbf{Temperature-scaling} & {\color[HTML]{000000} 0.18 (0.17)} & {\color[HTML]{000000} 0.22 (0.21)} & {\color[HTML]{000000} \textbf{7 (10)}} & {\color[HTML]{000000} 0.23 (0.23)} & {\color[HTML]{000000} 0.25 (0.25)} & {\color[HTML]{000000} 2 (2)} \\ \cline{2-8} 
\rowcolor[HTML]{FFF5E3} 
\cellcolor[HTML]{F0C88C} & \cellcolor[HTML]{FFE7C8}\textbf{Histogram Binning} & {\color[HTML]{000000} \textbf{0.04 (0.04)}} & {\color[HTML]{000000} \textbf{0.19 (0.19)}} & {\color[HTML]{000000} 5 (5)} & {\color[HTML]{000000} \textbf{0.05 (0.04)}} & {\color[HTML]{000000} \textbf{0.19 (0.19)}} & {\color[HTML]{000000} 2 (3)} \\ \cline{2-8} 
\rowcolor[HTML]{FFF5E3} 
\multirow{-4}{*}{\cellcolor[HTML]{F0C88C}\textbf{Llama-3.1-70B}} & \cellcolor[HTML]{FFE7C8}\textbf{Isotonic Regression} & {\color[HTML]{000000} \textbf{0.04 (0.04)}} & {\color[HTML]{000000} \textbf{0.19 (0.19)}} & {\color[HTML]{000000} 5 (5)} & {\color[HTML]{000000} \textbf{0.05 (0.04)}} & {\color[HTML]{000000} \textbf{0.19 (0.19)}} & {\color[HTML]{000000} 2 (3)} \\ \hline
\rowcolor[HTML]{FFF5E3} 
\cellcolor[HTML]{F0C88C} & \cellcolor[HTML]{FFE7C8}\textbf{Platt-scaling} & \textbf{0.06 (0.05)} & \textbf{0.19 (0.19)} & \textbf{6 (6)} & \textbf{0.03 (0.03)} & \textbf{0.23 (0.22)} & \textbf{3 (5)} \\ \cline{2-8} 
\rowcolor[HTML]{FFF5E3} 
\cellcolor[HTML]{F0C88C} & \cellcolor[HTML]{FFE7C8}\textbf{Temperature-scaling} & 0.23 (0.21) & 0.24 (0.24) & 5 (5) & 0.11 (0.11) & 0.24 (0.24) & 5 (5) \\ \cline{2-8} 
\rowcolor[HTML]{FFF5E3} 
\cellcolor[HTML]{F0C88C} & \cellcolor[HTML]{FFE7C8}\textbf{Histogram Binning} & \textbf{0.06 (0.05)} & \textbf{0.19 (0.19)} & \textbf{6 (6)} & 0.03 (0.05) & 0.23 (0.23) & \textbf{3 (5)} \\ \cline{2-8} 
\rowcolor[HTML]{FFF5E3} 
\multirow{-4}{*}{\cellcolor[HTML]{F0C88C}\textbf{DS-Coder-33B}} & \cellcolor[HTML]{FFE7C8}\textbf{Isotonic Regression} & \textbf{0.06 (0.05)} & \textbf{0.19 (0.19)} & \textbf{6 (6)} & 0.03 (0.05) & 0.23 (0.23) & \textbf{3 (5)} \\ \Xhline{2\arrayrulewidth}
\multicolumn{8}{l}{\footnotesize \textbf{Bold Number:} Best Result for that Model and Correctness Measure} \\
\multicolumn{8}{l}{\footnotesize Global-Scaling (Local-Scaling)} \\
\end{tabular}
}
\label{Tab:dcf_vul_ablation}
\end{table}
\begin{table}[!htbp]
\caption{Comparing Scalers for Fine-Grained Confidence Calibration using Minimum Token Probability on CR-Trans}
\centering
\resizebox{\columnwidth}{!}{%
\begin{tabular}{!{\vrule width 1.2pt}l!{\vrule width 1.2pt}l|ccc|ccc!{\vrule width 1.2pt}}
\Xhline{2\arrayrulewidth}
\rowcolor[HTML]{F0C88C} 
\cellcolor[HTML]{F0C88C} & \cellcolor[HTML]{F0C88C} & \multicolumn{3}{c|}{\cellcolor[HTML]{F0C88C}\textbf{EM}} & \multicolumn{3}{c!{\vrule width 1.2pt}}{\cellcolor[HTML]{F0C88C}\textbf{EP}} \\ \cline{3-8} 
\rowcolor[HTML]{FFE7C8} 
\multirow{-2}{*}{\cellcolor[HTML]{F0C88C}\textbf{Model}} & \multirow{-2}{*}{\cellcolor[HTML]{F0C88C}\textbf{Scaler}} & \textit{\textbf{ECE}} & \textit{\textbf{$\mathcal{B}$}} & \textit{\textbf{BC}} & \textit{\textbf{ECE}} & \textit{\textbf{$\mathcal{B}$}} & \textit{\textbf{BC}} \\ \Xhline{2\arrayrulewidth}
\cellcolor[HTML]{F0C88C} & \cellcolor[HTML]{FFE7C8}\textbf{Platt-scaling} & \cellcolor[HTML]{FFF5E3}{\color[HTML]{000000} 0.21 (0.10)} & \cellcolor[HTML]{FFF5E3}{\color[HTML]{000000} 0.25 (0.22)} & \cellcolor[HTML]{FFF5E3}{\color[HTML]{000000} \textbf{7 (8)}} & \cellcolor[HTML]{FFF5E3}{\color[HTML]{000000} 0.26 (0.18)} & \cellcolor[HTML]{FFF5E3}{\color[HTML]{000000} 0.25 (0.22)} & \cellcolor[HTML]{FFF5E3}{\color[HTML]{000000} \textbf{5 (7)}} \\ \cline{2-8} 
\rowcolor[HTML]{FFF5E3} 
\cellcolor[HTML]{F0C88C} & \cellcolor[HTML]{FFE7C8}\textbf{Temperature-scaling} & {\color[HTML]{000000} 0.09 (0.08)} & {\color[HTML]{000000} \textbf{0.21 (0.21)}} & {\color[HTML]{000000} 4 (5)} & {\color[HTML]{000000} \textbf{0.21 (0.13)}} & {\color[HTML]{000000} \textbf{0.23 (0.20)}} & {\color[HTML]{000000} 3 (5)} \\ \cline{2-8} 
\rowcolor[HTML]{FFF5E3} 
\cellcolor[HTML]{F0C88C} & \cellcolor[HTML]{FFE7C8}\textbf{Histogram Binning} & {\color[HTML]{000000} 0.20 (0.08)} & {\color[HTML]{000000} 0.24 (0.21)} & {\color[HTML]{000000} 6 (7)} & {\color[HTML]{000000} 0.23 (0.16)} & {\color[HTML]{000000} 0.24 (0.21)} & {\color[HTML]{000000} 5 (6)} \\ \cline{2-8} 
\rowcolor[HTML]{FFF5E3} 
\multirow{-4}{*}{\cellcolor[HTML]{F0C88C}\textbf{CodeLlama-70b}} & \cellcolor[HTML]{FFE7C8}\textbf{Isotonic Regression} & {\color[HTML]{000000} \textbf{0.20 (0.07)}} & {\color[HTML]{000000} 0.24 (0.21)} & {\color[HTML]{000000} 7 (7)} & {\color[HTML]{000000} 0.23 (0.16)} & {\color[HTML]{000000} 0.24 (0.21)} & {\color[HTML]{000000} \textbf{5 (7)}} \\ \hline
\rowcolor[HTML]{FFF5E3} 
\cellcolor[HTML]{F0C88C} & \cellcolor[HTML]{FFE7C8}\textbf{Platt-scaling} & \textbf{0.19 (0.12)} & \textbf{0.21 (0.21)} & \textbf{6 (10)} & 0.20 (0.08) & 0.20 (0.17) & \textbf{7 (8)} \\ \cline{2-8} 
\rowcolor[HTML]{FFF5E3} 
\cellcolor[HTML]{F0C88C} & \cellcolor[HTML]{FFE7C8}\textbf{Temperature-scaling} & 0.09 (0.16) & 0.24 (0.23) & 2 (4) & \textbf{0.14 (0.07)} & \textbf{0.18 (0.17)} & 5 (7) \\ \cline{2-8} 
\rowcolor[HTML]{FFF5E3} 
\cellcolor[HTML]{F0C88C} & \cellcolor[HTML]{FFE7C8}\textbf{Histogram Binning} & 0.19 (0.15) & \textbf{0.21 (0.21)} & 4 (9) & 0.20 (0.07) & 0.20 (0.17) & 6 (7) \\ \cline{2-8} 
\rowcolor[HTML]{FFF5E3} 
\multirow{-4}{*}{\cellcolor[HTML]{F0C88C}\textbf{Qwen2.5-72B}} & \cellcolor[HTML]{FFE7C8}\textbf{Isotonic Regression} & 0.19 (0.13) & \textbf{0.21 (0.21)} & \textbf{6 (10)} & 0.19 (0.07) & 0.20 (0.17) & 6 (8) \\ \Xhline{2\arrayrulewidth}
\multicolumn{8}{l}{\footnotesize \textbf{Bold Number:} Best Result for that Model and Correctness Measure} \\
\multicolumn{8}{l}{\footnotesize Global-Scaling (Local-Scaling)} \\
\end{tabular}
}
\label{Tab:cr_trans_ablation}
\end{table}

\begin{figure}[h]
    \centering
    \includegraphics[width=\columnwidth]{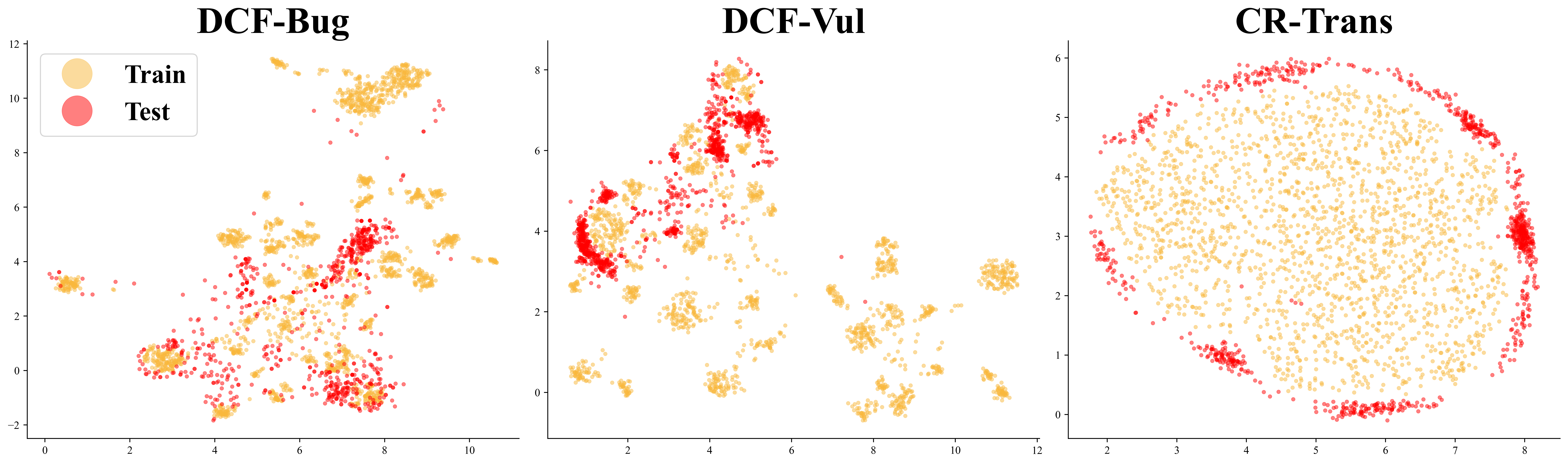} 
    \captionsetup{justification=centering} 
    \caption{UMAP Visualisations of Qwen3-Embedding-8B Representations of Inputs w/ CodeLlama-70B Outputs}
    \label{fig:umap}
\end{figure}

\textbf{Why is local Platt-scaling more essential for CR-Trans compared to DCF-Bug and DCF-Vul?}
As established by the main results, CR-Trans suffers from much higher error heterogeneity, compared to DCF-Bug and DCF-Vul, resulting in the reliance on local Platt-scaling to achieve reasonable calibration error.
Figure~\ref{fig:umap} compares UMAP visualisations of Qwen3-Embedding-8B
representations of examples across each of the ACR Tasks.
Each example consists of an input concatenated with its corresponding code revision generated by CodeLlama-70B.
We find that examples in CR-Trans are not only highly varied within the training set, but also exhibit covariate shift when comparing between training and test sets.
Intuitively, automated code refinement has the potential to cover a much larger and unexpected variation of $C_{pre}$ and $R_{nl}$ due to the fact that code reviews are human oriented.
The requested code revisions may address a wide range of issues, spanning across both functional issues and evolvability concerns~\cite{mantyala}, as well as any ad hoc issues in between.
This implies that not only is the data diverse in any collected calibration set, but it can also give rise to vastly different scenarios during deployment.
In contrast, the scope of both DCF-Bug and DCF-Vul is more limited than that of CR-Trans: DCF-Bug focuses specifically on bugs, while DCF-Vul targets vulnerabilities. 
In contrast, CR-Trans can theoretically capture both types of issues as subpopulations, depending on the focus of the human reviewer.
Additionally, both DCF-Bug and DCF-Vul are based on fixed static analysis rules $V_{nl}$, which means the variability in $C_{pre}$ is limited.
This implies that scenarios encountered during deployment will be relatively similar to those found in the training sets, meaning that there is limited opportunity for covariate shift.
In summary, the open-ended nature of automated code refinement induces a wider range of scenarios, increasing both variability and unpredictability of errors from the underlying LLM, thereby highlighting the need for local Platt-scaling with a backoff strategy.

\begin{figure}[h]
    \centering
    \includegraphics[width=\columnwidth]{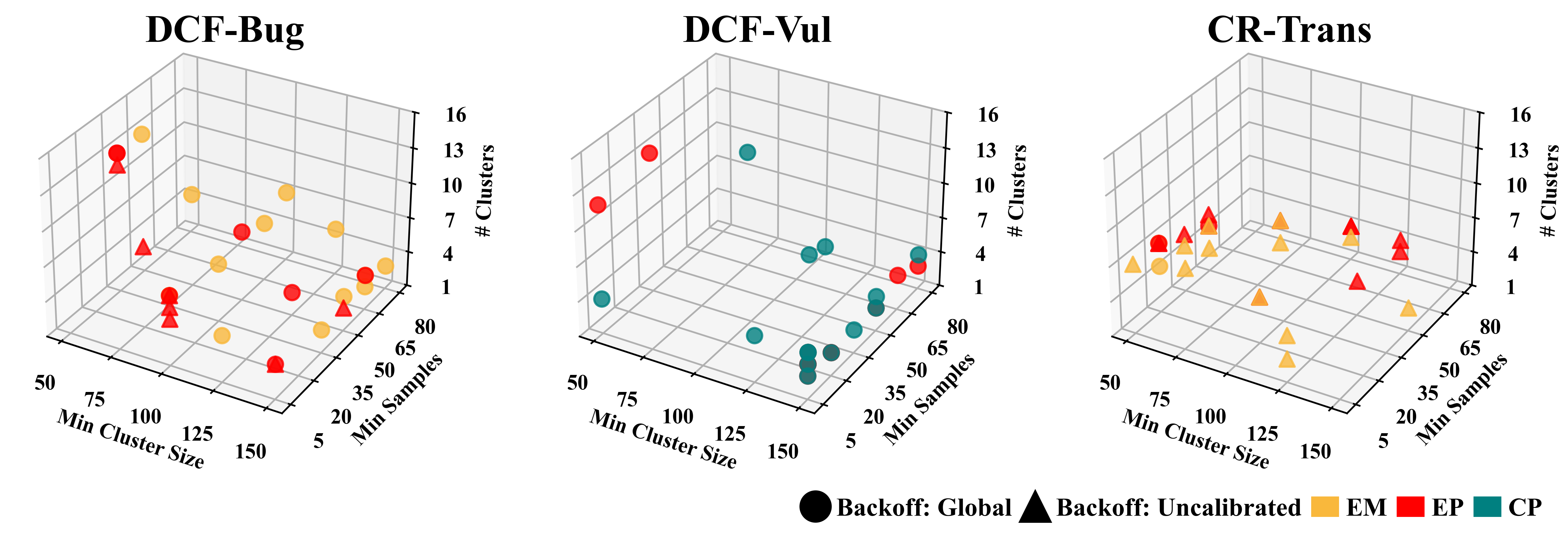} 
    \captionsetup{justification=centering} 
    \caption{Optimal Hyperparameters for Local Platt-Scaling w/ Minimum Token Probability}
    \label{fig:optimal_params}
\end{figure}

\textbf{How many clusters does local Platt-scaling require to achieve optimal confidence calibration?}
The Z-axis for the scatter plots in Figure~\ref{fig:optimal_params} shows the number of clusters induced by local Platt-Scaling under the most optimal hyperparameters for the top performing confidence score i.e., minimum token probability.
Interestingly, we find that more clusters are required in DCF-Bug, despite previous findings of modest error heterogeneity compared to DCF-Vul and CR-Trans.
For both DCF-Bug and DCF-Vul, the number of induced clusters range from three to 13.
For DCF-Bug, seven clusters were most frequently induced, whereas for DCF-Vul, the most common number was four.
For CR-Trans, the number of induced clusters range from three to seven, with the most common number being five.
The fact that conventional Platt-scaling, based on a single global cluster, achieves low calibration error with high bin coverage on DCF-Bug suggests that errors induced by different clusters are sufficiently similar to be captured by a shared $\sigma(z)$ function.
For this task, local Platt-scaling is only able to marginally reduce calibration error by fitting many slightly different calibrators.
In contrast, CR-Trans exhibited high calibration error under conventional Platt-scaling, which indicates that errors for the task do not neatly lie on a shared $\sigma(z)$ function.
For this task, local Platt-scaling is able to significantly reduce calibration error by fitting a few highly disparate calibrators.
We expect this to also be the case for the few clusters induced in DCF-Vul, as local Platt-scaling is able to significantly increase bin coverage in the task.

\textbf{To what extent do the optimal hyperparameters for local Platt-scaling vary?}
The X-axis and Y-axis for the scatter plots in Figure~\ref{fig:optimal_params} show the most optimal combination of hyperparameters for local Platt-scaling.
The former represents \textit{minimum cluster size}, whilst the latter represents \textit{minimum samples}.
The backoff strategy is denoted by the marker.
In general, we find that optimal hyperparameter combinations vary widely for DCF-Bug, minimally for DCF-Vul, and moderately for CR-Trans.
Whilst, there isn't a general preference for \textit{minimum samples}, we find that optimal values for \textit{minimum cluster size} in CR-Trans tends toward a smaller value (50-100), whilst the opposite is true for DCF-Vul (125-150).
In terms of backoff strategy, DCF-Vul only prefers the global calibrator.
Whilst this is the case against $EM$ for DCF-Bug, half of the cases against $EP$ prefer backing off to the uncalibrated confidence score.
For CR-Trans, all but four cases prefer the uncalibrated confidence score, indicating that outliers in the task deviate substantially from the shared $\sigma(z)$ function. 
This further corroborates previous findings of substantial error heterogeneity and covariate shift in automated code refinement.

\begin{figure}[h]
    \centering
    \includegraphics[width=\columnwidth]{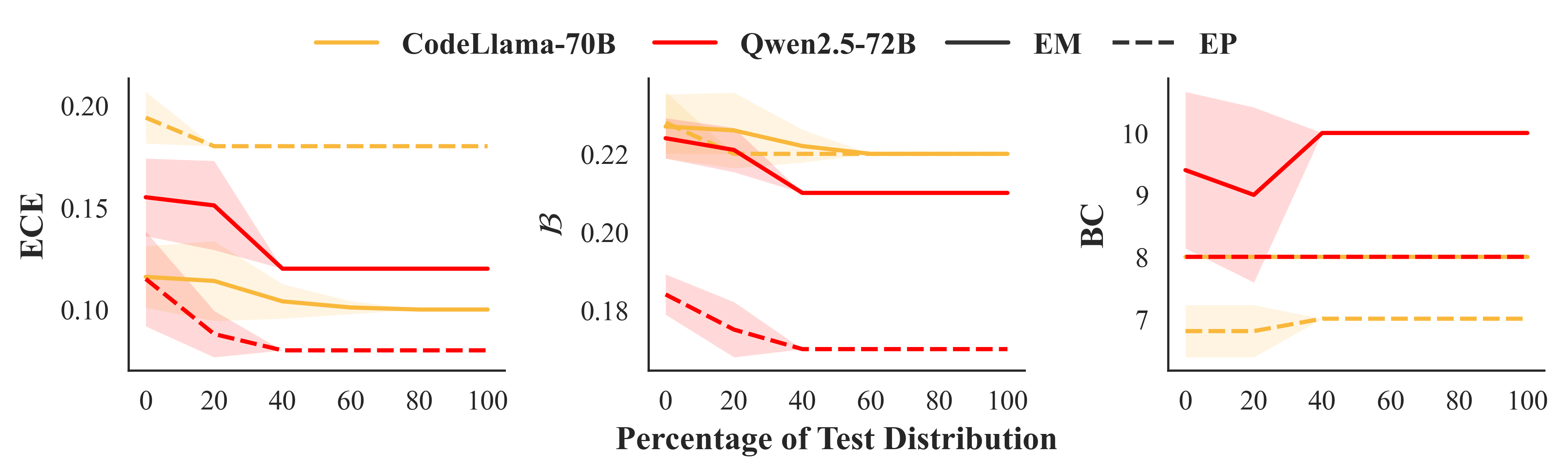} 
    \captionsetup{justification=centering} 
    \caption{Hyperparameter Search for Local Platt-Scaling w/ Minimum Token Probability using CR-Trans Test Samples}
    \label{fig:small_valid}
\end{figure}

\textbf{How much validation data is required to approximate the optimal hyperparameters for local Platt-scaling in automated code refinement?}
As established, the open-ended nature of CR-Trans may induce covariate shift, necessitating local Platt-scaling. 
Since the optimal hyperparameters are unknown a priori, we examine how many samples from the test distribution are required to obtain an accurate approximation.
Figure~\ref{fig:small_valid} shows hyperparameter search experiments for local Platt-scaling with minimum token probability using various percentages of the test distribution.
We focus on two of the most performant models in CR-Trans, CodeLlama-70B and Qwen2.5-72B.
We explore validation subsets between 0-100\% (step=20) of the full test set, where 0\% requires random hyperparameter selection.
The lines represent the mean calibration result across 10 resamples, whilst the shaded region represents the standard error.
In general we find that the optimal $ECE$, $\mathcal{B}$, and $BC$ can be consistently approximated within 20-40\% of the new test distribution, after which the results completely stabilise.
This indicates that optimal calibration can be achieved using only early samples from the new test distribution, and that the hyperparameter selection is stable.

\textbf{What is the recommended confidence calibration strategy for practical deployment in ACR tasks?}
For practitioners requiring well-calibrated confidence scores from their LLMs, selecting the most appropriate fine-grained strategy depends on the specific ACR task they are targeting and their tolerance for added latency.
For those targeting automated program repair in the setup of DCF-Bug, minimum token probability with global Platt-scaling is generally sufficient.
This prescription is also broadly applicable to those targeting vulnerability repair in the setup of DCF-Vul.
The local Platt-scaling method becomes necessary when achieving the lowest possible calibration error is critical, such that even a $\Delta0.02$ increase in ECE is intolerable.
Another scenario is in vulnerability repair, where the practitioner's focus is on ensuring a wider coverage of probability intervals for $EP$.
In both scenarios, the practitioner must accept increased latency.
Specifically, global Platt-scaling incurs only an additional logistic regression call per code revision, whereas local Platt-scaling further requires a forward pass through the Qwen3-Embedding-8B model and a cluster assignment before the logistic regression.
We note that, given the relatively small size of the embedding model, local Platt-scaling remains significantly faster than approaches like P(True)~\cite{kadavath2022language,tian-etal-2023-just,lin2022teaching} or stochastic sampling~\cite{chung2024sampling}.
For practitioners interested in automated code refinement in the style of CR-Trans, combining minimum token probability with local Platt-scaling is required; without it, the resulting confidence scores are severely miscalibrated.

\section{Case Study on Downstream Utility}
\label{sec:case_study}
To illustrate the practical benefits of fine-grained confidence calibration, we evaluate its impact on a representative downstream application: reliable automated code revision through abstention policies~\cite{ZablotskaiaPMN023}.
Specifically, we suppress code revisions whose associated confidence falls below a pre-selected threshold and measure the precision of the retained revisions.
Unlike calibration, which assesses general agreement between predicted confidence and empirical likelihood of correctness, abstention evaluates whether confidence estimates can reliably support selective deployment decisions. 
Since thresholding cumulatively retains all estimates above the given rate, a well-calibrated confidence score should provide developers with a reliable approximation of the lower-bound precision to expect.
To reflect realistic deployment scenarios, we ground the analysis on the best-performing model for each task.

\begin{figure}[!htbp]
    \centering
    \includegraphics[width=\columnwidth]{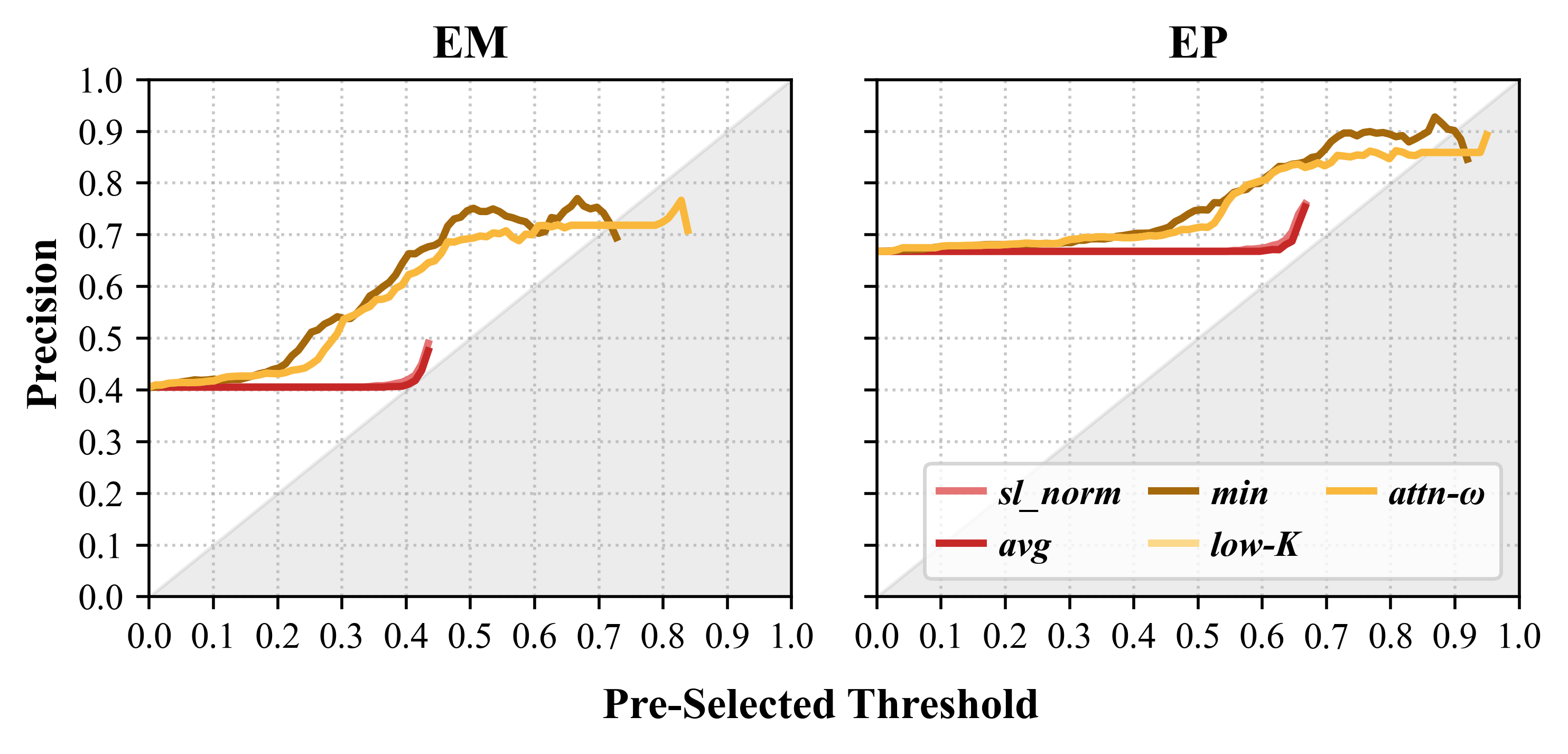} 
    \captionsetup{justification=centering} 
    \caption{Precision at Abstention Threshold for CodeLlama-70B in DCF-Bug}
    \label{fig:abstention_dcf_bug}
\end{figure}

\begin{figure}[!htbp]
    \centering
    \includegraphics[width=\columnwidth]{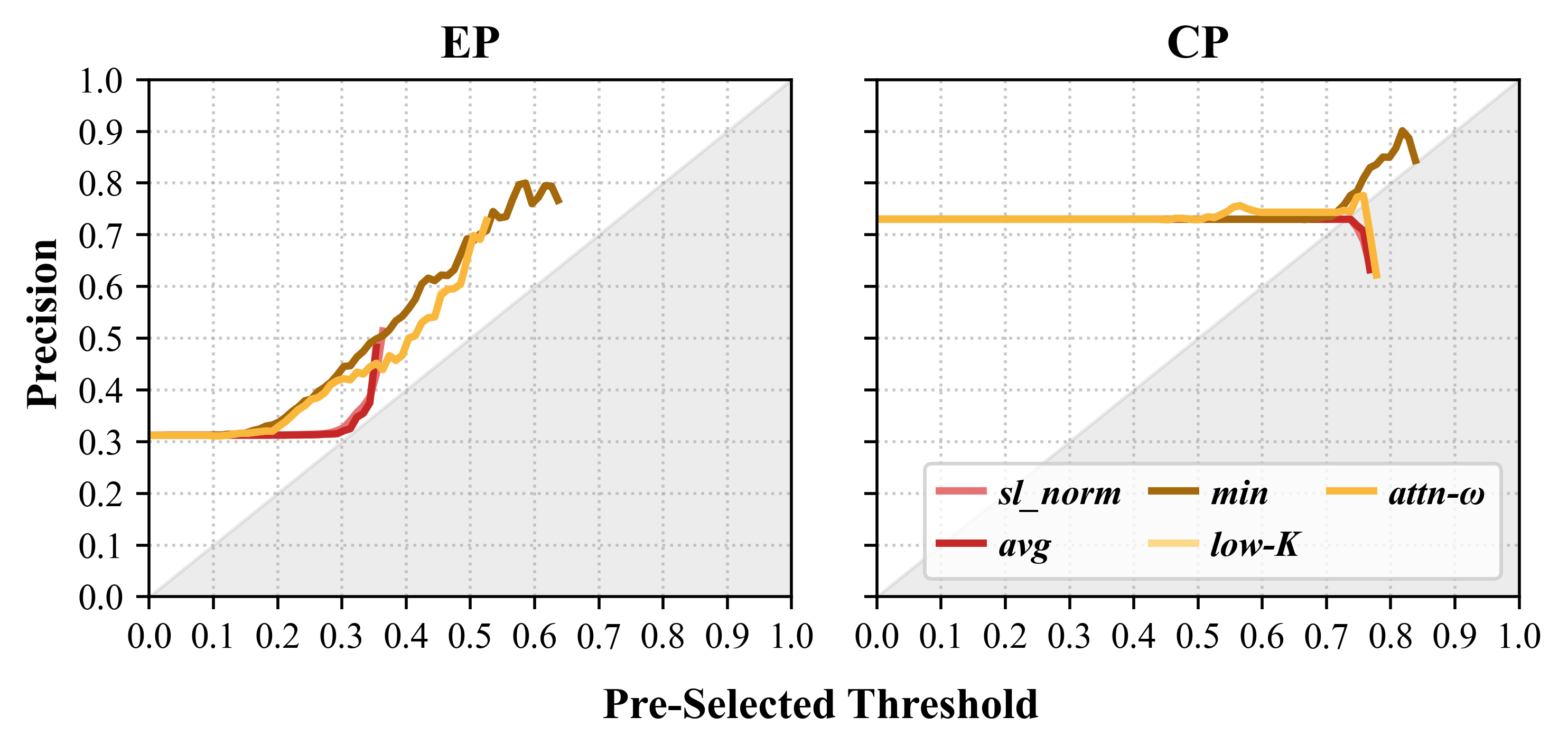} 
    \captionsetup{justification=centering} 
    \caption{Precision at Abstention Threshold for Llama-3.1-70B in DCF-Vul}
    \label{fig:abstention_dcf_vul}
\end{figure}

\begin{figure}[!htbp]
    \centering
    \includegraphics[width=\columnwidth]{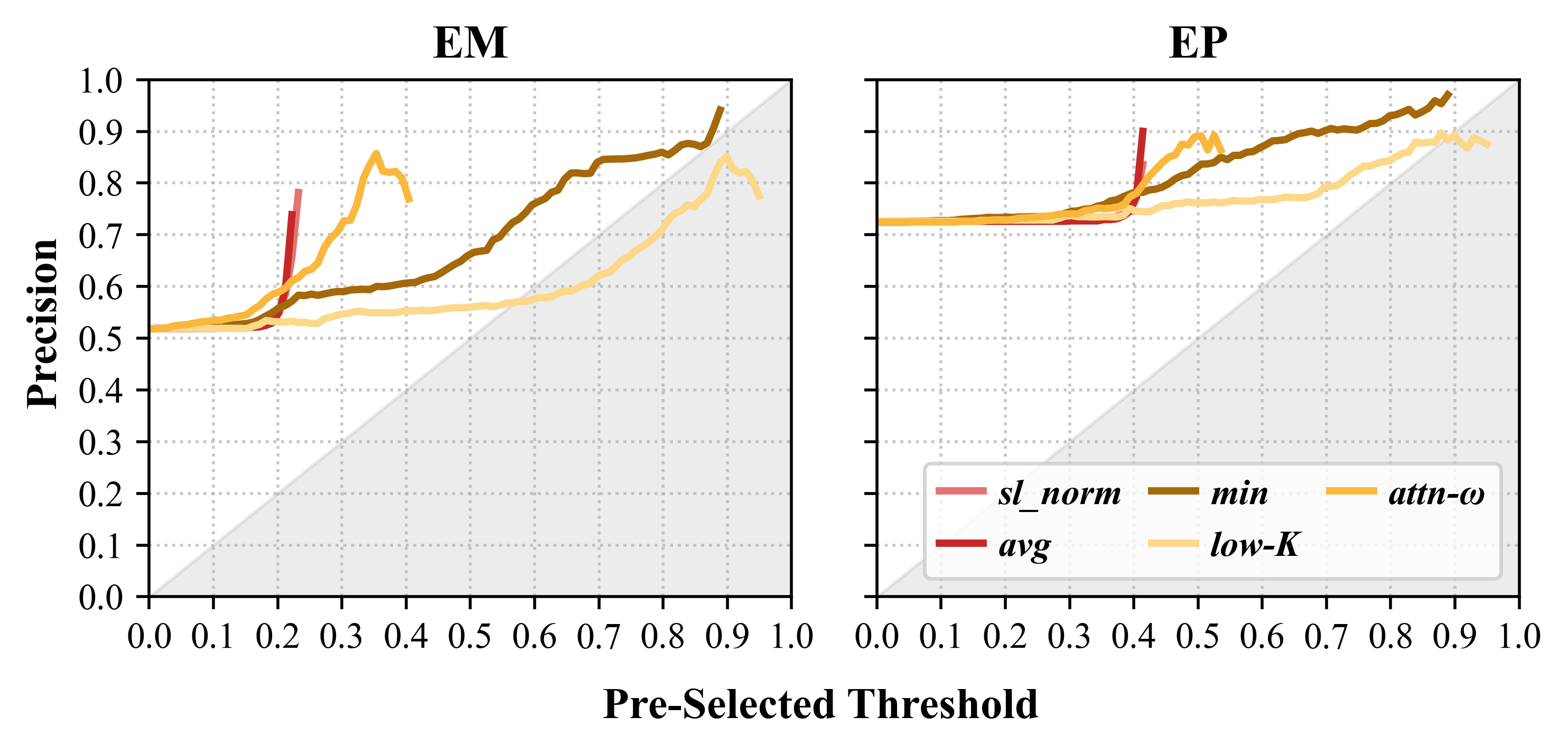} 
    \captionsetup{justification=centering} 
    \caption{Precision at Abstention Threshold for CodeLlama-70B in CR-Trans}
    \label{fig:abstention_cr_trans}
\end{figure}

Figure~\ref{fig:abstention_dcf_bug} compares the precision achieved at each simulated abstention threshold for code revisions generated by CodeLlama-70B for DCF-Bug. 
Similar trends were observed for both $EM$ and $EP$, where the sequence-level confidence scores were only able to maintain the approximate lower-bound precision up to thresholds of 0.44 and 0.68, respectively.
Subsequently, no code revisions were observed with confidence estimates exceeding these thresholds.
This highlights the limited practical utility induced by their low $BC$. 
In essence, simply never abstaining and statically assigning all code revisions with the model's overall accuracy as its confidence yields comparable results.
In contrast, using fine-grained confidence scores with local Platt-scaling can support thresholds up to 0.73 and 0.87-0.91, respectively; demonstrating wider and more useful operating ranges for threshold selection.

Figure~\ref{fig:abstention_dcf_vul} compares the precision achieved at each simulated abstention threshold for code revisions generated by Llama-3.1-70B for DCF-Vul.
In terms of $EP$ and $CP$, sequence-level confidence scores were only able to maintain the approximate lower-bound precision up to thresholds of 0.36-0.37 and 0.74, respectively.
In contrast, using fine-grained confidence scores with local Platt-scaling can support thresholds up to 0.54-0.65 and 0.77-0.85, respectively; minimum token probability offers the strongest advantage ($\Delta$0.11-0.28) for this task.

Figure~\ref{fig:abstention_cr_trans} compares the precision achieved at each simulated abstention threshold for code revisions generated by CodeLlama-70B for CR-Trans.
In terms of $EM$ and $EP$, sequence-level confidence scores were only able to maintain the approximate lower-bound precision up to thresholds of 0.23-0.24 and 0.42, respectively.
In contrast, using fine-grained confidence scores with local Platt-scaling can support thresholds up to 0.41-0.90 and 0.55-0.90, where minimum token probability consistently offers the widest operating ranges.

The results of this case study demonstrate that fine-grained confidence calibration, particularly minimum token probability combined with local Platt-scaling, provides practical downstream utility where coarse-grained conventional approaches prove inadequate. 
This is particularly relevant for developers with low tolerance for false-positive outputs from their automated software development tools~\cite{montes2025emotional,6606613,10.1145/2970276.2970347}.

\section{Threats to Validity}
\label{sec:threats}
We now discuss the threats to internal and external validity.

\textbf{Internal Validity.}
To mitigate dataset specific effects, we consider three different ACR tasks, each associated with distinct training and test sets. 
In addition, for each task we report results using two complementary correctness metrics.
To mitigate potential effects from data leakage, we either use test sets with non-permissive licenses or conduct semantic preserving code transformations coupled with natural language paraphrasing.
To address prior concerns about unstable results in small software engineering benchmarks~\cite{Spiess0PPRAJDA25} (120-164), we introduce larger training (1.6K–1.8K) and test sets (777–1K).
This is a prerequisite condition for reliable calibration calculation.
To ensure comparability and consistency across inference runs, we consider only zero-shot prompts with greedy decoding and deploy all models using native bfloat-16 precision.

\textbf{External Validity.}
To demonstrate the general applicability of our fine-grained calibration techniques across LLMs, we evaluate 14 open-source models drawn from three distinct series, spanning sizes from $\leq$8B to $\leq$72B parameters.
Given that many state-of-the-art models now have hundreds of billions to trillions of parameters~\cite{ren2023pangu,dubey2024llama,yang2025qwen3,liu2024deepseek}, we encourage researchers with sufficient computing resources to replicate our experiments using more powerful models.
Because our methods rely on token-level softmax probabilities with some requiring access to the attention mechanism, they are generally not applicable to most closed-source models that are only available through an API.
As such, we consider our fine-grained calibration methods as white-box approaches.
We focus on three main ACR tasks. 
Since our experiments are not exhaustive, replication studies on additional code revision tasks are needed before applying our approaches in production. 
For each task, we present only a specific problem setup due to limitations in the sizes of other recent benchmarks~\cite{repairbench,appatch}. 
We recommend follow-up studies for alternative task formats once sufficient non-contaminated data is available for facilitating reliable calculation of calibration error.
Whilst we explored three different correctness metrics, they remain imperfect proxies for true correctness due to the possibility of false positives and negatives. Therefore, future researchers should replicate our studies when more reliable proxies for correctness are available for these tasks.

\section{Conclusion}
\label{sec:conclusion}
In this study, we explored the use of fine-grained approaches for confidence calibration of LLMs in automated code revision tasks.
Our experiments encompassed three tasks, namely automated program repair, vulnerability repair, and automated code refinement; three correctness metrics, namely exact match, edit progress, and checks passed; and 14 different open-source models spanning across various sizes.
We find that fine-grained confidence scores, in particular the minimum token probability, consistently achieve both lower calibration error and greater coverage of probability intervals compared to traditional sequence-level confidence scores.
Additionally, we find that applying Platt-scaling with an ensemble of local calibrators can further improve over the conventional implementation that uses a single global calibrator.
For automated program and vulnerability repair, the selection between local and global Platt-scaling is contingent upon the user’s prioritisation of absolute calibration error reduction versus inference latency; in contrast, for automated code refinement, the application of local Platt-scaling is imperative to achieving an adequately calibrated confidence signal.
Looking forward, future research could further explore universal calibrators to support adaptive switching across different software engineering tasks~\cite{shen2024thermometer}, robust calibration for out-of-distribution scenarios~\cite{tomani2021post}, as well as confidence calibration across different reasoning paradigms such as chain-of-thought~\cite{cot}, tree-of-thoughts~\cite{yao2023tree}, and multi-step agentic setups~\cite{YaoZYDSN023}.
We hope this work encourages further research on confidence calibration in AI for software engineering and, more broadly, promotes the development of principled uncertainty quantification methods for reliable and trustworthy AI-driven software development tools.

\bibliographystyle{IEEEtran}
\bibliography{reference.bib}

\vfill
\end{document}